\RequirePackage{fix-cm}
\documentclass[twocolumn]{svjour3}          
\smartqed  
\usepackage{graphicx}
\usepackage{graphicx}
\usepackage{multirow}
\usepackage{pgfplotstable}
\usetikzlibrary{calc,math}
\usetikzlibrary{matrix,shapes,decorations}
\usetikzlibrary{shapes.geometric}
\usetikzlibrary{arrows, decorations.markings}
\usetikzlibrary{patterns}
\usepackage{adjustbox}
\usepackage{lipsum}
\usepackage{caption}
\usepackage{subcaption}
\usepackage{graphicx}
\usepackage{soul}
\usepackage{enumitem}
\usepackage{makecell}
\usepackage{microtype}
\usepackage{url}

\usepackage[utf8]{inputenc}
\usepackage{tcolorbox}
\usepackage{graphicx}
\usepackage{multirow,makecell}
\usepackage[ruled,vlined,linesnumbered]{algorithm2e}
\usepackage{subcaption}
\usepackage{booktabs}
\usepackage{pifont}
\usepackage{lipsum}
\usepackage{makecell}
\usepackage[para,online,flushleft]{threeparttable}
\usepackage{mathtools}
\usepackage{listings}
\usepackage{wrapfig}
\usepackage{balance}
\usepackage{adjustbox}
\usepackage{tabularx}
\usepackage{fancybox}
\usepackage[framemethod=tikz]{mdframed}
\usepackage{enumitem}
\usepackage{scalerel}
 
\usepackage[a-2b]{pdfx}
\usepackage{cleveref}

\usetikzlibrary{shadows}
\tikzset{every shadow/.style={opacity=1}}

\newcommand{\todo}[1]{\textcolor{orange}{\{\{#1\}\}}}

\newcommand{\BH}[1]{\textcolor{orange}{BH:~#1}}
\newcommand{\CL}[1]{\textcolor{brown}{CL:~#1}}
\newcommand{\MI}[1]{\textcolor{magenta}{MI:~#1}}

\newcommand{\review}[1]{#1}
\newcommand{\rev}[1]{\textcolor{black}{#1}}

\newcommand{\eat}[1]{}
\newcommand{\at}[1]{\protect\ensuremath{\mathsf{#1}}\xspace}
\newcommand{\stitle}[1]{\vspace{0.5ex}\noindent{\bf #1}}

\begin{document}
\title{Data Formats in Analytical DBMSs:\\Performance Trade-offs and Future Directions}




\author{Chunwei Liu
\and Anna Pavlenko
\and Matteo Interlandi
\and Brandon Haynes
}


\institute{Chunwei Liu \at
              MIT CSAIL \\
              \email{chunwei@csail.mit.edu}           
           \and
           Anna Pavlenko, Matteo Interlandi,  Brandon Haynes \at
              Microsoft
\email{annapa, 
mainterl,
brhaynes@microsoft.com}
}

\date{Received: date / Accepted: date}

\maketitle

\begin{abstract}
\eat{
\CL{GPT-3 writes the abstract from introduction:} \MI{These are really nice!}

\CL{v1}
In this paper, we evaluate the suitability of three popular formats for use as a storage format in a DBMS. We systematically identify and explore the high-level features that are important to support efficient querying in modern OLAP DBMSs, and evaluate the ability of each format to support these features. We find that each format has trade-offs that make it more or less suitable for use as a storage format in a DBMS, and identify opportunities for more holistic co-designing a unified in-memory and on-disk data representation.

\CL{v2}
This paper evaluates the suitability of three popular formats (Apache Arrow, Parquet, and ORC) to be used as a storage format in an analytical DBMS. We systematically identify and explore the high-level features that are important to support efficient querying in modern OLAP DBMSs, and evaluate the ability of each format to support encoding, compression, and transcoding for both real-world and synthetic datasets, and various data types. Our hope is that this study can be used as a guide for system developers using these formats, as well as provide the community with directions to pursue for improving these common open formats.

\MI{v3}}
This paper evaluates the suitability of Apache Arrow, Parquet, and ORC as formats for subsumption in an analytical DBMS. We systematically identify and explore the high-level features that are important to support efficient querying in modern OLAP DBMSs and evaluate the ability of each format to support these features. We find that each format has trade-offs that make it more or less suitable for use as a format in a DBMS and identify opportunities to more holistically co-design a unified in-memory and on-disk data representation. \rev{Notably, for certain popular machine learning tasks, none of these formats perform optimally, highlighting significant opportunities for advancing format design.} Our hope is that this study can be used as a guide for system developers designing and using these formats, as well as provide the community with directions to pursue for improving these common open formats. 

\end{abstract}

\newmdenv[shadow=true,shadowcolor=black,shadowsize=0pt,linewidth=1pt,skipabove=1pt]{highlightbox}


\section{Introduction}

Over the last decade, a number of new common and open formats have been proposed that purport to improve performance and ease interoperation across OLAP DBMSs and storage layers such as data lakes~\cite{10.1145/3035918.3056100}. 
Today, storage formats such as Parquet~\cite{parquet} and ORC~\cite{orc} are the cornerstone reference architectures for cloud-scale data warehousing systems~\cite{armbrust2021lakehouse}. 
At the same time, the in-memory format Apache Arrow~\cite{arrow} is widely considered to be the default means of interoperation across different data systems~\cite{rodriguez2021zero}, and several systems are even exploring how to leverage it end-to-end~\cite{Dremio}. 
Each of these in-memory and storage formats 
attempt to minimize disk, memory, and IO costs,
and each applies a wide variety of optimizations to maximize analytic read performance.


Though the benefits for common open formats are now well established~\cite{ivanov2020impact},
there has been less exploration of the relative benefits of these formats for \textit{direct subsumption} in an analytical DBMS as a native format. 
This is despite the robust discussion in database community about the relative merits of these formats for this purpose (e.g.,~\cite{abadiArrow,wesarrow,zeng2023empirical}).
One reason for this is that each format
makes design choices that optimize for its use as a common and open format, and these choices often conflict with longstanding analytical DBMS techniques.
For example,
we increasingly see
a push to directly leverage Arrow as an end-to-end, in-memory format within a DBMS~\cite{li2020mainlining}. 
At the same time, DBMS in-memory columnar formats typically
encode data~\cite{vertipaq,8187108} to minimize space and reduce memory requirements. However, Apache Arrow 
by default provides no encoding support, 
leaving it at odds with typical DBMS design.
On the other hand, on-disk formats such as Parquet arrange data in a form that is much closer to that found in modern columnar DBMSs (e.g., by employing run-length encoding mixed with dictionary encoding and bit packing).
However,
Parquet is widely leveraged 
only as storage format and exposes no
dedicated in-memory representation.  Instead, developers 
bring Parquet data into memory and convert it to the Arrow format, which, as stated above, is suboptimal for a columnar DBMS. 
Finally, a format such as ORC at first glance appears to offer  the best of both worlds, 
since it provides both an efficiently encoded data format and a related in-memory representation.  
Nevertheless, Arrow and Parquet are considered the standard nowadays because of their popularity in terms
of activity in open-source projects and support from big data frameworks and large-scale query providers~\cite{jainanalyzing}.

Given this environment,
the goal of this paper is 
to evaluate these three formats, explore their trade-offs, and evaluate their performance as candidates for direct subsumption in an analytical DBMS. 
Three main challenges exist in subsuming an in-memory format such as Arrow or traditionally on-disk formats such as Parquet or ORC.  First, a DBMS needs to be able to efficiently (de)serialize and (de)compress on-disk data to and from an in-memory representation.  For this, efficiency directly depends on format's compression ratio, decompression speed, and transcoding performance.
These trade-offs can be subtle and the line between an ``in-memory'' or ``on-disk'' format is often blurry.  For example, in some cases a DBMS could improve performance by writing Arrow to disk or directly operating on Parquet in memory, avoiding transcoding costs and taking advantage of the features offered by each format. \rev{Additionally, for specialized machine learning tasks, such as retrieval augmented generation (RAG) and $k$-nearest neighbor search, current formats require significant improvements across all dimensions to support these tasks effectively.}


Second, prior work has established that it is highly advantageous for a DBMS to ``push down'' computation as far as possible (e.g., to disk coprocessors or into the compressed domain) and to do so over as many data types as possible~\cite{gracia2019lamda,yang2021flexpushdowndb}.
Computation pushdown subsumes a number of related techniques.  \textit{Column pruning} and \textit{data skipping} respectively enable a DBMS to avoid decompressing columns or rows that do not contain data relevant to a query answer (e.g., when executing a range query by skipping data regions that do not contain data within the range).  Techniques such as \textit{direct querying}
enable a DBMS to retrieve query answers without an expensive decode or decompression step \cite{abadi2008column,abadi2006integrating,zhang2022compressdb}.
As we show in~\autoref{tab:format}, the ability for common in-memory and on-disk formats to support these techniques is uneven; to maximize performance a DBMS should optimize for the resulting trade-offs.

Finally, to maximize performance, modern DBMSs leverage modern techniques such as 
vectorized execution (e.g., SIMD)~\cite{li2013bitweaving,hentschel2018column,jiang2018boosting,wang2019accelerating,jiang2021good,liu2021decomposed}
and query compilation~\cite{gandiva}.
Here again the ability for a format to support these techniques is uneven.  For example, Parquet can vectorize operations over some query types whereas Arrow is inherently unencoded and more amenable to vectorized execution.  

In summary, in this paper we present the first detailed, empirical evaluation of three popular and increas\-ingly-adopted formats and evaluate their suitability to be used as a native format in a DBMS. Our hope is that this study can be used as a guide for system developers using these formats, as well as provide the community with directions to pursue for improving these common open formats.

\begin{table*}[t]
\centering
\caption{A comparison of the features found in common open columnar data formats.}
\label{tab:format}
\scriptsize{
\begin{tabular}{lcccc}
\toprule
\textbf{Feature} & \textbf{Arrow} & \textbf{Feather} & \textbf{Parquet} & \textbf{ORC} \\ \midrule
\textbf{Encoding Methods} & DICT & DICT & DICT(-RLE), RLE, BP, Delta & DICT, RLE, BP, Delta \\ \midrule
\textbf{Compression Codecs} & None & Zstd, LZ4 & Gzip, Snappy, Zstd, LZ4, (LZO) & Snappy, Zlib, LZ4 \\ \midrule
\textbf{Skipping} & Chunk-level & None & Record-level & Chunk-level \\ \midrule
\textbf{Direct Query} & None & None & None & None \\ \midrule
\textbf{Primary Purpose} & In-Memory Compute & On-Disk Storage & On-Disk Storage & On-Disk Storage \\ \midrule
\textbf{Representative Systems} & Dremio, Spark, Pandas & Pandas & Spark, Hive, Presto & Hive, Presto \\ 
\bottomrule
\end{tabular}
}
\end{table*}

Our contributions include:



\begin{itemize}

    

    
    \item We succinctly summarize the design nuances and distinctions of three widely-adopted open columnar formats: Apache Arrow, Parquet and ORC~(\autoref{sec:background-formats}).
    

    \item We systematically identify and explore the high-level features that are important to support efficient querying in modern OLAP DBMSs~(\autoref{sec:methodology}).

    \item For each format, we evaluate its ability to support efficient encoding, compression, and transcoding (\autoref{sec:eval-compression}) for both real-world, synthetic datasets, and various data types\rev{, including scalars and nested vectors}.

    \item We benchmark the ability of each format to support select-project (SP) operations found near the leaves of query plans.  We evaluate these in isolation~(\autoref{sec:eval-micro}) and in combination (\autoref{sec:eval-api}) using TPC-DS query plan fragments and over various data types.
    


    \item We evaluate the ability for each format to take advantage of recent trends such as vectorization, query compilation, and direct querying~(\autoref{sec:eval-opt}).

    \item \rev{We demonstrate that 
    each format's ability to operate on machine learning vector embeddings in terms of compression and (de)serialization is inadequate
    and suggest improvements 
    to better-balance compression performance and task accuracy.}

    \item We identify key opportunities to holistically co-design a unified in-memory and on-disk data representation. 
    
    
 
\end{itemize}
\vspace{-2ex}

\section{Background: Compression and Data Encoding}
\label{sec:background-encoding}

Data systems employ compression algorithms to reduce on-disk or in-memory data sizes and improve bandwidth utilization \cite{graefe1990data}. 
Conventional compression has traditionally focused on minimizing file size. 
This focus on size alone, while appropriate for storage, overlooks DBMS query execution performance~\cite{roth1993database,paparrizos2021vergedb,jiang2021good}. 
Conversely, in an analytical columnar DBMS, compressed size is usually balanced with the ability to query directly on the compressed data. 

\vspace{-1.5ex}
\subsection{Compression}
\label{sec:background-compression} 
Because of their generality, byte-oriented compression techniques (e.g., Gzip \cite{deutsch1996Gzip}, Snappy \cite{wiki:Snappy}, and Zlib \cite{gailly2004Zlib}) are widely used to reduce data size~\cite{abadi2006integrating,paparrizos2021vergedb}. They treat the input values as a byte stream and compress them sequentially. Byte-oriented compression is applicable to all data types and, in general, exhibits good compression ratios~\cite{jiang2021good}.  However, these methods are computationally intensive~\cite{abadi2006integrating}. A data block needs to be fully decompressed before individual  values can be accessed. 
This often introduces unnecessary overhead for query execution. 

\vspace{-1.5ex}
\subsection{Encoding}
\label{sec:background-encoding}

For decades, many data engines used row-oriented storage formats for OLTP query workloads~\cite{abadiArrow}. As more complex OLAP workloads became common, \textit{columnar storage formats} began to predominate \cite{abadi2008column}.
Columnar databases store data of the same type together \cite{shi2020column}, allowing systems to leverage \textit{lightweight compression}, also referred to as \textit{encoding}~\cite{abadi2006integrating}. Encoding methods such as \textit{dictionary encoding}, \textit{run-length encoding}, and \textit{bit-packed encoding} are typically designed to compress a specific type of data, enabling efficient compression and better record-level access relative to the general-purpose compression approaches described in the previous section.

Some encoding methods also support direct querying and data skipping to improve query performance \cite{abadi2006integrating,jiang2021good,liu2022fast}. 
Systems such as Redshift~\cite{gupta2015amazon} and SQL Server~\cite{larson2011sql} support many lightweight compression approaches that reduce the storage cost of data; at the same time, they apply compression-specific optimizations to improve query execution performance. Previous research \cite{liu2019mostly,jiang2020pids,liu2024adaedge,boncz2020fsst,liu2021decomposed} has also demonstrated that, for specific datasets, good encoding achieves a
comparable compression ratio with far fewer CPU cycles than does byte-oriented compression algorithms.

We then provide an overview of several widely-used encoding algorithms discussed in this paper and highlight their applicable scenarios.


\stitle{Bit-Packed Encoding (BP)} works on numerical data. It finds the minimal number of bits needed to represent values and removes superfluous leading zeros. It works best when the target numbers have similar bit-width. 

\stitle{Dictionary Encoding (DICT)} works on all data types. It encodes each distinct entry with an integer key and bit-packs the integers. Dictionary encoding works best when the dataset has small cardinality and many repetitions. Queries on dictionary encoded data can be applied either on the fully decoded data or directly in the encoded domain after query rewriting using dictionary translation.

\stitle{Run-Length Encoding (RLE)} works on data with many consecutive repetitions. It replaces a run of the same value with a pair consisting of the value and how many times it is repeated. 

\begin{figure}
    \centering
    \includegraphics[trim={0 0 0 0},clip,scale=0.50]{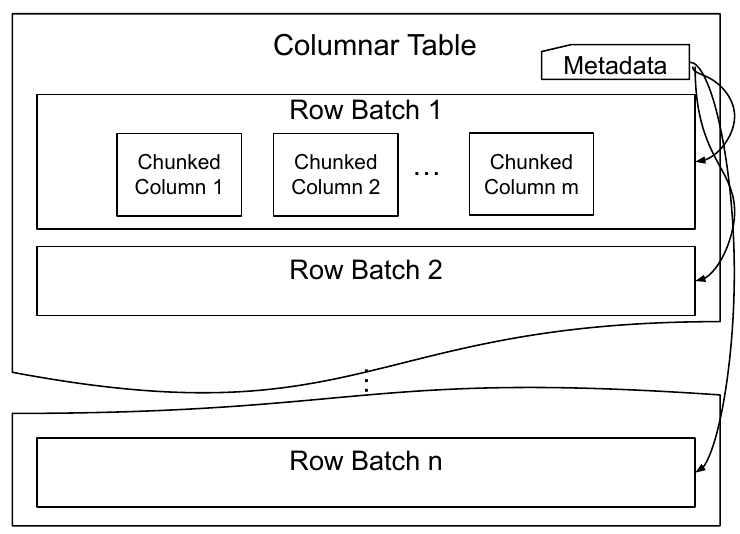}\vspace{-2ex}
    \caption{Columnar format layout.}
    \vspace{-5ex}
    \label{fig:arrow}
\end{figure}

\begin{figure*}
\begin{subfigure}[b]{0.33\textwidth}
\centering
\includegraphics[scale=0.60]{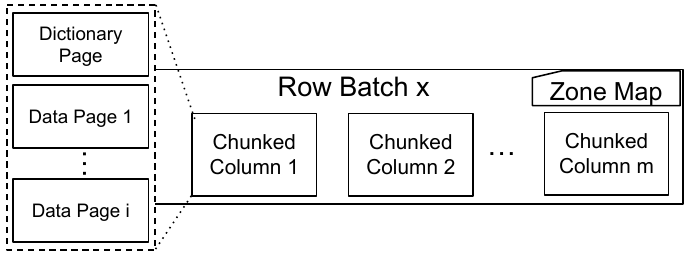}
\end{subfigure}
\hspace{2em}
\begin{subfigure}[b]
{0.66\textwidth}
    \centering
    \includegraphics[trim={0 0 0 0},clip,scale=0.60]{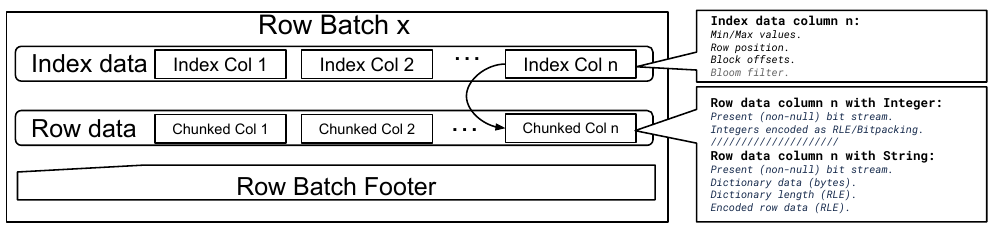}
\end{subfigure}
\\
  \begin{minipage}[t]{0.4\textwidth}
\centering
\caption{A Parquet row batch.}
    \label{fig:parquet-rg}
  \end{minipage}
  \begin{minipage}[t]{0.59\textwidth}
    \vspace{3ex}
    \caption{An ORC row batch.}
    \label{fig:orc-stripe}
  \end{minipage}
\end{figure*}

\stitle{Hybrid Encodings} 
are derived from the above encoding techniques. Dictionary run-length encoding (\textbf{DICT-RLE}) applies RLE  on the dictionary encoded keys to further compress data. Bit-packed and run-length hybrid encoding is used as a default implementation for the Parquet RLE encoder. Hybrid encoding usually achieves better compression performance
at the cost of performance.

\section{Columnar Open Formats}
\label{sec:background-formats}

 In big data environments today, there are many optimized data formats for columnar data storage and computation, as we show in \Cref{tab:format}. Interestingly, these data formats share the same basic underlying design. Therefore, we begin by providing a ``generic'' architecture that summarizes the substantial commonality in modern columnar data format design (\autoref{sec:generic-columnar}). Then, we drill into the idiosyncrasies found in Arrow, Parquet and ORC (Sections~\ref{sec:arrow}, \ref{sec:parquet}, and \ref{sec:orc}, respectively). To ease the parsing for non-familiar readers, we take the liberty of providing a unified naming convention. Table ~\ref{tab:name-align} has the mapping between our naming and each format. 

\begin{table}
\vspace{1ex}
\small
\caption{Column format name convention mapping.}
\label{tab:name-align}
\centering
\begin{tabular}{lll}
\toprule
\textbf{} & \textbf{Row Batch} & \textbf{Chunked Column}         \\ \midrule
\textbf{Arrow}   & Record Batch         & Chunked Array         \\ \midrule
\textbf{Parquet} & Row Group            & Column Chunk            \\ \midrule
\textbf{ORC}     & Stripe               & Row Column            \\ \bottomrule
\end{tabular}
\vspace{-3ex}
\end{table}



\subsection{Open Columnar Formats 101}
\label{sec:generic-columnar}

\autoref{fig:arrow} summarizes a generic columnar format design.
Columnar storage formats physically 
arrange data such that
all the records belonging to the same column are stored sequentially.
To achieve better data access at scale, columnar formats partition columns into \emph{chunks}. 
\emph{Chunked columns} are not created arbitrarily; instead, row-level alignment is attained by first splitting a table horizontally into \emph{row batches} where, within each batch, rows are then partitioned  into column chunks.
Metadata about the row batches (e.g., their location, number, length, compression algorithm, etc.) are stored either into the footer or in the preamble of the file.

\subsection{Apache Arrow (Feather)}
\label{sec:arrow}

Apache Arrow~\cite{arrow} is a columnar data structure supporting efficient in-memory computing.
Arrow can represent both flat and hierarchical data.
Arrow is designed to be complementary to on-disk columnar data formats such as Parquet and ORC, and in fact it shares with them the same design depicted in \autoref{fig:arrow}. 
On-disk data files are decompressed, decoded, and loaded into Arrow in-memory columnar arrays.
Each row batch has a default size of 64K rows. 
Arrow column chunks have a \emph{present} bit-vector signaling whether a value is null (or not), and, for strings, optionally a dictionary.

The Arrow columnar format has some compelling properties: random access is $\mathcal{O}(1)$ for entries in the same chunked column, and each value cells are sequential in memory, so it's efficient to iterate over. 
Arrow also defines a binary serialization protocol for converting a collection of row batches that can be used for messaging, interprocess communication (IPC), and writing  blobs into storage. Deserializing an Arrow blob has effectively zero cost. 

Closely related to the Arrow format, \textit{Arrow Feather} \cite{feather} is a column-oriented binary disk-based format, leveraging the same IPC as the in-memory Arrow format.
Additionally, Feather adds dictionary encoding (for strings) and compression (Zstd, LZ4).
Datasets stored in Arrow Feather are loaded in-memory as \textit{Arrow Tables}.


\vspace{-1ex}
\subsection{Parquet}
\label{sec:parquet}








 Parquet~\cite{parquet} is a columnar-oriented storage format inspired by the nested data storage format outlined in Google Dremel \cite{melnik2010dremel}. 
 Parquet integrates many efficient compression and encoding approaches to achieve space-efficiency.
 A Parquet file is structured almost exactly as described in \autoref{sec:generic-columnar}; however, as illustrated in \autoref{fig:parquet-rg}, each column chunk is partitioned into a \textit{dictionary page}
 and series of \textit{data pages}. Its file footer additionally contains \textit{zone maps} (e.g., min, max, and number of NULLs) at the row batch, chunked column, and data page level.  This enables efficient data skipping.
 Row batches have a recommended size of 512-1024 MB.
 Parquet applies dictionary encoding per data page and falls back to plain encoding when a dictionary grows larger than a predefined threshold.
 Parquet is designed to be space and IO-efficient at the expense of CPU utilization for decoding. It does not provide any data structures for in-memory computing. 

\vspace{-1ex}
\subsection{ORC}
\label{sec:orc}
Optimized Row Columnar (ORC) \cite{orc} is a storage format designed for read-heavy analytical workloads. 
ORC files are organized as in \autoref{fig:arrow} where
 the default row batch size is 250 MB. 
 Differently than Parquet, as is shown in \Cref{fig:orc-stripe},
 ORC organizes columns into an \emph{index} that contains min/max values, bloom filters, etc., and \emph{row data} with a \emph{present} bit-vector indicating NULL entries.
 The chunked columns in the row data are formatted based on the encoding type.

 ORC exposes a corresponding in-memory format, which contains a row-level index and NULL bit-vector data structures 
 for fast querying and NULL checks. ORC supports dictionary encoding (at the row batch level) for string data. Similar to Parquet, ORC falls back to plain encoding when the 
 number of distinct values is greater than a threshold (e.g., for Hive, 80\% of the records).

\begin{table}[h]
\centering
\scriptsize
\caption{Default encoding by format and data type. 
}
\label{tab:enc-compare}

\begin{tabular}{lccc}
\toprule
\textbf{} & \textbf{Parquet} & \textbf{Arrow} & \textbf{ORC} \\ 
\midrule
\textbf{Integer} & DICT(-RLE)$^\ast$ & None & RLE \\ 
\midrule
\textbf{Double} & DICT(-RLE)$^\ast$  & None & None  \\ 
\midrule
\textbf{String/Binary} & DICT(-RLE)$^\ast$ & DICT & DICT-RLE \\
\midrule
\textbf{Decimal} & DICT(-RLE)$^\ast$ & None & Int-RLE$^\dagger$ \\
\midrule
\rev{\textbf{Vector (Nested)}} &
\rev{Dremel~\cite{melnik2010dremel}} & \rev{${\sim}$Drill~\cite{diffDrillArrow}} & \rev{List$^\ddagger$~\cite{orc}} \\
\bottomrule
\vspace{2pt}
\end{tabular}
{\raggedright $\ast$ Parquet uses DICT in the latest C++ API, while DICT-RLE was used in its legacy Java API. \rev{$\dagger$ Int-RLE refers to the encoding where the decimal value is scaled to an integer and then encoded with RLE encoding. $\ddagger$ ORC encodes a vector using its List type, which includes a presence stream, a  length stream with RLE, and an element value stream that encodes based on the underlying type.}\par}
\end{table}

\subsection{Discussion}
Overall, Parquet and ORC provide the most comprehensive compression support for common data types, whereas Arrow Feather supports the fewest. ORC provides more auxiliary information for query execution (e.g., its zone map and support for bloom filters). Arrow Feather applies the same compression type to all arrays in the same record batch, whereas Parquet is more granular and allows compression to vary across column chunks. 
This flexibility enables intelligent encoding and compression selection based on the data features or workload characteristics \cite{jiang2021good}. As summarized in \autoref{tab:enc-compare}, each format applies different default encoding strategies. 

In terms of data access, both Arrow and ORC require data to be fully loaded into dedicated in-memory data structures (an Arrow Table 
 or ORC \texttt{ColumnVectorBatch}, respectively) before further
query execution can begin. On the other hand, Parquet exposes a \textit{streaming API} that allows pipelining data parsing and query execution, leading to more optimization opportunities.
However, Parquet does not itself provide any dedicated in-memory data structures.

\section{Methodology}
\label{sec:methodology}

In the subsequent sections, we benchmark the performance of Arrow, Parquet, and ORC over (non-nested) relational data.
This is not strictly an apples-to-apples comparison because each format was developed with a different use case in mind: Arrow eases the sharing of in-memory data across systems, Parquet is a generic on-disk format, and ORC is a storage format for relational big data systems.
Nevertheless, this comparison is important for evaluating the design choices (e.g., encoding method, compression, implementation decisions) made by each format and to understand the limitations and opportunities when using these formats in analytical DBMSs.

\subsubsection*{Dimensions.} To most fairly compare the formats, we evaluate each format across the following dimensions:
\begin{description}[leftmargin=0.25cm]
    \item{\textbf{1. Compression ratio.}} Each format applies different encoding methods and supports different compression algorithms.  The final achievable compression ratio is a result of these decisions, and so we evaluate each format using the variously supported encoding and compression algorithms (\Cref{sec:eval-ratio}).
    \item{\textbf{2. Transcoding throughput.}}
    While compression ratio alone is sufficient if we care only about minimizing disk or memory usage, this comes at the cost of having to compress, convert, and decompress (i.e., transcode) the data when accessing it (\Cref{sec:eval-transcode}).
    \item{\textbf{3. Data access.}} For each data type, what are the costs of accessing them? Data is often accessed by column (i.e., projected) or filtered using a predicate. 
    We evaluate the costs of each format when applying simple data access operations (\Cref{sec:eval-micro}).
    \item{\textbf{4. End-to-end evaluation over subexpressions.}}
    Since we care about the performance of each format when evaluating analytical queries, we explore the performance of each format over a set of query subexpressions drawn from TPC-DS (\Cref{sec:eval-api}).
    \item{\textbf{5. Advanced features.}}
    Given the many trade-offs baked into each format, we explore the extent to which we can extend them to support novel features such as computation pushdown into the encoded domain and hardware acceleration (\Cref{sec:eval-opt}).
\end{description}


\begin{table*}[h]
\caption{Evaluation overview and key results.} 
\small
\label{tab:eval-overview}
\begin{tabular}{llll}
\toprule
\textbf{Evaluation dimension}   & \textbf{Best Overall} & \textbf{Key Advantage}                                                       & \textbf{Section}                                        \\ 
\midrule
\textit{Compression ratio}        & Parquet            & Comprehensive encoding and compression support                           & \ref{sec:eval-ratio}                  \\ 
\midrule
\textit{Compression throughput}   & Arrow Feather         & Fast serialization                                                    & \ref{sec:eval-compression-overhead}   \\ 
\midrule
\textit{Decompression throughput} & Arrow Feather         & Fast deserialization                                                  & \ref{sec:eval-decompression-overhead} \\ 
\midrule
\rev{\textit{Vector transcoding}} & \rev{Arrow Feather}         & \rev{Fast (de)serialization}                                                  & \ref{sec:vector_trans} \\ 
\midrule
\textit{Projection evaluation}         & Parquet, ORC     & Fine-grained skipping while loading data                             & \ref{sec:eval-project}                \\ 
\midrule
\textit{Predicate evaluation}          & ORC                   & Fine-grained loading, dedicated in-memory representation & \ref{sec:eval-filter}                 \\ 
\midrule
\textit{Bitmap evaluation}             & ORC                   & Fine-grained loading, dedicated in-memory representation & \ref{sec:eval-bitvector-evaluation}   \\ 
\midrule
\textit{Subexpression evaluation}
  & ORC                   & \begin{tabular}[c]{@{}l@{}}Fine-grained loading, dedicated in-memory representation\\ and efficient skipping\end{tabular} & \ref{sec:eval-api}                    \\ 
\midrule
\textit{Direct querying}    & Parquet               & In-memory mapping with data skipping and direct querying                                                     & \ref{sec:eval-opt}      \\ 
\midrule
\textit{Vectorized execution}               & Parquet               & \makecell{In-memory mapping with data skipping, direct querying, SIMD}                                                 & \ref{sec:eval-opt}           \\ 
\bottomrule
\end{tabular}
\end{table*}



To summarize our findings, in \Cref{tab:eval-overview} we show the structure of our experiments and the overall best format for each dimension. 


\subsubsection*{Setup.}
All experiments are performed on an Azure Standard D8s v3 (8 vCPUs, 32 GiB memory), premium SSD LRS, and Ubuntu 18.04.
We test Apache Arrow 5.0.0, ORC 1.7.2, the Apache Parquet Java API version 1.9.0, \rev{and PyArrow 17.0.0}.  Where needed, we use the Apache Arrow C++ library to write in-memory Arrow tables to disk.
%
We perform experiments using (i) the TPC-DS dataset at scale 10, \review{(ii) the Join Order Benchmark (JOB) \cite{job}, (iii) the Public BI Benchmark (BI) \cite{public_bi_benchmark}, (iv) real-world datasets drawn from public data sources including GIS, machine learning, and financial datasets (CodecDB)~\cite{jiang2021good},} \rev{and (v) popular retrieval
augmented generation (RAG) and embedding-related datasets 
drawn from the Hugging Face community-curated collection \cite{huggingfacedatasets:online}.} 

%
%
%
For all the experiments, we report numbers when the system caches are cold by default. For selected experiments, we also report numbers when caches have been warmed up, i.e., to simulate frequently accessed datasets. Unless stated otherwise, 
we use each format's default settings.
Different results could certainly be obtained if dataset-specific parameter tuning were applied to each format.
\review{However, such fine-grained configuration tuning is beyond the scope of the paper and is left as future work.}

\section{Compression and Transcoding}
\label{sec:eval-compression}


In this section, we evaluate the compression performance of Arrow, Parquet, and ORC (\Cref{sec:eval-ratio}) and the related costs for transcoding data from compressed to in-memory formats (\Cref{sec:eval-transcode}).



\subsection{Encoding \& Compression Performance}
\label{sec:eval-ratio}

We begin by exploring the compression performance of each format through three sets of experiments. In the first experiment (\Cref{sec:eval-compression-real}) we evaluate how each format's supported encodings 
perform over a set of real-world datasets.
The last two experiments leverage \eat{synthetic} TPC-DS \cite{nambiar2006making} to illustrate
the performance of each format when compression is applied on top of encodings.
For the synthetic experiments on TPC-DS, we begin by evaluating
compression algorithms over the full dataset (\Cref{sec:eval-compression-tpcds}), and then explore 
how compression performance varies by data type (\Cref{sec:eval-compression-column}). \rev{Finally, we assess the compression performance of each format on real-world embedding datasets (\Cref{sec:embedding}).}

\subsubsection{Encoding Performance over Real-World Datasets.}
\label{sec:eval-compression-real}

\begin{table}[]
\centering
\vspace{1ex}
\caption{{Total size (in GB) by  format for columns in the CodecDB, BI, and JOB datasets. 
We serialize each column separately into each format and group their compressed size by data type. The raw dataset is in CSV format. For Arrow, we report with dictionary encoding enabled (Arrow DICT) and disabled (the default). We copy the file size (marked with~\textbf{*}) from the Arrow default column for CR computation as there is no dictionary support for integer and float types.
 }}
\vspace{-2ex}
\scriptsize{
\begin{tabular}{|l|@{\hskip 4pt}c@{\hskip 4pt}|@{\hskip 4pt}c@{\hskip 4pt}|@{\hskip 4pt}c@{\hskip 4pt}|@{\hskip 4pt}c@{\hskip 4pt}|@{\hskip 4pt}c@{\hskip 4pt}|@{\hskip 4pt}c@{\hskip 4pt}|}
\hline
\multirow{2}[1]{*}{}\textbf{Data} & \multirow{2}[1]{*}{}\textbf{\#} &
\multirow{2}[1]{*}{}\textbf{Raw} &\multirow{2}[1]{*}{}\textbf{Parquet} & \multirow{2}[1]{*}{}\textbf{ORC} & \multirow{2}[1]{*}{}\textbf{Arrow} & \multirow{2}[1]{*}{}\textbf{Arrow}  \\ 
\textbf{Type} & \textbf{Cols.} & \textbf{Size} & \textbf{Size} & \textbf{Size} & \textbf{Size} & \textbf{(DICT)} \\
\hline
\textbf{Integer} & 12k & \phantom{0}57.3 & \phantom{0}9.8 & \phantom{0}13.5 & \phantom{0}59.3 & \phantom{0}{{$59.3^*$}} \\ 
\hline
\textbf{Float} & \phantom{0}7k & \phantom{0}58.8 & 24.0 & \phantom{0}58.2 & \phantom{0}59.8 & \phantom{0}{{$59.8^*$}} \\ \hline
\textbf{String} & 13k & 373.5 & 31.0 & \phantom{0}62.2 & 403.4 & 118.3 \\ \hline
\textbf{Total} & 31k & 489.7 & 64.7 & 133.9 & 522.5 & 237.4 \\ \hline
\multicolumn{3}{|l|}{\textbf{Compression Ratio}} & 0.13 & \phantom{0}0.27 & \phantom{0}1.07 & \phantom{0}0.48 \\ \hline
\end{tabular}
}
\label{tab:bi-codec-job}
\vspace{-2ex}
\end{table}

\begin{table*}[]
\centering
\caption{\rev{Average and stddev compression ratios on each dataset by data type.}}
\label{tab:comp_avg_sd_sep}
\small{
\begin{tabular}{|>{\color{black}}c|>{\color{black}}c|>{\color{black}}c>{\color{black}}c|>{\color{black}}c>{\color{black}}c|>{\color{black}}c>{\color{black}}c|>{\color{black}}c>{\color{black}}c|}
\hline
\multirow{2}{*}{\textbf{Dataset}} & \multirow{2}{*}{\textbf{Type}} & \multicolumn{2}{>{\color{black}}c|}{\textbf{Parquet}} & \multicolumn{2}{>{\color{black}}c|}{\textbf{ORC}} & \multicolumn{2}{>{\color{black}}c|}{\textbf{Arrow}} & \multicolumn{2}{>{\color{black}}c|}{\textbf{ArrowDict}} \\ \cline{3-10} 
 &  & \multicolumn{1}{>{\color{black}}c|}{\textbf{AVG}} & \textbf{STD} & \multicolumn{1}{>{\color{black}}c|}{\textbf{AVG}} & \textbf{STD} & \multicolumn{1}{>{\color{black}}c|}{\textbf{AVG}} & \textbf{STD} & \multicolumn{1}{>{\color{black}}c|}{\textbf{AVG}} & \textbf{STD} \\ \hline
\multirow{3}{*}{\textbf{CODEC}} & String & \multicolumn{1}{>{\color{black}}c|}{0.256} & 0.359 & \multicolumn{1}{>{\color{black}}c|}{0.258} & 0.31 & \multicolumn{1}{>{\color{black}}c|}{1.585} & 0.692 & \multicolumn{1}{>{\color{black}}c|}{0.99} & 0.889 \\ \cline{2-10} 
 & Float & \multicolumn{1}{>{\color{black}}c|}{0.351} & 0.294 & \multicolumn{1}{>{\color{black}}c|}{1.27} & 1.099 & \multicolumn{1}{>{\color{black}}c|}{1.383} & 1.286 & \multicolumn{1}{>{\color{black}}c|}{} &  \\ \cline{2-10} 
 & Integer & \multicolumn{1}{>{\color{black}}c|}{0.291} & 0.27 & \multicolumn{1}{>{\color{black}}c|}{0.273} & 0.16 & \multicolumn{1}{>{\color{black}}c|}{1.241} & 0.964 & \multicolumn{1}{>{\color{black}}c|}{} &  \\ \hline
\multirow{3}{*}{\textbf{BI}} & String & \multicolumn{1}{>{\color{black}}c|}{0.074} & 0.148 & \multicolumn{1}{>{\color{black}}c|}{0.119} & 0.227 & \multicolumn{1}{>{\color{black}}c|}{1.391} & 0.474 & \multicolumn{1}{>{\color{black}}c|}{0.699} & 0.538 \\ \cline{2-10} 
 & Float & \multicolumn{1}{>{\color{black}}c|}{0.335} & 0.217 & \multicolumn{1}{>{\color{black}}c|}{1.558} & 0.886 & \multicolumn{1}{>{\color{black}}c|}{1.588} & 0.884 & \multicolumn{1}{>{\color{black}}c|}{} &  \\ \cline{2-10} 
 & Integer & \multicolumn{1}{>{\color{black}}c|}{0.2} & 0.162 & \multicolumn{1}{>{\color{black}}c|}{0.249} & 0.173 & \multicolumn{1}{>{\color{black}}c|}{1.567} & 0.522 & \multicolumn{1}{>{\color{black}}c|}{} &  \\ \hline
\multirow{2}{*}{\textbf{JOB}} & String & \multicolumn{1}{>{\color{black}}c|}{0.872} & 1.741 & \multicolumn{1}{>{\color{black}}c|}{0.923} & 1.227 & \multicolumn{1}{>{\color{black}}c|}{2.607} & 3.316 & \multicolumn{1}{>{\color{black}}c|}{2.557} & 4.791 \\ \cline{2-10} 
 & Integer & \multicolumn{1}{>{\color{black}}c|}{0.713} & 2.186 & \multicolumn{1}{>{\color{black}}c|}{0.432} & 1.038 & \multicolumn{1}{>{\color{black}}c|}{1.687} & 4.485 & \multicolumn{1}{>{\color{black}}c|}{} &  \\ \hline
\multirow{3}{*}{\textbf{ALL}} & String & \multicolumn{1}{>{\color{black}}c|}{0.207} & 0.343 & \multicolumn{1}{>{\color{black}}c|}{0.222} & 0.307 & \multicolumn{1}{>{\color{black}}c|}{1.535} & 0.677 & \multicolumn{1}{>{\color{black}}c|}{0.915} & 0.871 \\ \cline{2-10} 
 & Float & \multicolumn{1}{>{\color{black}}c|}{0.342} & 0.255 & \multicolumn{1}{>{\color{black}}c|}{1.427} & 0.999 & \multicolumn{1}{>{\color{black}}c|}{1.494} & 1.09 & \multicolumn{1}{>{\color{black}}c|}{} &  \\ \cline{2-10} 
 & Integer & \multicolumn{1}{>{\color{black}}c|}{0.247} & 0.267 & \multicolumn{1}{>{\color{black}}c|}{0.261} & 0.18 & \multicolumn{1}{>{\color{black}}c|}{1.408} & 0.837 & \multicolumn{1}{c|}{} &  \\ \hline
\end{tabular}}
\vspace{3ex}
\end{table*}

\begin{figure*}
     \centering\hspace{-2ex}
     \begin{subfigure}[t]{0.33\textwidth}
         \centering     
         {\includegraphics[width=\textwidth]{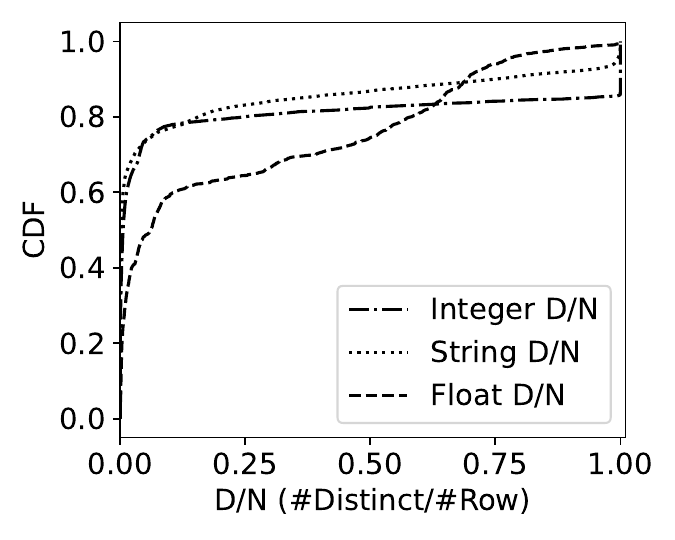}}
         \caption{\rev{CODEC (${\sim}18k$ columns)}}
         \label{fig:codec_cn}
     \end{subfigure}
     \begin{subfigure}[t]{0.33\textwidth}
         \centering
         {\includegraphics[width=\textwidth]{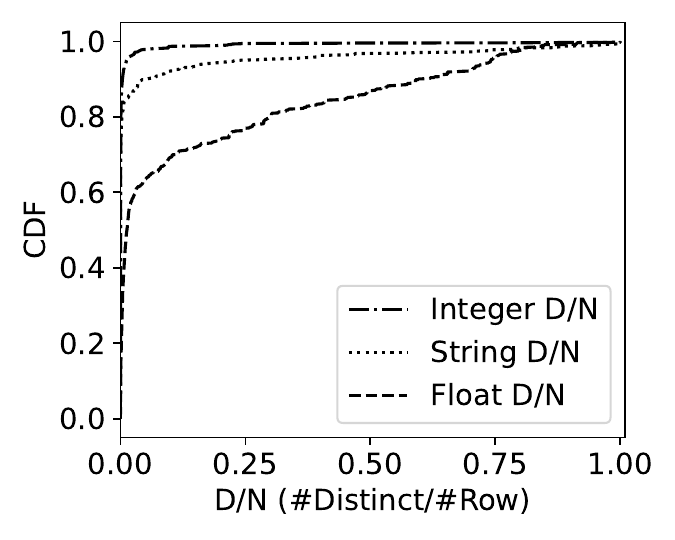}}
         \caption{\rev{BI (${\sim}13k$ columns)}}
         \label{fig:bi_cn}
     \end{subfigure}
     \begin{subfigure}[t]{0.33\textwidth}
         \centering
         {\includegraphics[width=\textwidth]{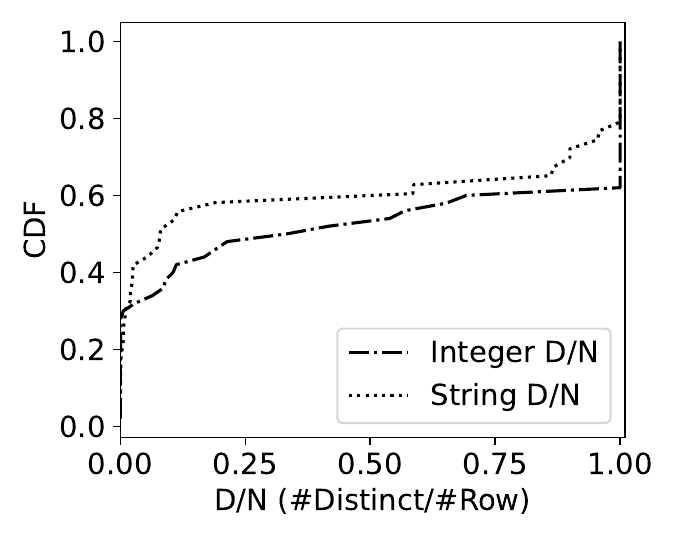}}
         \caption{\rev{JOB (${\sim}100$ columns)}}
         \label{fig:job_cn}
     \end{subfigure}
        \caption{\rev{Ratio of number of distinct values (\#Distinct) to the number of rows (\#Rows) in the CodecDB, Public BI and JOB datasets. The spikes for integer types near $D/N=1$ in the CODEC and JOB datasets occur because of primary key columns, which contain no duplicate values.
        }}\vspace{-2ex}
\label{fig:cn}
\end{figure*}

\begin{figure*}
     \centering\hspace{-2ex}
     \begin{subfigure}[t]{0.33\textwidth}
         \centering
         {\includegraphics[width=\textwidth]{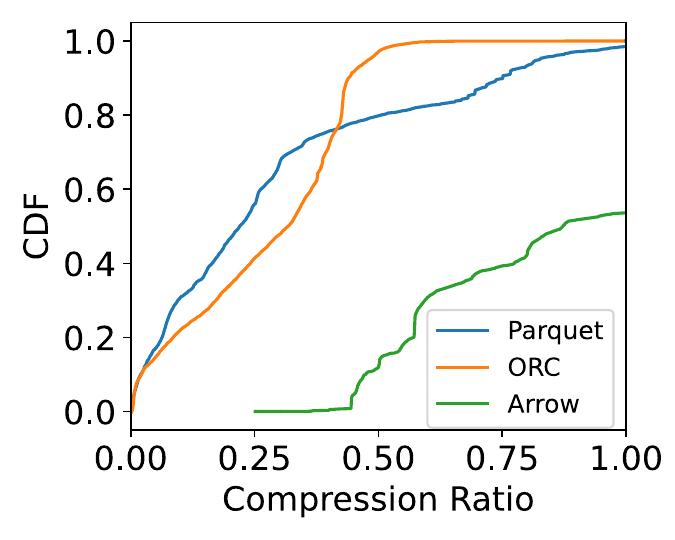}}
         \caption{\rev{Integer (Int32, Int64)}}
         \label{fig:codecinteger}
     \end{subfigure}
     \begin{subfigure}[t]{0.33\textwidth}
         \centering
         {\includegraphics[width=\textwidth]{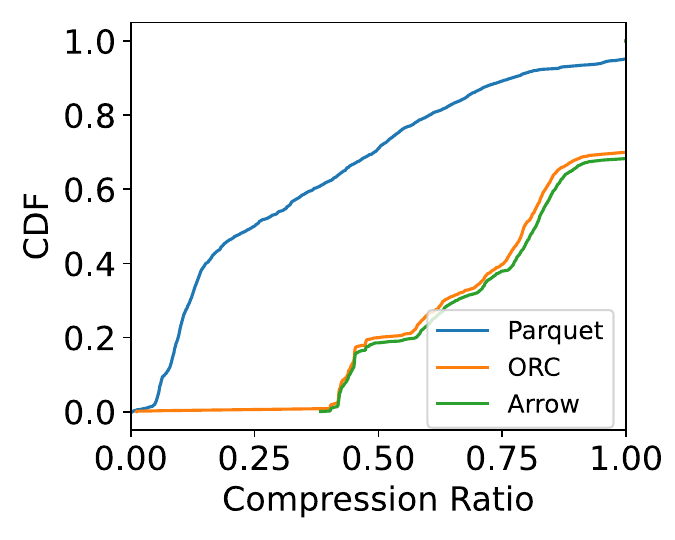}}
         \caption{\rev{Float (Float, Double)}}
         \label{fig:codec_comp-mem}
     \end{subfigure}
     \begin{subfigure}[t]{0.33\textwidth}
         \centering
         {\includegraphics[width=\textwidth]{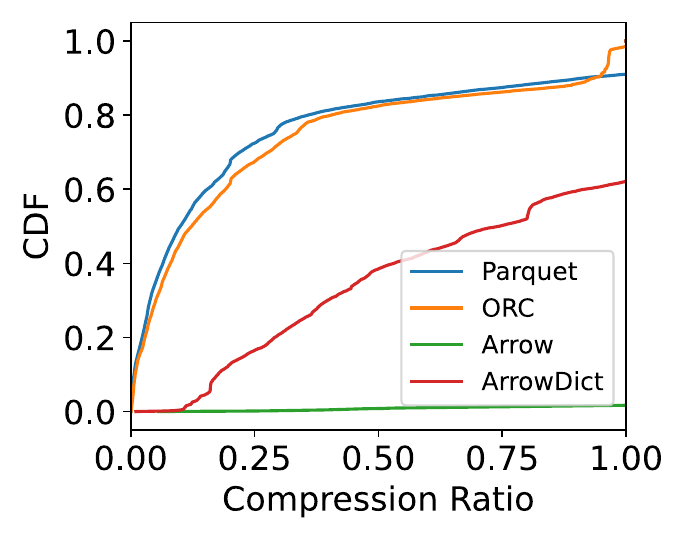}}
         \caption{\rev{String (Binary)}}
         \label{fig:codecstring}
     \end{subfigure}
        \caption{\rev{Compression ratios on the CodecDB real-world datasets with ${\sim}18k$ columns. The figures show the effective compression ratio (CR) in the range $[0,1]$. The CDF lines do not always reach $1.0$ at $CR=1.0$ because of underperforming compression on some columns.
        }} 
\label{fig:codec_cr}
\end{figure*}

\begin{figure*}
     \centering\hspace{-2ex}
     \begin{subfigure}[t]{0.33\textwidth}
         \centering
         {\includegraphics[width=\textwidth]{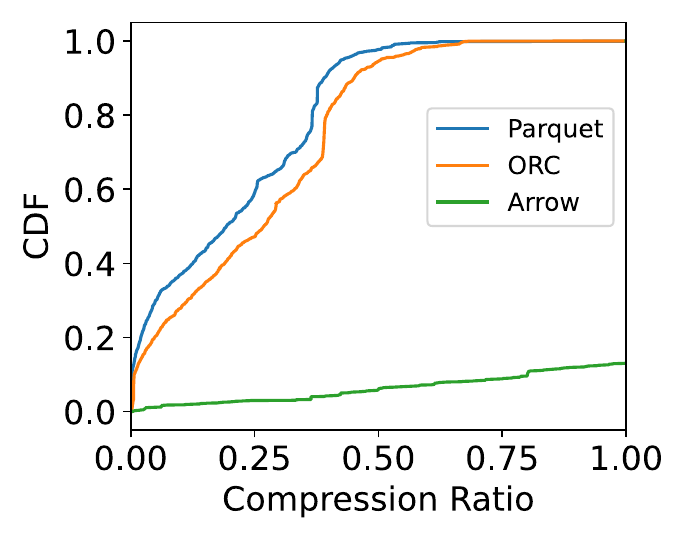}}
         \caption{\rev{Integer ({\tt int32}, {\tt int64})}}
         \label{fig:biinteger}
     \end{subfigure}
     \begin{subfigure}[t]{0.33\textwidth}
         \centering
         {\includegraphics[width=\textwidth]{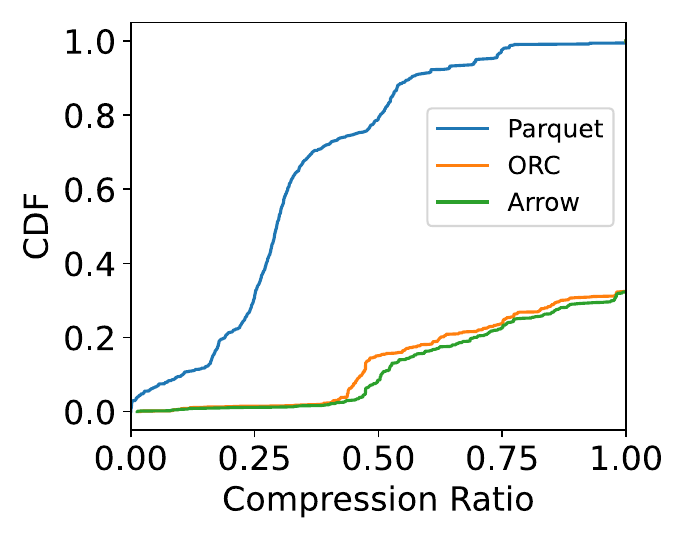}}
         \caption{\rev{Float (float, double)}}
         \label{fig:bicomp-mem}
     \end{subfigure}
     \begin{subfigure}[t]{0.33\textwidth}
         \centering
         {\includegraphics[width=\textwidth]{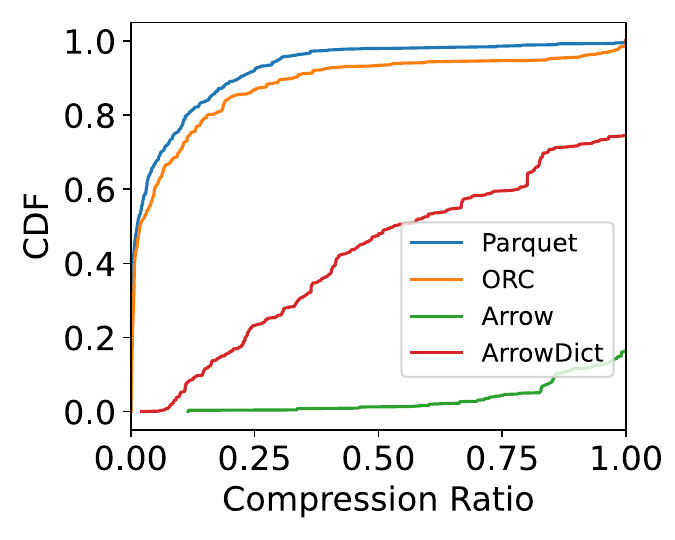}}
         \caption{\rev{String (binary)}}
         \label{fig:bistring}
     \end{subfigure}
        \caption{\rev{Compression ratio performance on the BI dataset over ${\sim}13k$ columns. }}
\label{fig:bi_cr}
\end{figure*}

\begin{figure*}
     \centering\hspace{-2ex}
     \begin{subfigure}[t]{0.33\textwidth}
         \centering
         {\includegraphics[width=\textwidth]{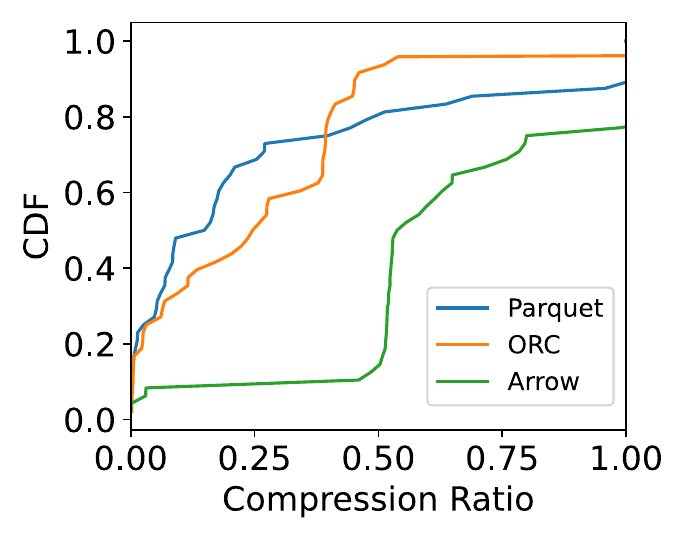}}
         \caption{\rev{Integer ({\tt int32}, {\tt int64})}}
         \label{fig:jobinteger}
     \end{subfigure}
     \hspace{6em}
     \begin{subfigure}[t]{0.33\textwidth}
         \centering
         {\includegraphics[width=\textwidth]{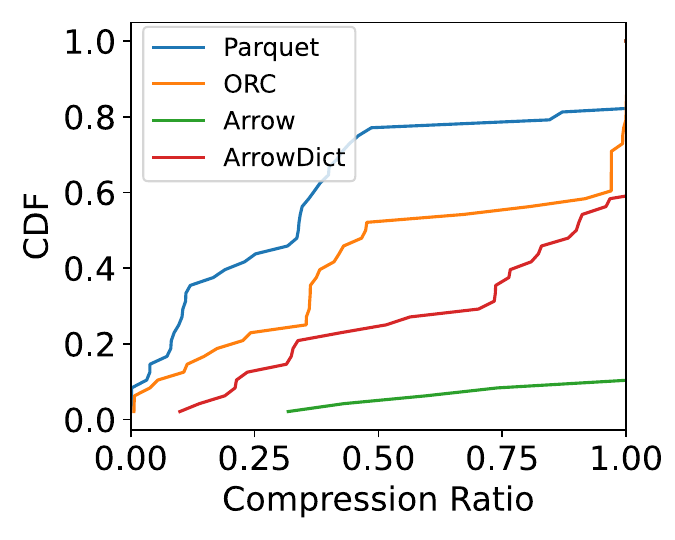}}
         \caption{\rev{String (binary)}}
         \label{fig:jobstring}
     \end{subfigure}
        \caption{\rev{Compression ratio performance on the JOB dataset with approximately ${\sim}100$ columns. (There are no floating-point columns in this dataset.). 
        }}
\label{fig:job_cr}
\end{figure*}

\begin{figure*}
    \centering
        \begin{subfigure}[b]{0.325\textwidth}
            \includegraphics[width=\textwidth]{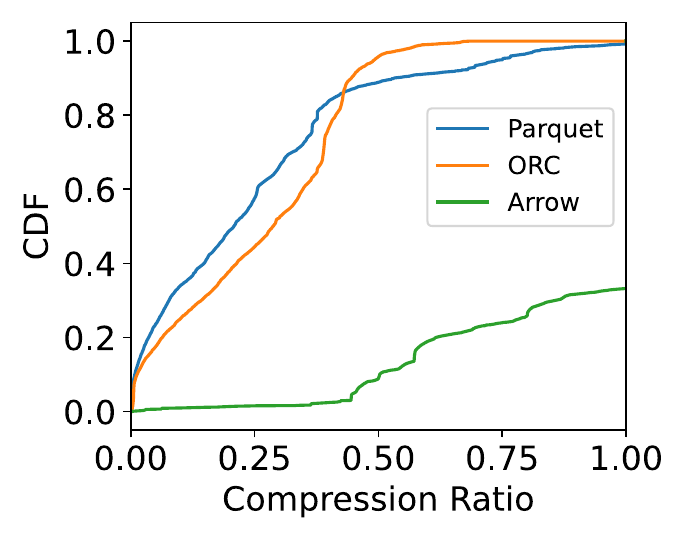}
            \caption{Integers}
            \label{fig:real_int}
        \end{subfigure}
        \begin{subfigure}[b]{0.325\textwidth}
            \includegraphics[width=\textwidth]{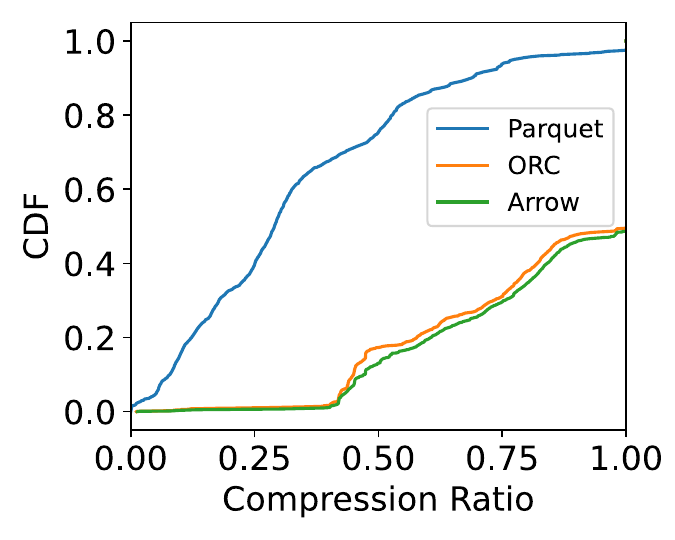}
            \caption{Floats}
            \label{fig:real_float}
        \end{subfigure}
        \begin{subfigure}[b]{0.325\textwidth}
            \includegraphics[width=\textwidth]{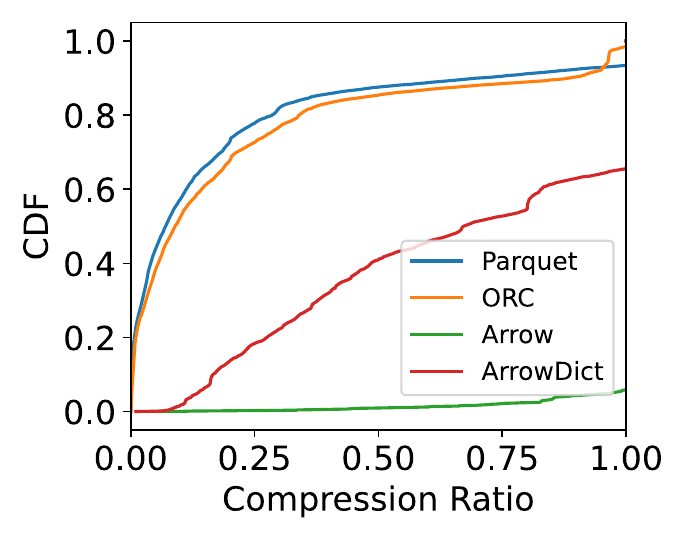}
            \caption{Strings}
            \label{fig:real_string}
        \end{subfigure}
        \caption{Column compression ratio CDFs over the CodecDB, BI and JOB datasets. 
        }
        \label{fig:cr_cdf}
\end{figure*}

\begin{figure}
  \centering
        \includegraphics[width=0.7\columnwidth]{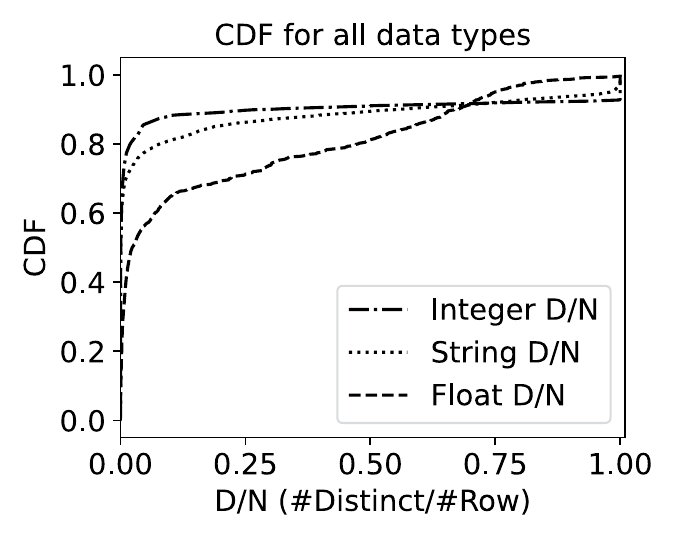}
        \caption{Distinct value CDFs.}
        \label{fig:real_cn}
\end{figure}

In this experiment, we group each data column by data type, 
convert each column one-by-one into each format, 
and finally aggregate the statistics of the compressed columns. \Cref{tab:bi-codec-job} shows the overall compression performance and statistics, \rev{grouped by data type,} over the ${\sim}31$k columns in the CodecDB, Public BI and JOB, datasets.  
\rev{\Cref{tab:comp_avg_sd_sep} provides a detailed view of the compression ratio for each dataset and data type.
To avoid confusion, we do not apply any further compression after default encoding techniques are applied
 (we will explore how each format behaves when compression is enabled in the following sections). 
}

\rev{We begin by showing detailed CDFs over the ratio of the number of distinct values to number of distinct rows (\Cref{fig:cn}).
As we can see, each dataset has a distinct CDF shape, confirming that our analysis covers data with varying characteristics.}
%
%
\rev{Next,
\autoref{fig:codec_cr} shows the compression ratio (CR) performance on the CodecDB real-world dataset.
For integers in this dataset (\autoref{fig:codecinteger}), dictionary+RLE encoding in Parquet performs well for low distinctness ratios, while RLE alone in ORC works relatively well for high distinctness ratios for integer columns. This is the reason there is a crossover point between Parquet and ORC. For floats (\autoref{fig:codec_comp-mem}), Parquet outperforms both ORC and Arrow Feather due to its use of dictionary encoding. These formats perform similarly as they both use a plain encoding. For strings (\autoref{fig:codecstring}), both Parquet and ORC fall back to plain encoding on some columns when dictionary encoding takes up more space relative to plain encoding. But the two formats work differently in this case---Parquet's plain encoding introduces a higher space cost for saving the string length values, whereas ORC's plain encoding uses RLE for string length values. However, Parquet's dictionary encoding is more effective than ORC because of the extra layer of RLE for the dictionary-encoded keys. That is why Parquet works better in terms of total compressed size while ORC works better in terms of the number of effective compression ($CR<1$) on the CodecDB dataset.}

\rev{\autoref{fig:bi_cr} shows compression performance on the BI dataset.  In this datasets we observe that both the integer and string columns have a lower distinctness ratio than did CodecDB, so Parquet's dictionary encoding has better compression performance in this figure.  This is also seen in \autoref{tab:comp_avg_sd_sep}.}
\rev{
We next show compression performance over the JOB dataset in \autoref{fig:job_cr}.
For integers (\autoref{fig:jobinteger}, the Arrow line increases significantly at around 0.5 because Arrow uses plain encoding for integer values. The original text values of many integer columns (with millions of rows) take 8 bytes on average (e.g., "4167491\textbackslash n"), while the plain encoded integer values consume 4 bytes, resulting in ${\sim}0.5$ compression ratio.  String compression (\autoref{fig:jobstring}) does not exhibit this behavior.}

\rev{
We finally show the compression ratio CDFs for each data type in \Cref{fig:cr_cdf} where we focus on the effective compression ratio range $[0, 1]$. Overall, Parquet exhibits the best compression performance due to its efficient encoding methods.
\Cref{fig:real_cn} further illustrates the CDFs for each data type, taking into account the number of distinct values in each column.  
This analysis highlights the impact of value distinctiveness on compression performance, particularly for formats that rely on dictionary encodings.
}

\textbf{Discussion.} \review{Overall, as we can see in \Cref{tab:bi-codec-job}, overall, Parquet performs the best over the whole dataset and is able to reduce 
the size of the column data to about 13\% of the original.}
\review{ORC is able to compress the dataset to $\sim$27\%.} \review{By contrast, Arrow Feather---with default settings---exhibits a 7\% \textit{increase} in size compared with the raw text file. We found that this overhead is introduced by the format's metadata, which adds a four-byte length prefix to each variable binary entry (i.e., the string ``abc'' consumes seven bytes in total). 
It also pads numerical data types.
On the other hand, with DICT enabled, Arrow Feather compresses string columns by 68\% and the whole dataset by 52\%.
}
\rev{This is further supported by \autoref{fig:bistring}, where the string columns in the BI dataset have a lower distinctness ratio than those in the CodecDB dataset.
On the other hand, string columns in the JOB dataset have a relatively high number of distinct values, and thus dictionary encoding (i.e., Parquet and ORC) performs relatively poorly (also shown in \Cref{tab:comp_avg_sd_sep}).}

For integers, ORC exhibits varying compression performance relative to Parquet. 
ORC achieves a better compression ratio for the CodecDB and JOB datasets (which contain a relatively higher number of distinct values), while it is worse for the BI dataset (which has a lower number of distinct values). This is because ORC applies RLE for integer columns (see \Cref{tab:enc-compare}), which performs better for columns with fewer distinct values, whereas Parquet applies DICT-RLE, which is slightly worse. 
Because of this, we observe
a crossover point for the Parquet and ORC CDFs in \Cref{fig:real_int}.

\review{For floats, as we can see from \Cref{fig:real_float}, Parquet outperforms ORC and Arrow Feather because of dictionary encoding. ORC
and Arrow Feather perform similarly as they both use plain encoding. For strings, Parquet and ORC outperform Arrow. Interestingly, both Parquet and ORC fall back to plain encoding
on some columns when dictionary encoding takes up larger space than plain encoding, but their dictionary encoding work differently:
Parquet’s plain encoding introduces a higher space cost for saving the string length values, while ORC’s plain encoding uses
RLE for string length values. However, Parquet’s dictionary encoding is more effective than ORC because of the extra layer of
RLE for the dictionary-encoded keys. That is why Parquet works better in terms of total compressed size (see Tables \ref{tab:bi-codec-job} and \ref{tab:comp_avg_sd_sep}) while ORC works
better in terms of the effectiveness (compression ratio $<1$; see \Cref{fig:real_string}).
}

\eat{\CL{In the experiments above, we used DICT-RLE for Java version Parquet for better encoding supports. The following two subsection, we use DICT for C++ version Parquet (no way to enable DICT-RLE on C++ version).}}
\subsubsection{Compression Performance.}
\label{sec:eval-compression-tpcds}

\begin{figure}
\centering
\includegraphics[trim={0 4.5cm 0 0},clip,width=\columnwidth]{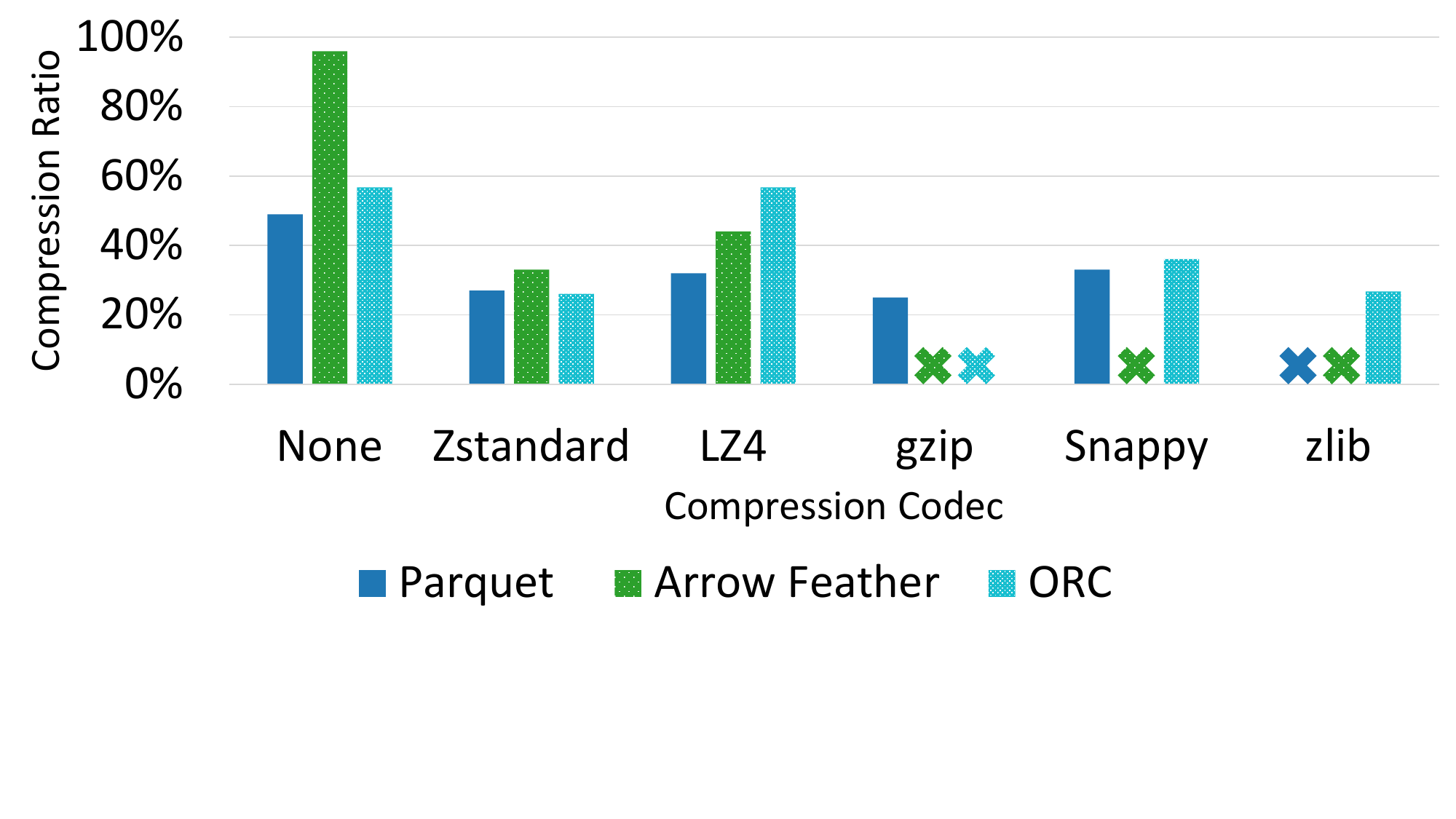}
\vspace{-3ex}
\caption{Compression ratio (compressed size / original CSV size) on TPC-DS (smaller is better). Uncompressed (None in the figure) only encodes using default settings. Not all formats support all compression algorithms.}
\vspace{-1ex}
\label{fig:tpcds-cr}
\end{figure}

For this experiment, we report the compression ratio of each format on the full TPC-DS dataset when different compression algorithms are applied.
We evaluate Zstandard (Zstd) at level 1 (we evaluate other levels later in this section), LZ4, Gzip, Snappy, and Zlib compression algorithms, and compare them against an uncompressed variant where data is encoded using the default settings. 
The results of this experiment are shown in \Cref{fig:tpcds-cr}.
In the uncompressed case, Parquet is about 2$\times$ better than Arrow Feather because Arrow Feather does not apply any encoding.  However, when compression is enabled, Arrow performs within $\sim$30\% of Parquet.
ORC achieves a similar compression ratio as Parquet, except under LZ4.  In this case, ORC automatically disables compression because it detects that the LZ4 compressed data size is greater than the original data size.

Finally, we observe that different compression algorithms yield different compression ratios. For example, increasing Zstd's level from 5 to 9 yields more aggressive compression and achieves smaller sizes. However, this gain is minimal ($< 1.5\%$) while the compression time increases by ${\sim}3\times$ for Arrow Feather and ${\sim}2\times$ for Parquet. 
We will show in \Cref{sec:eval-transcode} how decompression costs are also impacted by the choice of compression algorithm.

\begin{figure}
     \centering
     \begin{subfigure}[b]{\columnwidth} 
         \centering
         \includegraphics[trim={0 5.5cm 0 3.3cm},clip,width=\columnwidth]{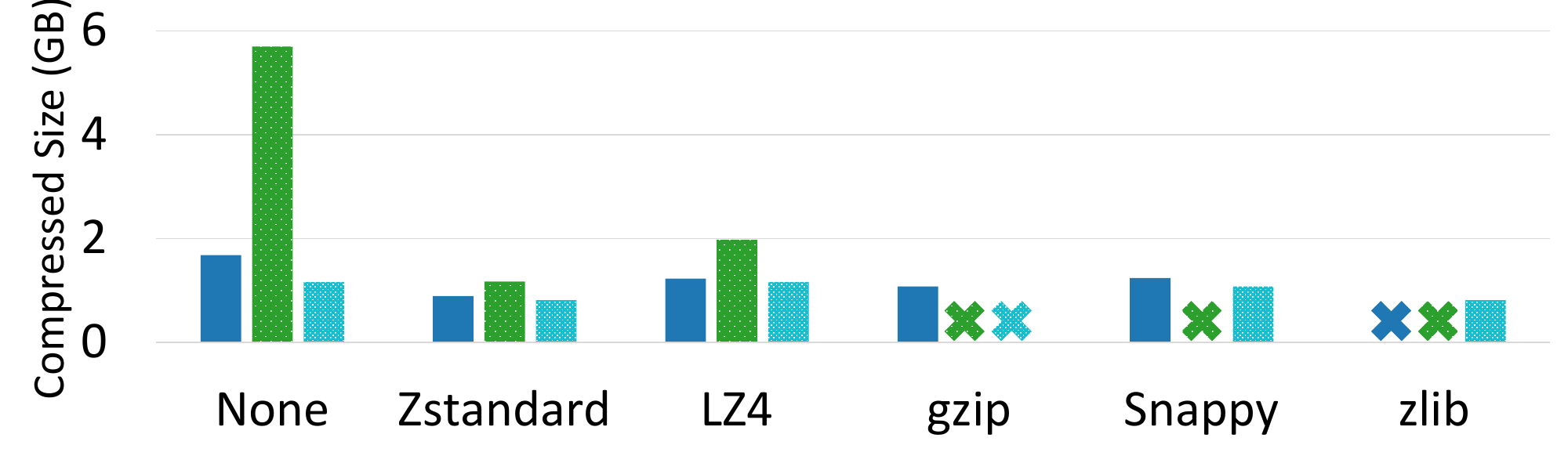}
         \vspace{-2ex}
         \caption{Integers (\texttt{int64} and \texttt{int32})}
         \label{fig:int-comp}
     \end{subfigure}
     
     \vspace{4ex}
     \begin{subfigure}[b]{\columnwidth} 
         \centering
         \includegraphics[trim={0 4.5cm 0 3.3cm},clip,width=\columnwidth]{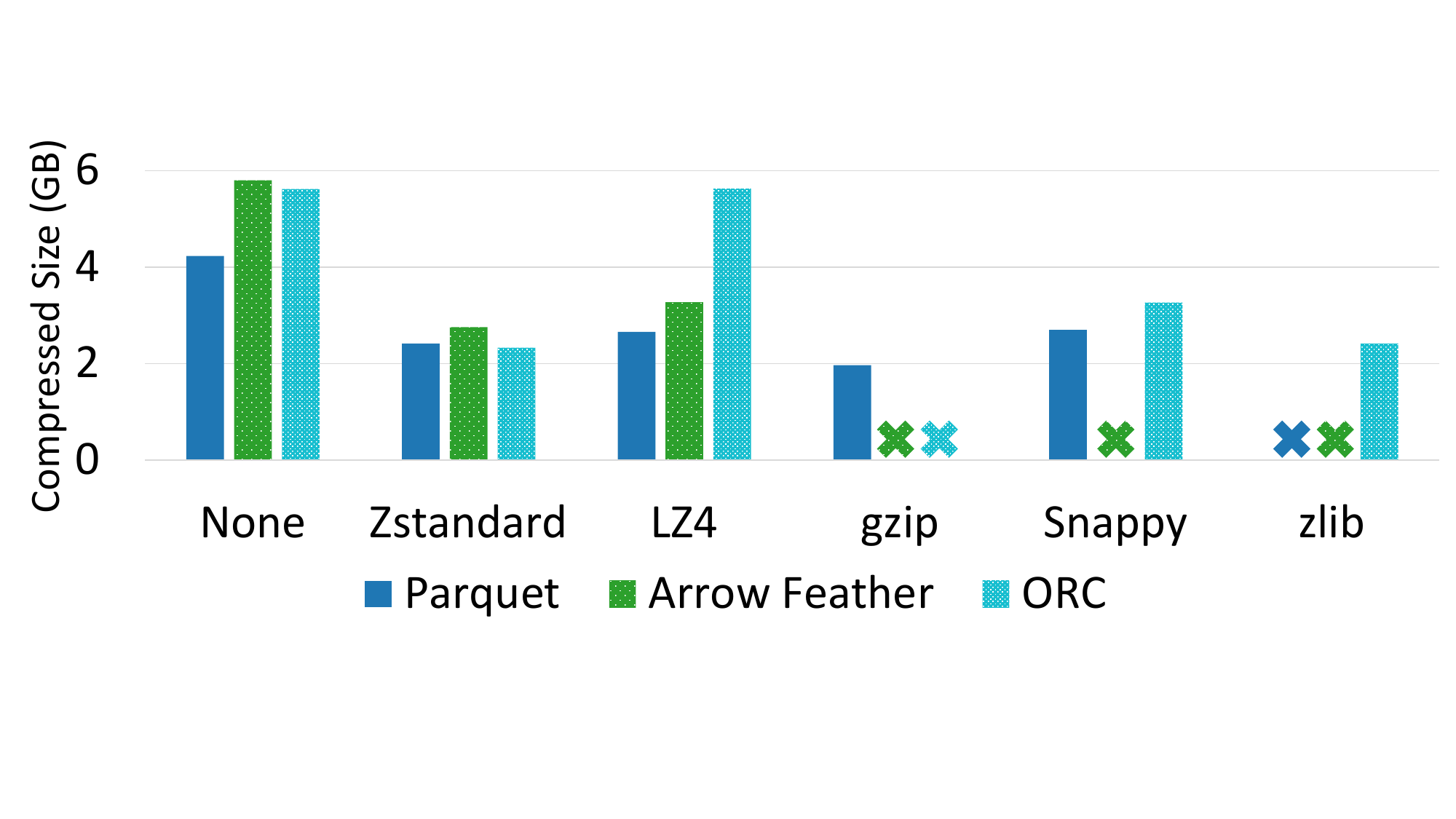}
         \caption{Doubles}
         \label{fig:double-comp}
     \end{subfigure}
        \caption{Total size on disk after compressing the numeric columns in TPC-DS.}
        \label{fig:type-comp-numeric}
\end{figure}

\begin{figure}
     \begin{subfigure}[b]{\columnwidth} 
         \centering
         \includegraphics[trim={0 4.5cm 0 3.3cm},clip,width=\textwidth]{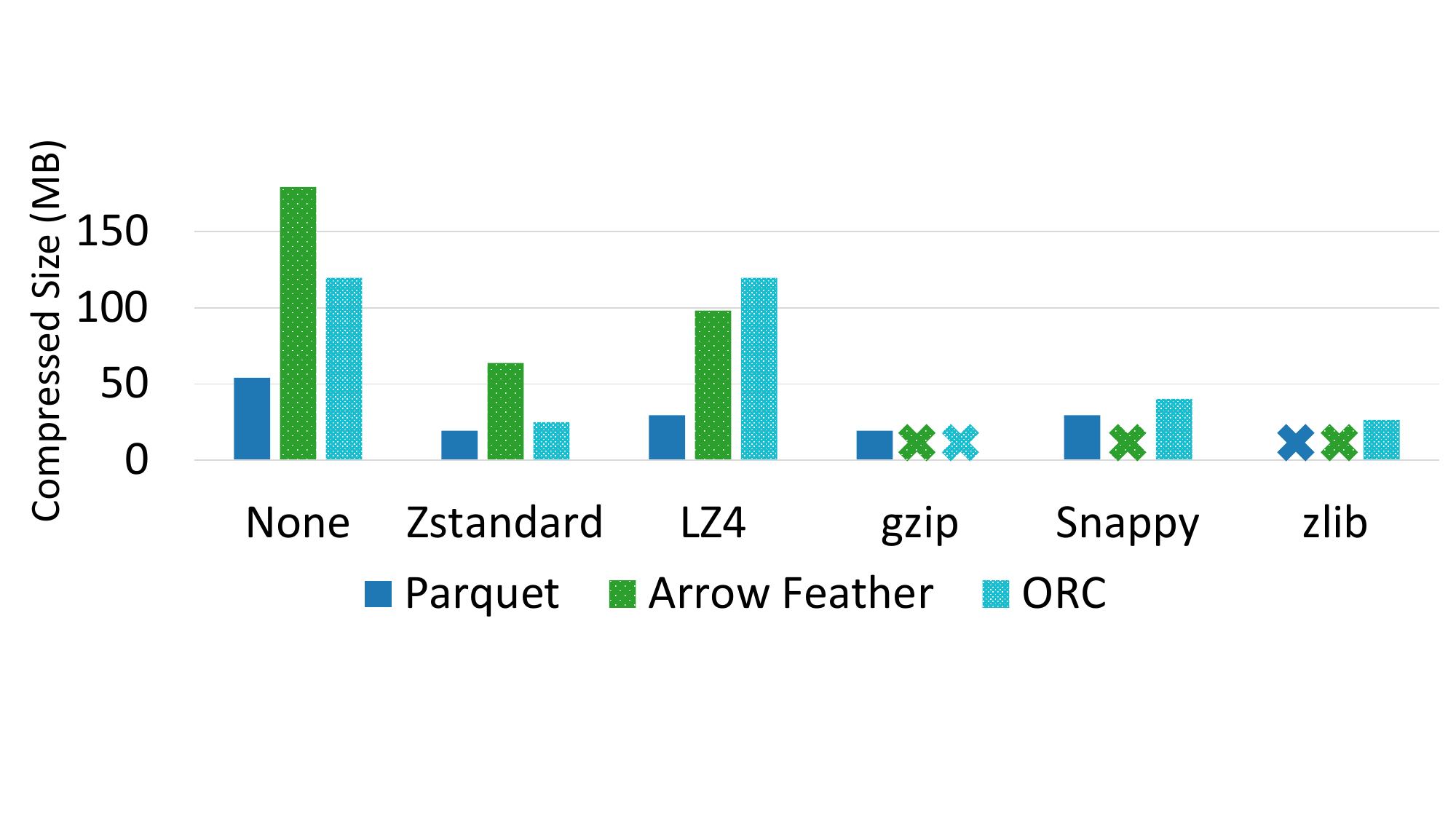}
     \end{subfigure}
        \caption{Total size on disk after compressing the string columns in TPC-DS.}
        \label{fig:type-comp-strings}
\end{figure}

\subsubsection{Compression Performance by Data Type.}
\label{sec:eval-compression-column}


In this experiment, we look at the performance over various column types in the TPC-DS dataset.
Specifically, we evaluate compression performance on integer (both \texttt{int32} and \texttt{int64}), double, and string data types.  We extract all columns of a given data type, compress them, and report the aggregate sizes by type.
The results are in Figures ~\ref{fig:type-comp-numeric} and ~\ref{fig:type-comp-strings}.

First, consider \Cref{fig:int-comp} which shows the aggregate compression performance on the integer columns. 
ORC achieves slightly better compression performance than Parquet.
This is because Parquet applies DICT and switches to plain encoding for some of the columns, whereas ORC always applies RLE. 
Arrow Feather does not encode by default.  This leads to the worst compression ratio when compression is disabled. Nevertheless, all three data formats perform similarly when compression is enabled, except for LZ4, where Arrow Feather is almost 50\% worse because it lacks encoding support for integers (we observe a similar result in \Cref{fig:tpcds-cr}).

Next, \Cref{fig:double-comp} shows the aggregate compression performance for the double columns. Parquet also applies DICT to this data type, whereas ORC and Arrow Feather do not encode at all. Because of this, Arrow and ORC have very similar performance both in the uncompressed and compressed setting, whereas Parquet is slightly better. The ORC outlier for LZ4 happens for the same reason as discussed in~\Cref{sec:eval-compression-tpcds}.

Finally, \autoref{fig:type-comp-strings} shows compression performance on string columns (both variable- and fixed-length). 
By default, Arrow does not encode this type, whereas ORC and Parquet apply DICT.
Among all formats, Parquet has the best compression performance, followed by ORC and Arrow. ORC produces larger compressed sizes than Parquet, because: (i) ORC has a smaller default block size and thus pays more dictionary overhead per row batch; and (ii) 
it more frequently falls back to plain encoding because of its row batch-level dictionary encoding (versus the chunk-level used in Parquet).
Again, LZ4 ORC disables compression because it offers no benefit. 



\begin{figure*}
     \centering\hspace{-2ex}
     \begin{subfigure}[t]{0.33\textwidth}
         \centering
         \includegraphics[trim={0.4cm 3cm 0.9cm 3.5cm},clip,width=\textwidth]{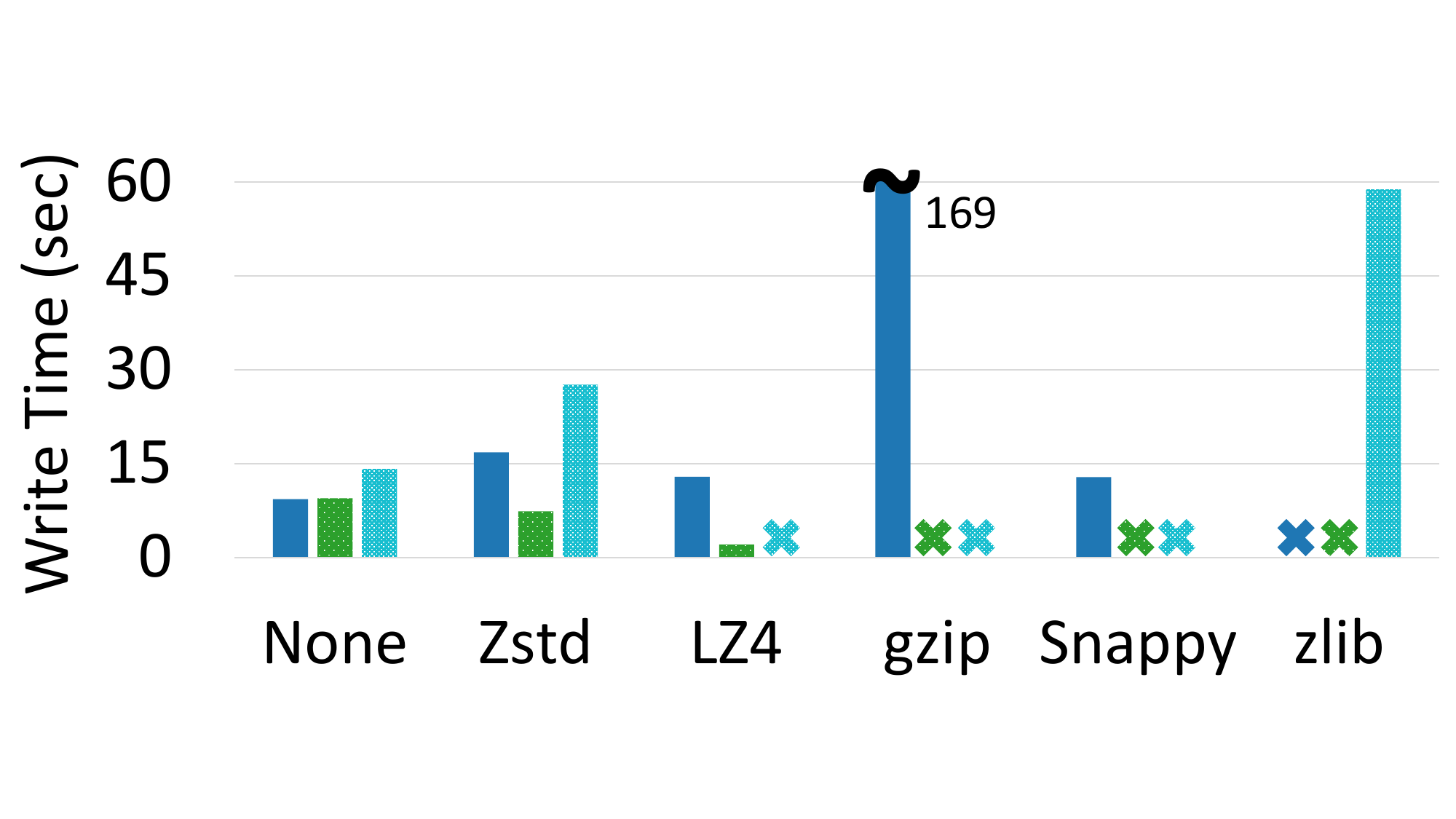}
         \vspace{1.25em}\vspace{-1ex}
         \caption{Writing to disk}
         \label{fig:comp-disk}
     \end{subfigure}
     \begin{subfigure}[t]{0.33\textwidth}
         \centering
         \includegraphics[trim={0.4cm 3cm 0.9cm 3.5cm},clip,width=\textwidth]{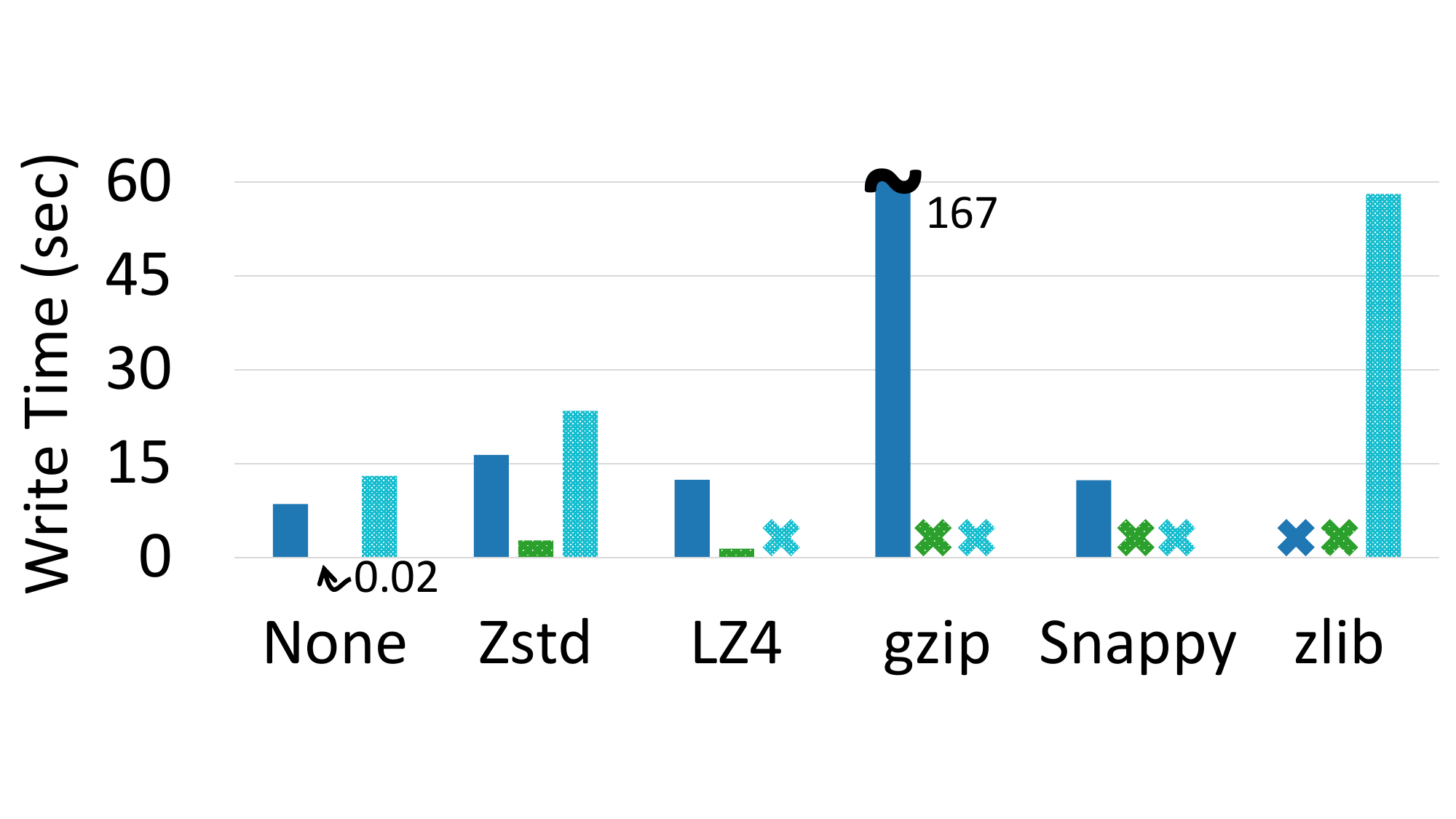}
         \\
         \includegraphics[trim={0.9cm 0 0.9cm 17cm},clip,width=\textwidth]{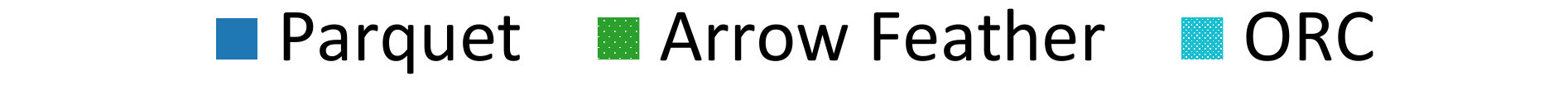}\vspace{-1ex}
         \caption{Writing to null device}
         \label{fig:comp-mem}
     \end{subfigure}
     \begin{subfigure}[t]{0.33\textwidth}
         \centering
         \includegraphics[trim={0.4cm 3cm 0.9cm 3.5cm},clip,width=\textwidth]{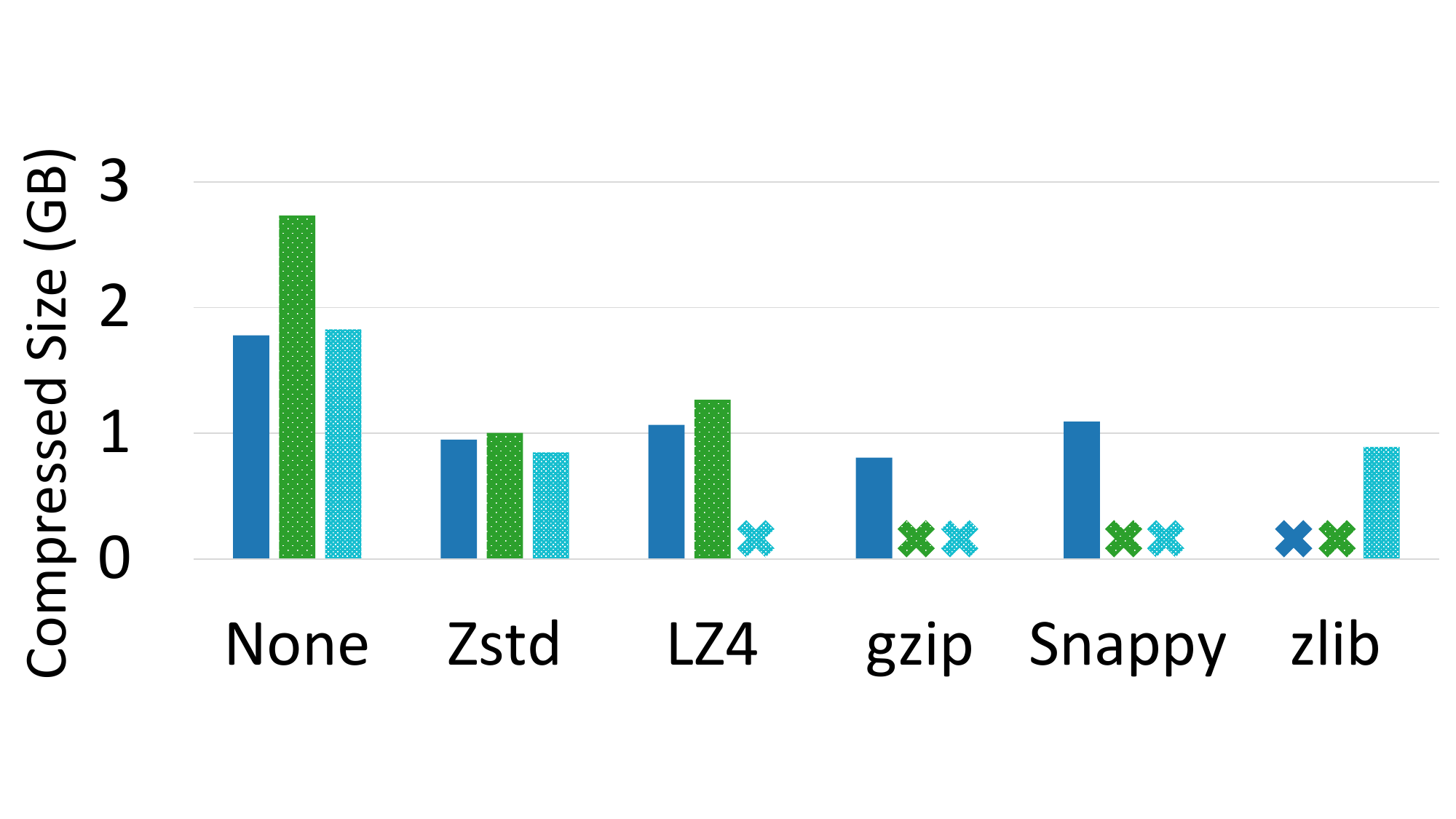}
    \vspace{1.25em}\vspace{-1ex}
         \caption{Compressed size on disk}
         \label{fig:comp-size}
     \end{subfigure}
        \caption{Write time from an Arrow in-memory table to each format stored either on disk or in memory. 
}
\label{fig:cthr}
\end{figure*}

\subsubsection{\rev{Compression Performance over Embedding Datasets.}}
\label{sec:embedding}

\rev{
We next explore compression performance of each format on vector-oriented datasets.
To do so, we used the Hugging Face Dataset portal~\cite{huggingfacedatasets:online} as a source of retrieval-augmented generation (RAG) and other embedding datasets. 
For the RAG datasets, to evaluate different scales, we selected two high-performance embedding models: \textit{UAE-Large-V1} \cite{uaeli2023angle} and \textit{SFR-Embedding-Mistral} \cite{SFRAIResearch2024}, with respective parameter sizes of 335 and 7100 million.
We additionally assembled a diverse set of 38 unique embedding datasets having at least 500 monthly downloads.
These datasets feature dimensions ranging from 256 to 4096, and record counts ranging from thousands to millions.}

\rev{We load these datasets into in-memory Numpy arrays and subsequently utilize PyArrow \cite{pyarrow} to convert them into Parquet, Arrow Feather, and ORC.  We then
apply compression as described in the previous section.
}

\rev{
We evaluate two data layouts for the vectors.
First, we decompose each $n$-dimensional vector into $n$ float values and store each value in a separate column
(i.e., a \textit{columnar layout}).  Second, we persist each vector into a single nested column containing $n$ sub-elements 
(i.e., a \textit{nested layout}; see \Cref{tab:enc-compare}).
}

\begin{figure}
     \begin{subfigure}[b]{\columnwidth} 
         \centering
         {\includegraphics[width=\textwidth,trim={0 7cm 0 0},clip]{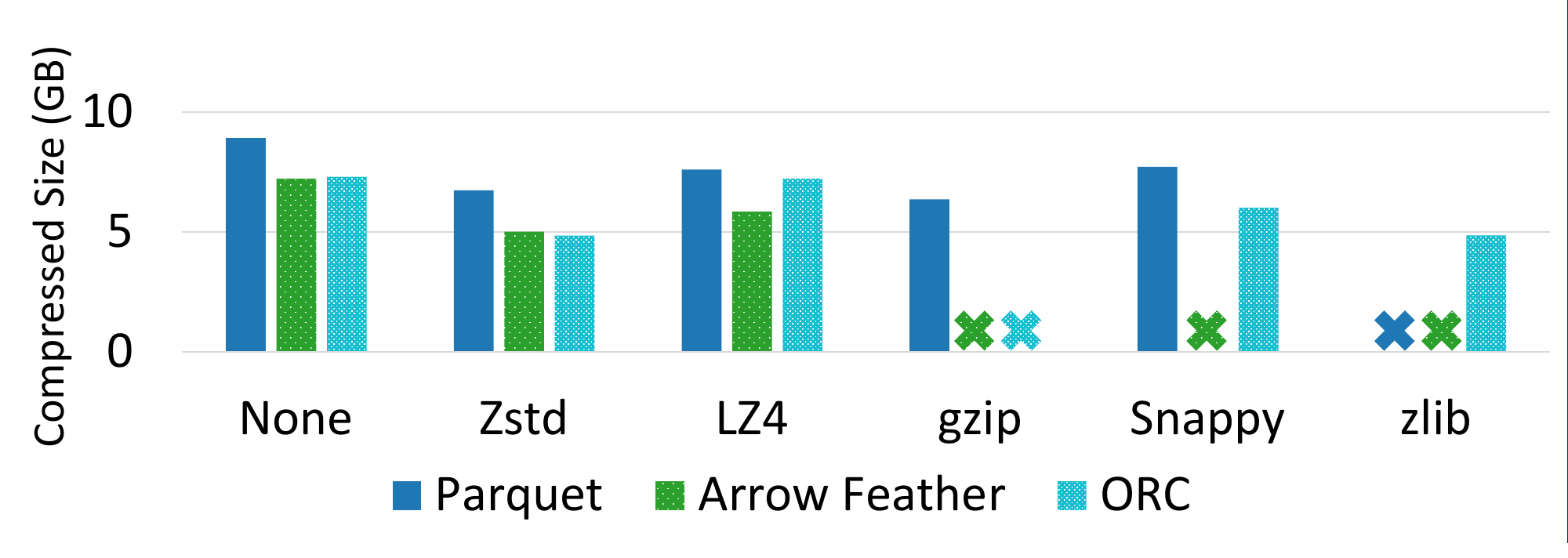}}
     \end{subfigure}
        \caption{\rev{Total size on disk after compressing embedding datasets using columnar layout (i.e., each $n$-dimensional vector is mapped into $n$ columns). 
        The CSV baseline (not shown in figure) consumes 18 GB.
        }}
        \label{fig:cr_values}
\end{figure}

\begin{figure}
     \begin{subfigure}[b]{\columnwidth} 
         \centering
                  {\includegraphics[width=\textwidth,trim={0 7cm 0 0},clip]{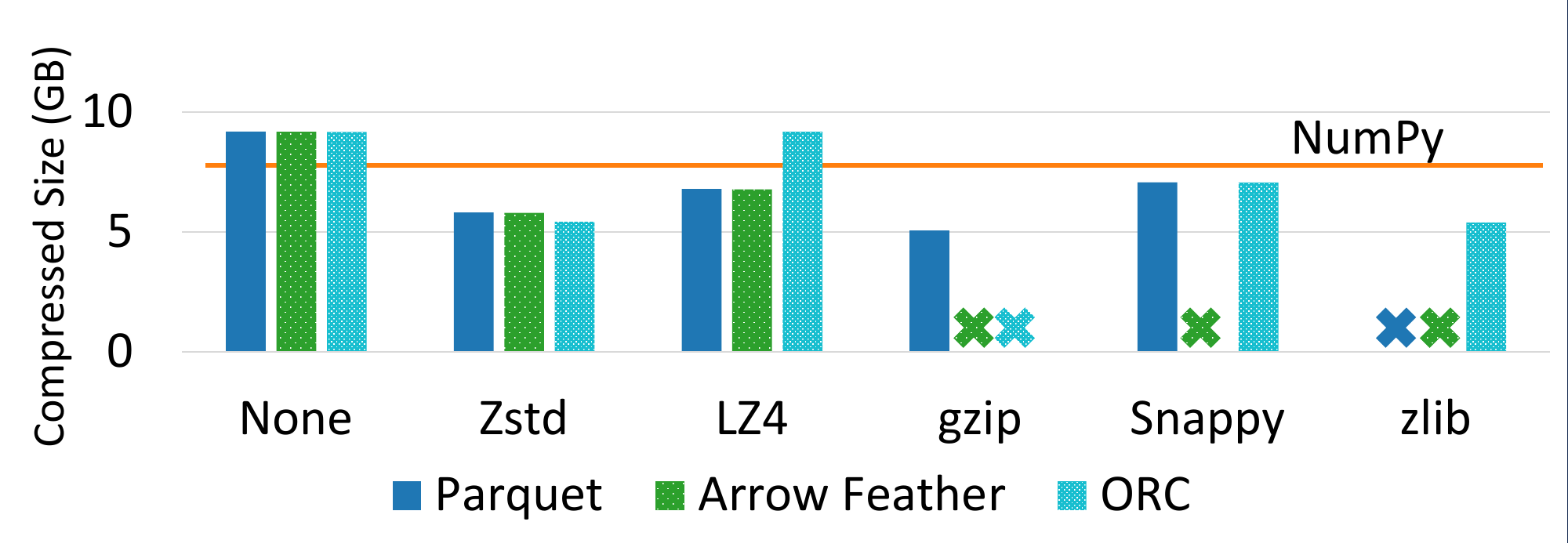}}
     \end{subfigure}
        \caption{\rev{Total size on disk after compressing embedding datasets using the nested layout (i.e., each vector is serialized as a single nested column). The ``NumPy'' baseline shows the size of data serialized using the NumPy uncompressed standard binary format.
        }}
        \label{fig:cr_vectors}
\end{figure}

\rev{
\Cref{fig:cr_values} and \Cref{fig:cr_vectors} respectively show results for the columnar and nested layouts.
Considering that embeddings comprise high-dimensional and high-entropy vectors, 
we see distinct behaviors across formats. For the columnar layout, Parquet struggles with the
large number of columns,
rendering its dictionary encoding less effective compared to the plain encoding used by Arrow Feather and ORC.
Both Arrow Feather and ORC exhibit similar performance except with LZ4 
for the same reasons discussed in \Cref{sec:eval-compression-tpcds}.}

\rev{
For the nested layout (\Cref{fig:cr_vectors}),
neither format effectively compresses the embeddings dataset. With Zstd and Gzip enabled, ORC performs slightly better (but still ${<}10\%$ improvement) because of its better compression implementation.
}

\stitle{\rev{Discussion.}}
\rev{In summary, current formats do not adequately handle compression for widely used embedding datasets. 
Surprisingly, when we compare each format's uncompressed representation (the ``None'' category in \Cref{fig:cr_vectors}) to the uncompressed NumPy baseline (the orange line), we find that all formats produce \textit{larger outputs} than this baseline.  
This is due to overhead introduced by the nested layout encodings and the hierarchical metadata inherent to each format.}

\rev{When 
comparing compressed size over both the columnar and nested layouts,
Arrow and ORC produce larger files with the nested layout because of their additional per-record, metadata-related bits. On the other hand, Parquet produces smaller file sizes with the nested layout because its dictionary encoding is less effective on float types under the columnar layout.
}


\rev{In addition to the high-dimensional and high-entropy nature of embedding vectors, their high-precision further limits compression performance across all formats. This raises a pertinent question: is such high precision truly necessary for downstream AI tasks, especially those relying on high-dimensional distance calculations like RAG and $k$NN? Liu~\cite{liu2022fast} addresses this by evaluating a 1NN task on the UCR dataset collection, using various precision levels for the input float numbers. They report the 1NN accuracy for all 128 datasets, normalizing each dataset's accuracy by its corresponding full-precision accuracy. The results indicate that relative accuracy decreases gracefully as precision diminishes. Nonetheless, the performance remains remarkable, demonstrating the potential for reduced precision in machine learning tasks. This opens up the possibility that precision-controlled encoding, such as BUFF~\cite{liu2021decomposed}, could be a suitable option for embedding encoding in columnar formats.}


\subsection{Transcoding Overhead}
\label{sec:eval-transcode}

In practice, storage formats are converted into (or from) an in-memory presentation on reads (writes).
We now evaluate the overheads in transcoding (i.e., decompressing, converting, and compressing) each format. 
Specifically, \autoref{sec:eval-compression-overhead} explores the time required to compress and serialize each format from a common in-memory representation, while \autoref{sec:eval-decompression-overhead} evaluates the overhead of loading data, i.e., deserializing and decompressing each format into an in-memory representation amenable to query execution. \rev{In \Cref{sec:vector_trans}, we finally evaluate both the serialization and deserialization cost over embedding datasets.}


\subsubsection{Compression Overhead.}
\label{sec:eval-compression-overhead}

Our first experiment in this section explores how long it takes to serialize (and compress) the data from an in-memory representation to each disk format.  
For this experiment we use the \texttt{catalog\_sales}  TPC-DS table.
\review{The {\tt catalog\_sales} is a large (${\sim}14$M rows) and wide ($34$ columns) table containing integers and doubles. Its raw data size is 3GB.
}
All formats support serializing from an Arrow Table, and so we adopt it as our common in-memory representation.

\Cref{fig:cthr} shows the runtimes when:
(i) writing to disk (\ref{fig:comp-disk}); (ii) writing to the null device, which avoids any I/O overhead (\ref{fig:comp-mem}); as well as (iii) the data sizes per format (\ref{fig:comp-size}). 
We omit the LZ4 and Snappy bars for ORC as the Apache Arrow C++ library has limited compression support for the ORC format.
Starting with \Cref{fig:comp-disk}, we can see that Arrow Feather is the most efficient format in terms of compression and serialization runtime because it does not encode data. On the other hand, Arrow Feather’s lack of encoding leads to almost a 50\% larger footprint on disk (\Cref{fig:comp-size}). 
Interestingly, ORC compression time is $50\%$
slower than Parquet with comparable or slightly better compression
ratio on disk (up to 15\% better). \review{We think that this is because of better Parquet support in Arrow; both projects share the same codebase and data structures}. 

Finally, we isolate the compression overhead in \Cref{fig:comp-mem} by 
avoiding disk I/O by writing to the null device.
Here we can see a decrease in runtime for all the formats, although of different magnitudes.
Arrow Feather has the biggest difference, thanks to its inherent zero-copy implementation in Arrow.
The compression time for Parquet and ORC does not change substantially because the encoding and compression operations dominate the  total runtime.

\subsubsection{Decompression Overhead (i.e., table scan).}
\label{sec:eval-decompression-overhead}

\eat{
\begin{table}[]
\centering
\caption{Overheads (in seconds) for decompressing the TPC-DS \texttt{catalog\_sales} table from on disk formats (resides on a \textbf{real disk})  into in-memory Arrow.\CL{I redo the experiments, as the number changes 10-20\% compared with my previous run. Not sure if network or disk usage changes over time.} \CL{ORC supports two compression level settings (1 for SPEED and 3 for COMPRESSI/ON) for ZSTD. I would suggest to keep ZSTD-1 for all experiment for ORC ZSTD number.\MI{My understanding is that ORC only support ZSTD-1 and ZSTD-3, correct?}} \CL{Yes.}}
\label{tab:trancoding}
\begin{tabular}{|c|c|c|c|}
\hline
\textbf{Compression} & \textbf{Parquet} & \textbf{ORC} & \textbf{Arrow} \\ \hline
\textbf{None} & 6.704 & 7.068 & 6.917 \\ \hline
\textbf{ZSTD-1} & 8.050 & 12.968 & 6.315 \\ \hline
\textbf{ZSTD-5} & 7.676 & - & 5.974 \\ \hline
\textbf{ZSTD-9} & 7.716 & - & 5.892 \\ \hline
\textbf{LZ4} & 6.726 & 7.166 & 4.386 \\ \hline
\textbf{Zlib} & - & 16.069 & - \\ \hline
\end{tabular}
\end{table}
}

\begin{figure}
\centering
\includegraphics[trim={0 4cm 0 3.3cm},clip,width=\columnwidth]{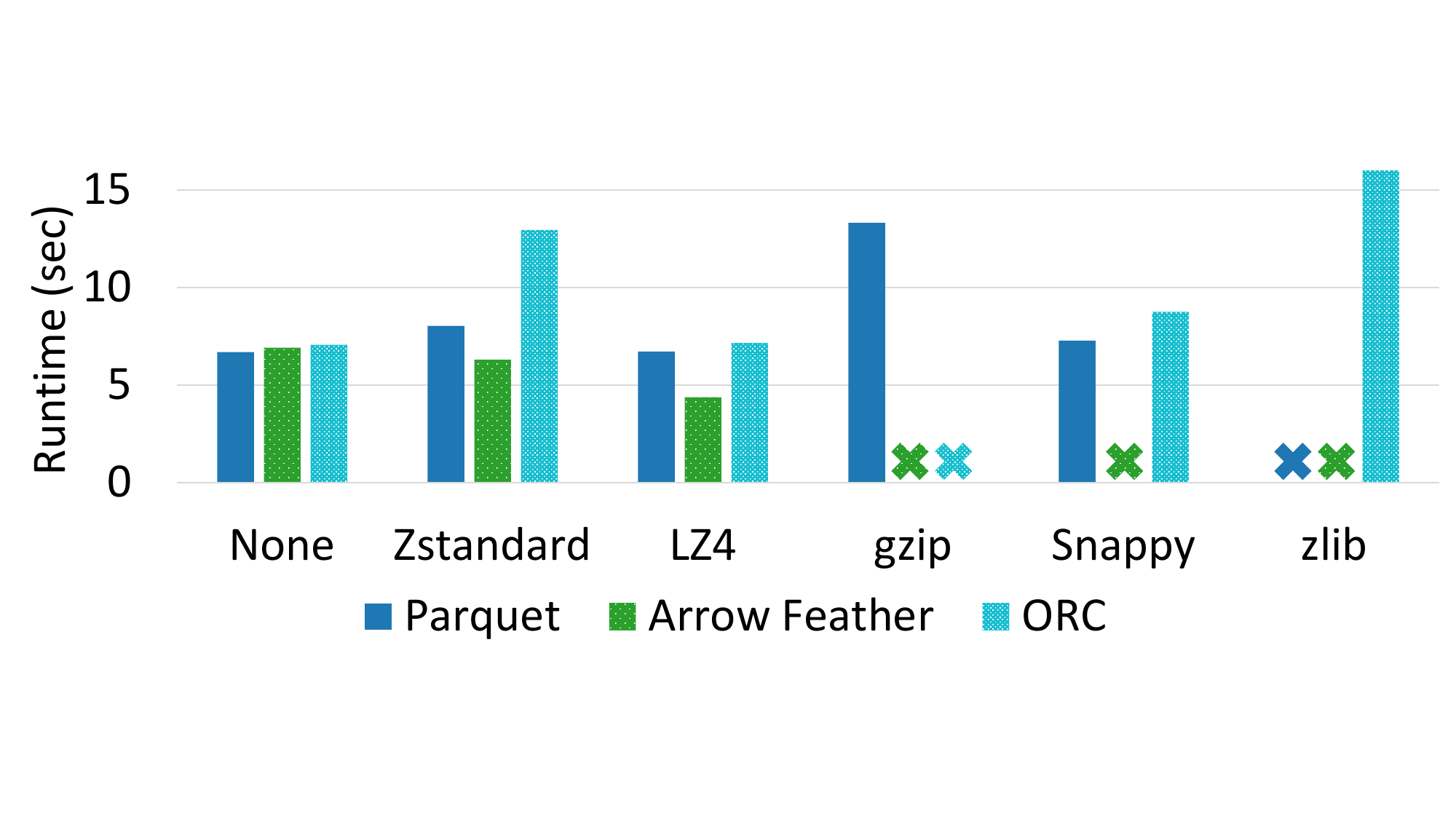}
\caption{Runtime (in seconds) for decompressing the TPC-DS {\tt catalog\_sales} table from the on-disk formats into in-memory Arrow.} 
\label{fig:decompression-overhead-disk}\vspace{-4ex}
\end{figure}

\eat{
\begin{table}[]
\centering
\caption{Overheads (in seconds) for decompressing the TPC-DS catalog\_sales table from on disk formats (resides on a \textbf{ramdisk}) into in-memory Arrow.}
\label{tab:trancodingram}
\begin{tabular}{|c|c|c|c|}
\hline
\textbf{Compression} & \textbf{Parquet} & \textbf{ORC} & \textbf{Arrow} \\ \hline
\textbf{None} & 5.319 & 6.083 & 1.573 \\ \hline
\textbf{ZSTD-1} & 7.351 & 10.551 & 5.616 \\ \hline
\textbf{ZSTD-5} & 7.159 & - & 5.371 \\ \hline
\textbf{ZSTD-9} & 7.122 & - & 5.306 \\ \hline
\textbf{LZ4} & 5.832 & 6.123 & 3.523 \\ \hline
\textbf{Zlib} & - & 15.486 & - \\ \hline
\end{tabular}
\end{table}
}

In this experiment we investigate the overhead of loading the {\tt catalog\_sales} TPC-DS table from disk into memory.
Our goal with this experiment is to simulate the overheads involved when a query processor is required to load and transform
a compressed dataset into a plain in-memory format amenable to query execution. 
We start from data on disk in the Parquet, ORC, or Arrow Feather formats, and we report the time required to load the data and convert it into the Arrow in-memory format. 
The results are shown in \autoref{fig:decompression-overhead-disk}.

Interestingly, loading compressed data has 30\% less overhead than the uncompressed case for Arrow under LZ4.
This is because LZ4 requires less disk I/O (since the file on disk is smaller; see \Cref{fig:comp-size}) while also providing ``fast enough'' decompression relative to the other compression methods.
For the other cases, Arrow always exhibits the best performance: this is expected since it does not require decoding the data and its on-disk compressed size is reasonable. 
Parquet is slightly worse than Arrow because of the cost of decoding data, while ORC has the worst performance (it is particularly bad for Zstd and zlib). We think that this is
due to decompression settings 
such as block size, buffer size, etc.
In general, for formats that heavily leverage encoding (i.e., Parquet and ORC) data compression leads to a heavy penalty on read performance.

\begin{figure}
\centering
\includegraphics[trim={0 4cm 0 3.3cm},clip,width=\columnwidth]{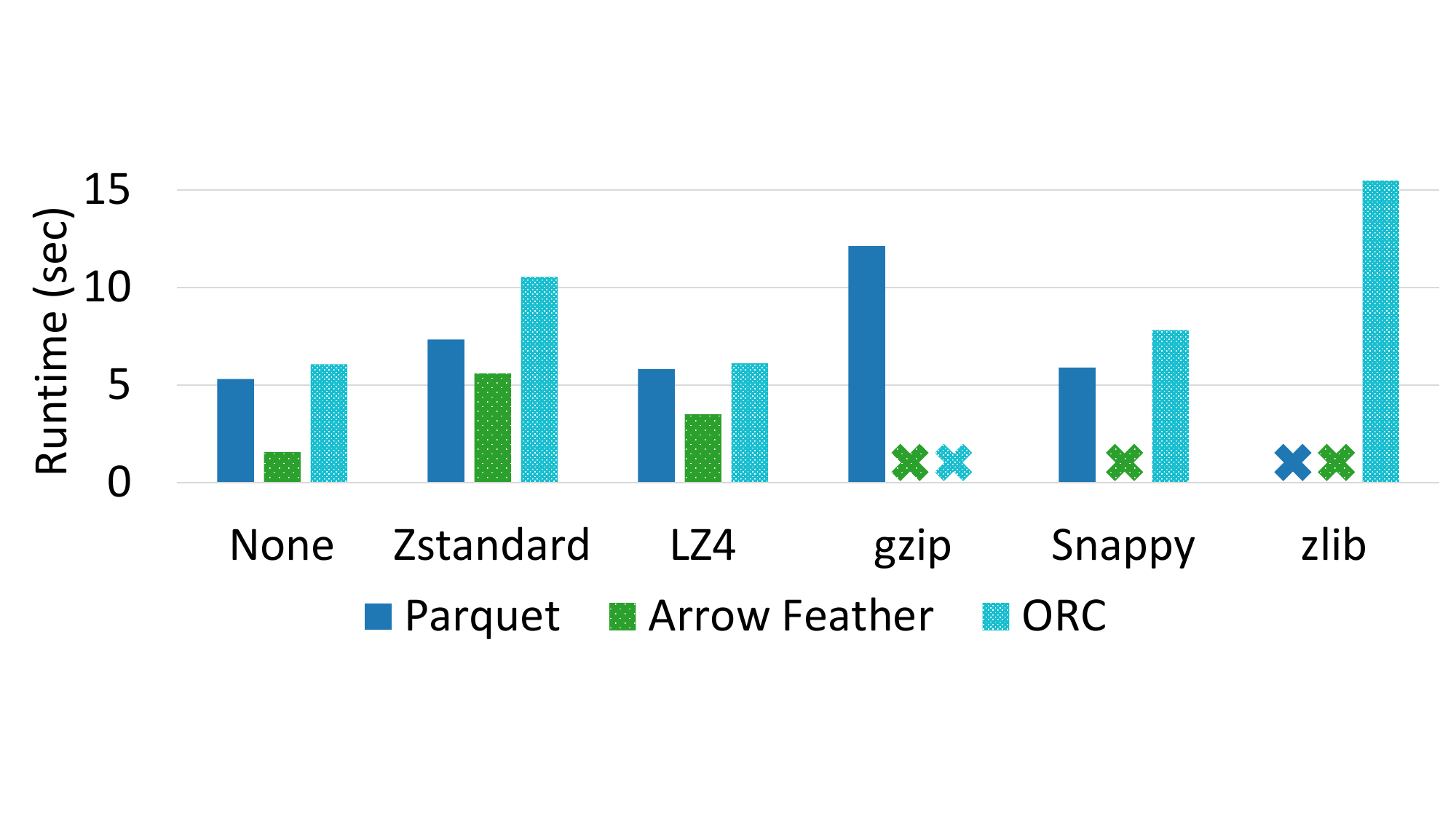}
\vspace{-2ex}
\caption{Runtime (in seconds) for decompressing the TPC-DS {\tt catalog\_sales} table from the formats in memory (ramdisk) into in-memory Arrow.} 
\label{fig:decompression-overhead-mem}\vspace{-5ex}
\end{figure}

To isolate disk I/O from compression overheads, we load each compressed dataset onto a memory-resident disk mounted on \texttt{tmpfs}.
As we can see from \Cref{fig:decompression-overhead-mem}, in all cases the runtimes decrease, especially for Arrow without compression. This result is intuitive because, for uncompressed data, the data size is much larger and disk bandwidth is saturated. Conversely, decompression is CPU-bound and not substantially impacted by the cost of bringing data into memory. Combined with previous compression experiments in \Cref{fig:cthr}, this shows the benefits of Arrow as a fast inter-process format, when disk I/O and size are not the bottleneck.

\subsubsection{Transcoding Cost over Embedding dataset}
\label{sec:vector_trans}

\begin{figure}
     \begin{subfigure}[b]{\columnwidth} 
         \centering
                  {\includegraphics[width=\textwidth,trim={0 7cm 0 0},clip]{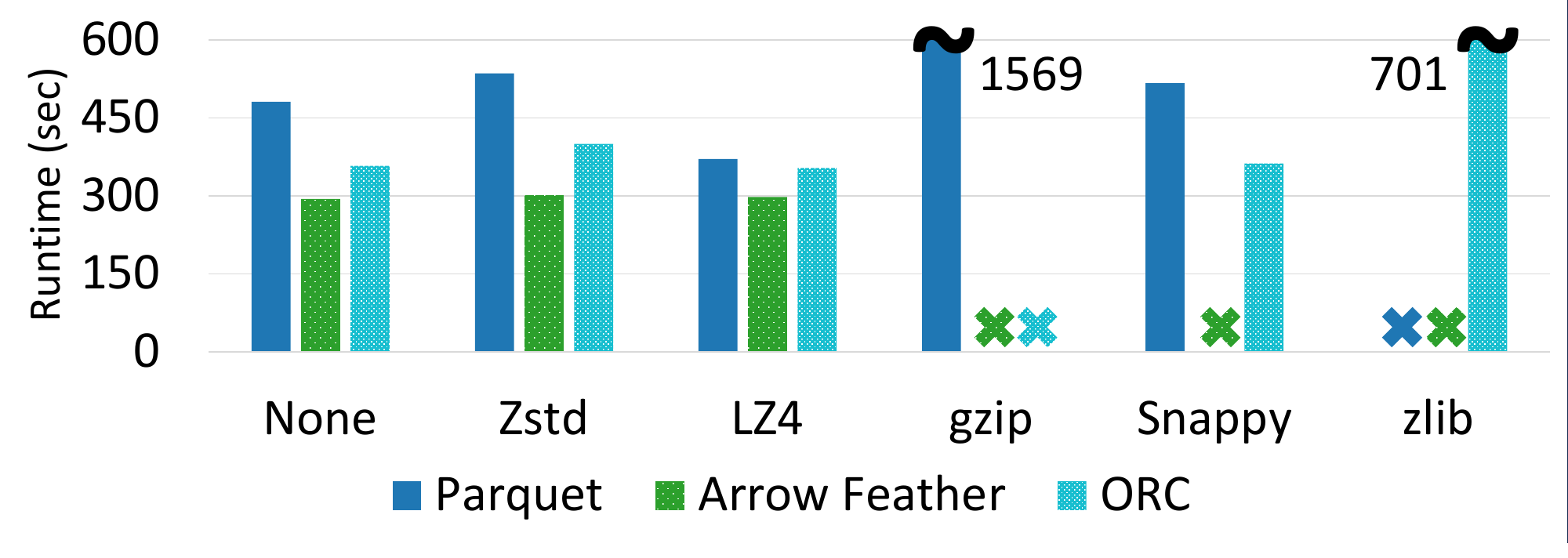}}
     \end{subfigure}
        \caption{\rev{Write time
        using a columnar layout (i.e., a $n$-dimensional vector is represented as $n$ columns) to each format on disk.
        The CSV baseline (not shown in figure) took 1,982 seconds to write. 
        }}
        \label{fig:compression_time_values}
\end{figure}

\begin{figure}
     \begin{subfigure}[b]{\columnwidth} 
         \centering
                  {\includegraphics[width=\textwidth,trim={0 7cm 0 0},clip]{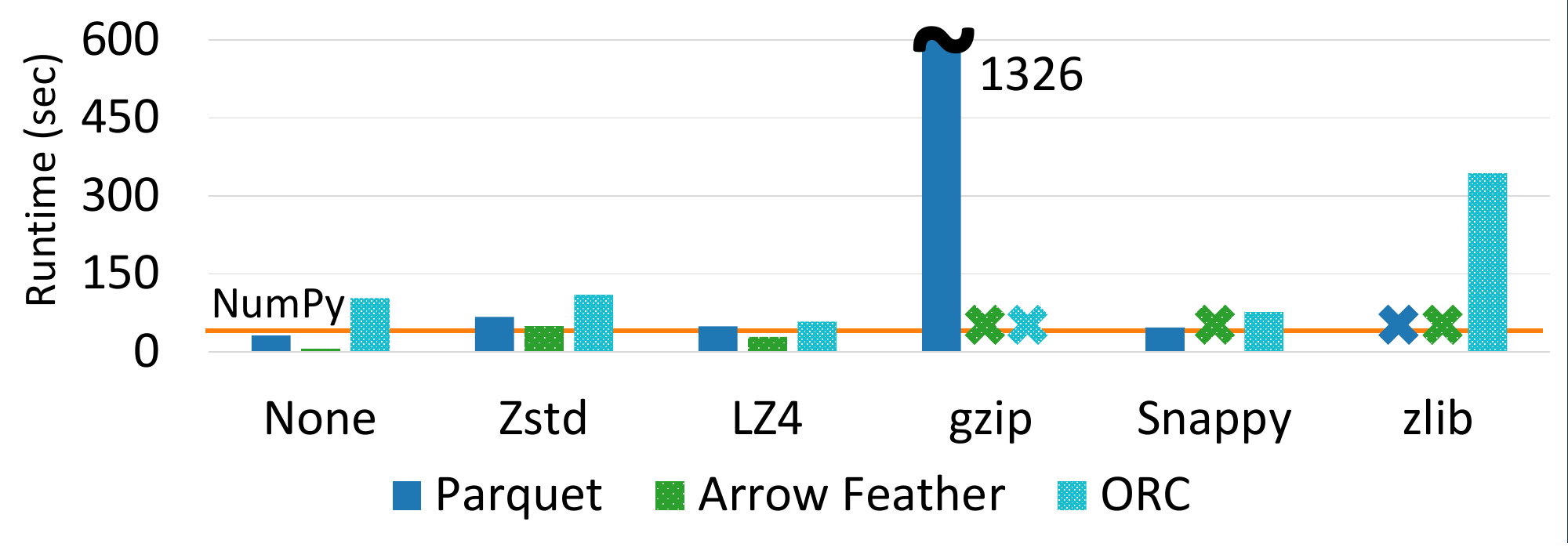}}
     \end{subfigure}
        \caption{\rev{Write time
        for vectors using the nested layout to each format on disk.  The ``NumPy'' line shows time to written vectors serialized using the NumPy uncompressed standard binary format.}}
        \label{fig:compression_time_vectors}
\end{figure}

\rev{We next evaluate the compression times using the columnar and nested layouts described in \Cref{sec:embedding}.  The respective results are shown in \autoref{fig:compression_time_values} and \autoref{fig:compression_time_vectors}. ORC and Arrow Feather, which utilize plain encoding, demonstrate greater efficiency compared to Parquet's dictionary encoding. 
 This results in shorter compression times under the columnar layout. Arrow Feather consistently outperforms the others due to its more efficient serialization. For the nested layout, where all formats must encode and compress vector embeddings, Parquet's Dremel-based encoding surpasses ORC's encoding in efficiency. Arrow remains the top performer overall, benefiting from its native serialization efficiency. When comparing the columnar and nested layouts between the same format,
 each format shows better efficiency with the nested layout.  This is due to more specialized encoding and the simplified  writing logic when emitting a single nested column.} 

\begin{figure}
     \begin{subfigure}[b]{\columnwidth} 
         \centering
                  {\includegraphics[width=\textwidth,trim={0 7cm 0 0},clip]{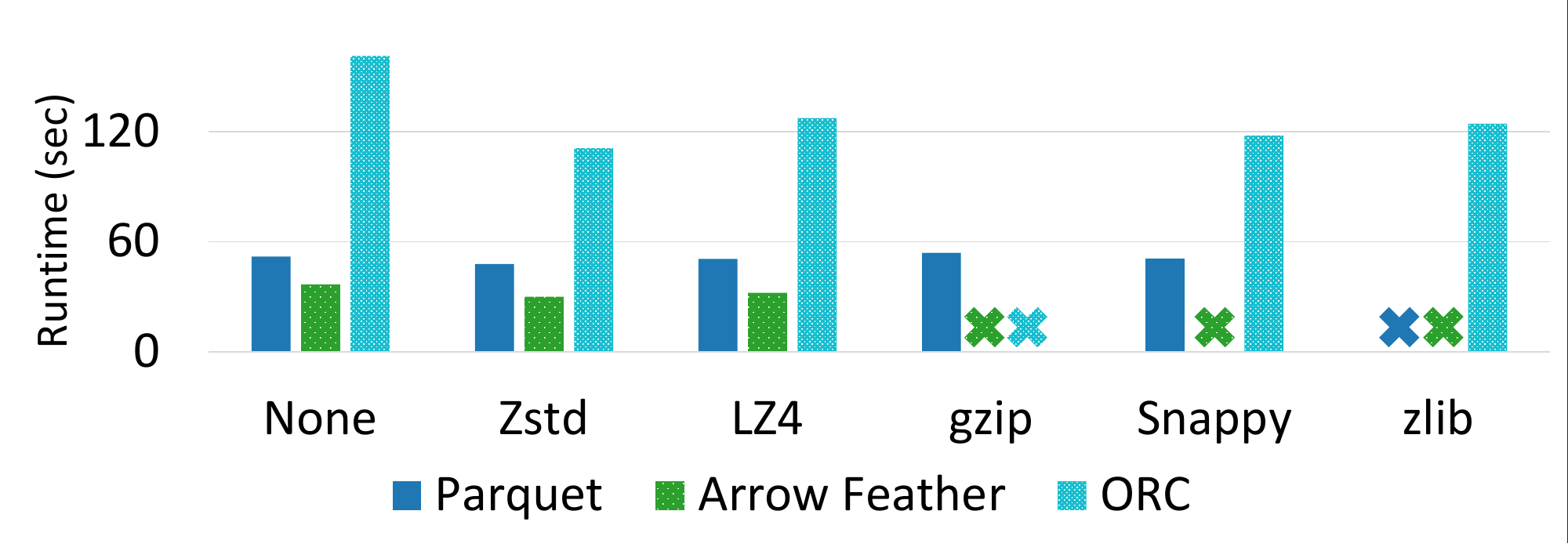}}
     \end{subfigure}
        \caption{\rev{Runtime (in seconds) for decompressing the table from the on-disk formats
into an in-memory table using the columnar layout.}}
        \label{fig:read_values}
\end{figure}

\begin{figure}
     \begin{subfigure}[b]{\columnwidth} 
         \centering
                  {\includegraphics[width=\textwidth,trim={0 7cm 0 0},clip]{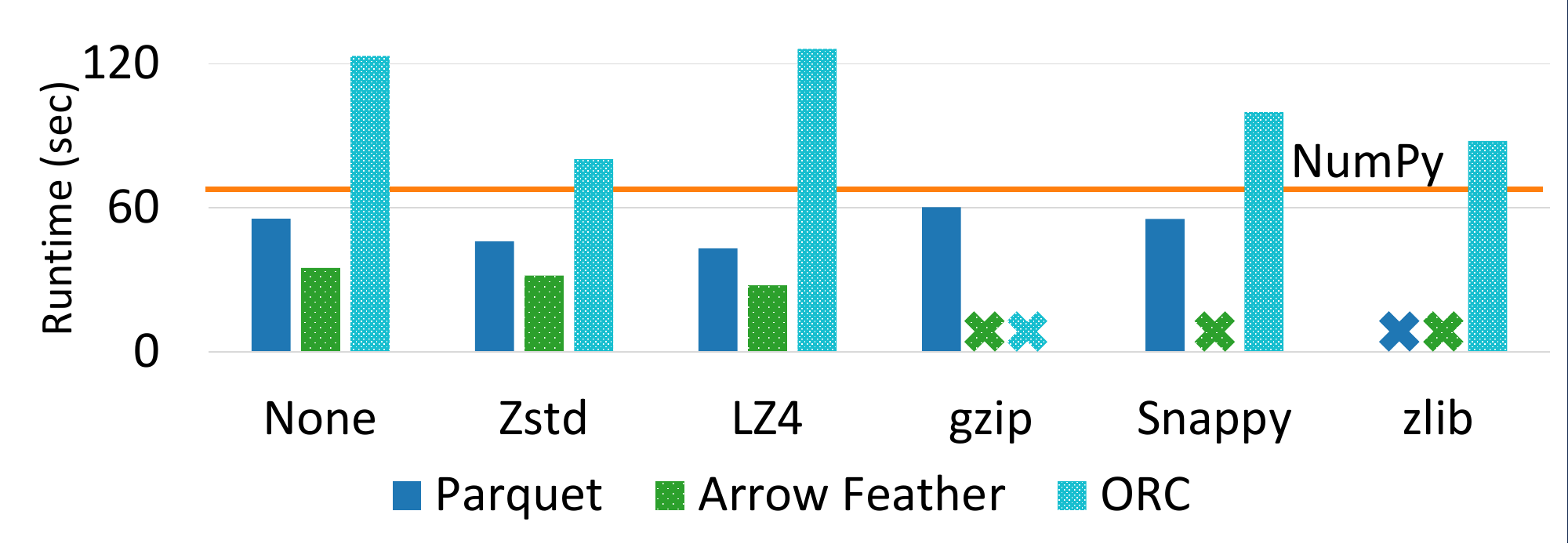}}
     \end{subfigure}
        \caption{\rev{Runtime (in seconds) for decompressing the table from the on-disk formats
into an in-memory table using the nested layout.  The ``NumPy'' line shows time to decompress data serialized using the NumPy uncompressed standard binary format.}}
        \label{fig:read_vectors}
\end{figure}

\rev{We next evaluate the performance of loading embedding datasets from disk into memory. The datasets are scanned into in-memory NumPy arrays, and we record the scan time for each file. We then report the runtime for loading the data in two layouts (columnar and nested) and respectively show the results in 
\autoref{fig:read_values} and \autoref{fig:read_vectors}.}

\rev{Arrow Feather consistently outperforms the other formats in both cases, whereas ORC shows the poorest performance. Specifically, the deserialization time for ORC is typically 2--3 times longer than for Parquet and 3--4 times longer than for Arrow Feather. The nested layout generally requires more load time due to more complex reading logic, which involves checking the definition and repetition levels for decoding the Parquet and Arrow and referring to the present and length streams for  ORC. By contrast, the columnar layout benefits from bulk-reading logic and greater parallelism, since the reads can be distributed across multiple column readers.}

\rev{In our comparison of vector embedding serialization and deserialization performance across these formats, it is evident that all formats fall short of the NumPy data science format in terms of write efficiency. This is notable despite the generally poor compression performance observed in these formats. Among them, Arrow Feather distinguishes itself with its superior read performance relative to NumPy, attributed to its more efficient serialization capabilities. However, it still lacks robust encoding support. Parquet, on the other hand, demonstrates read performance comparable to that of NumPy. In stark contrast, ORC consistently exhibits the poorest performance for both reading and writing operations. These findings underscore the importance for open columnar formats to better support machine learning use cases. Enhancing these formats through more efficient encoding techniques and optimized serialization and deserialization processes is crucial to meet the evolving demands of modern data-centric workflows.}

\review{To summarize, encoding and compression choices greatly impact performance, with formats like Parquet and ORC targeting size on disk, while Arrow targets raw read performance. To optimize both size and performance, formats should be carefully tuned to the workload and use case, and workload-aware compression selection is crucial. It remains an open question of how much computation can be pushed into the encoded space to minimize the decoding step while maximizing the compression ratio.}

\begin{figure*}
     \centering
     \begin{subfigure}[t]{0.32\textwidth}
         \centering
         \includegraphics[trim={0 3.9cm 0 3.5cm},clip,width=\columnwidth]{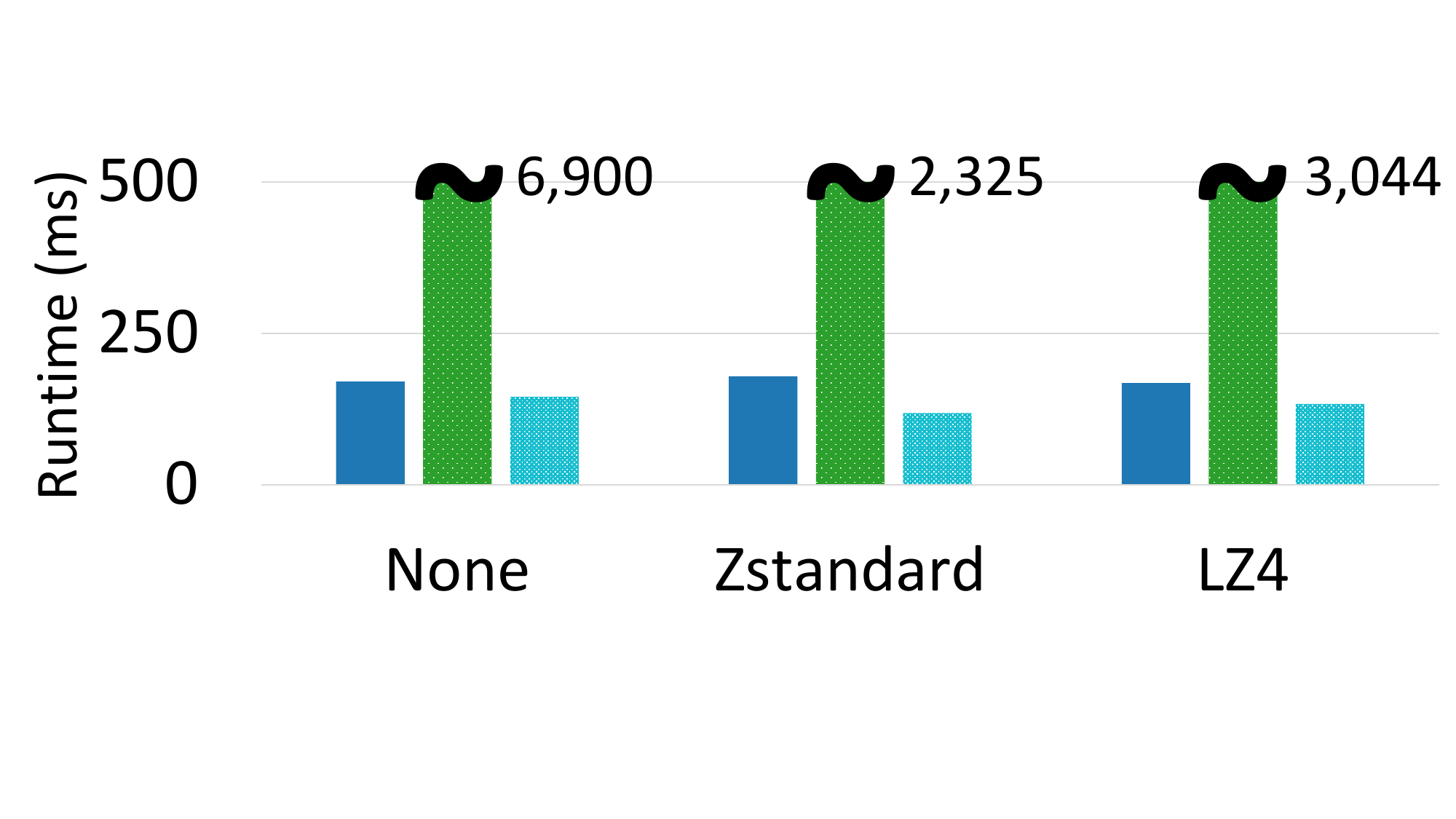}
         \vspace{0.5em}
         \caption{Integers}
        \vspace{-3em}
        \label{fig:projection-int}
    \end{subfigure}
     \begin{subfigure}[t]{0.32\textwidth}
         \centering
         \includegraphics[trim={0 3.9cm 0 3.5cm},clip,width=\columnwidth]{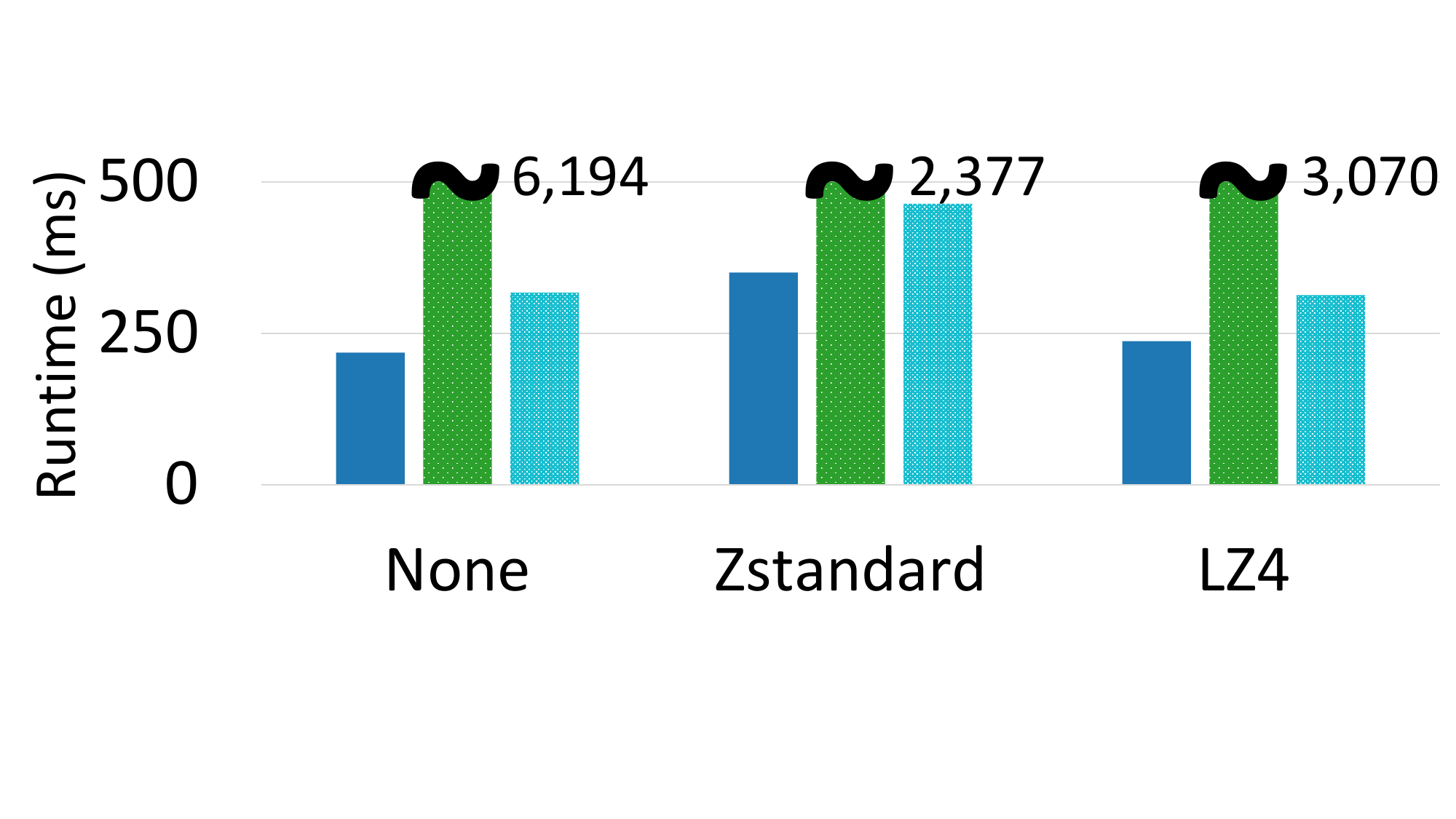}
         \vspace{0.5em}
         \caption{Doubles}
        \vspace{-3em}
         \label{fig:projection-double}
     \end{subfigure}
     \begin{minipage}[t]{0.32\textwidth}
         \centering
         \includegraphics[trim={0 3.9cm 0 3.5cm},clip,width=\textwidth]{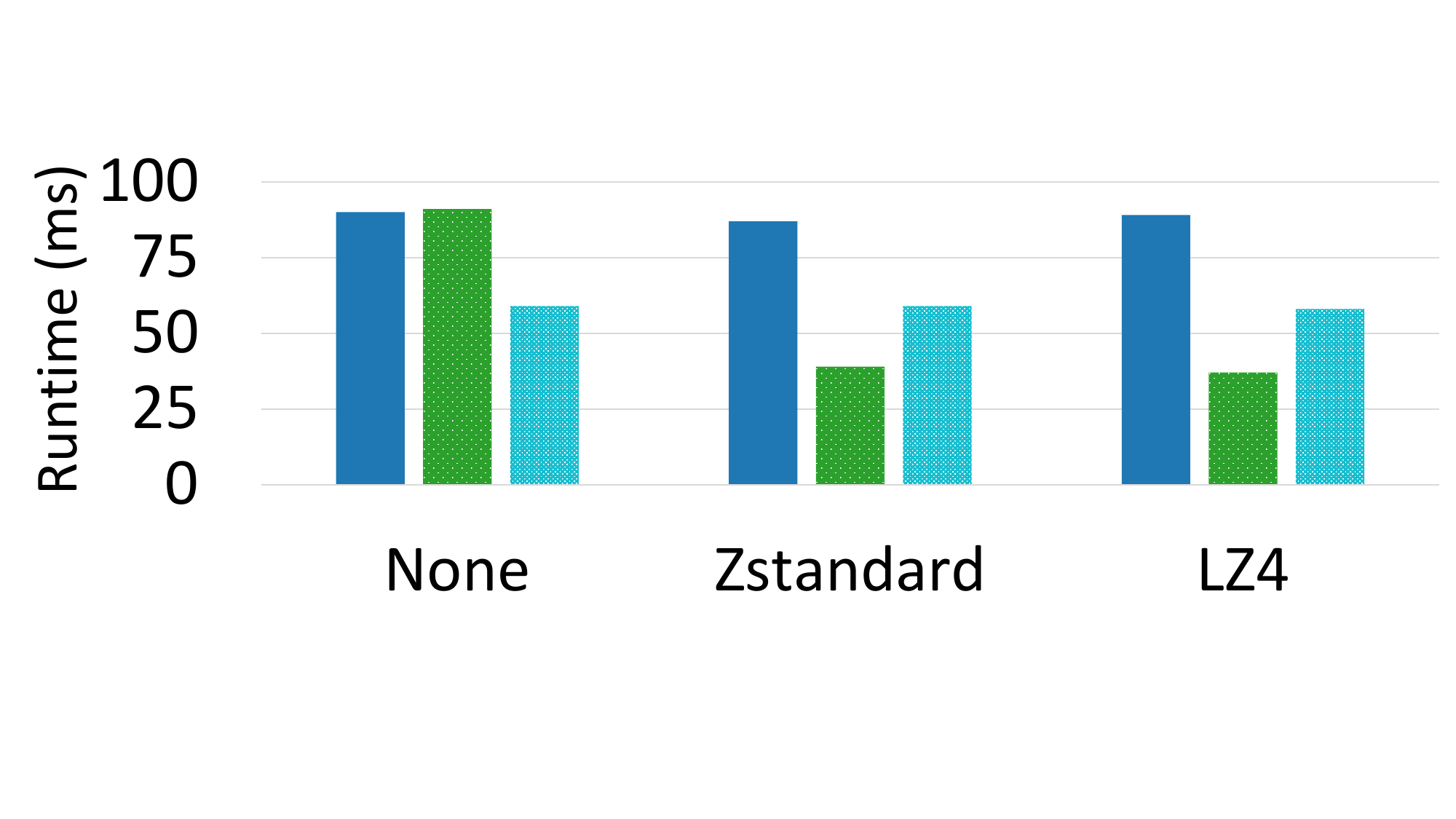}
    \end{minipage}
     \\
     \hspace{11em}
     \includegraphics[trim={5cm 0 5cm 17cm},clip,width=0.25\textwidth]{figures-formatted/figure5-legend.pdf}
     \hfill
     \includegraphics[trim={5cm 0 5cm 17cm},clip,width=0.25\textwidth]{figures-formatted/figure5-legend.pdf}
     \hspace{1em}
     ~
\\
  \begin{minipage}[t]{0.66\textwidth}
    \caption{{Projecting numeric types on the {\tt catalog\_sales} table.}}
    \label{fig:projection-numeric}
  \end{minipage}
  \begin{minipage}[t]{0.33\textwidth}
    \caption{Projecting strings on the {\tt customer\_demographic} table.}
    \label{fig:projection-string}
  \end{minipage}
\end{figure*}

\section{Data Access Microbenchmarks}
\label{sec:eval-micro}

\begin{figure*}
     \begin{subfigure}[b]{0.33\textwidth}
         \centering
         \includegraphics[trim={0 4.5cm 0 0},clip,width=\textwidth]{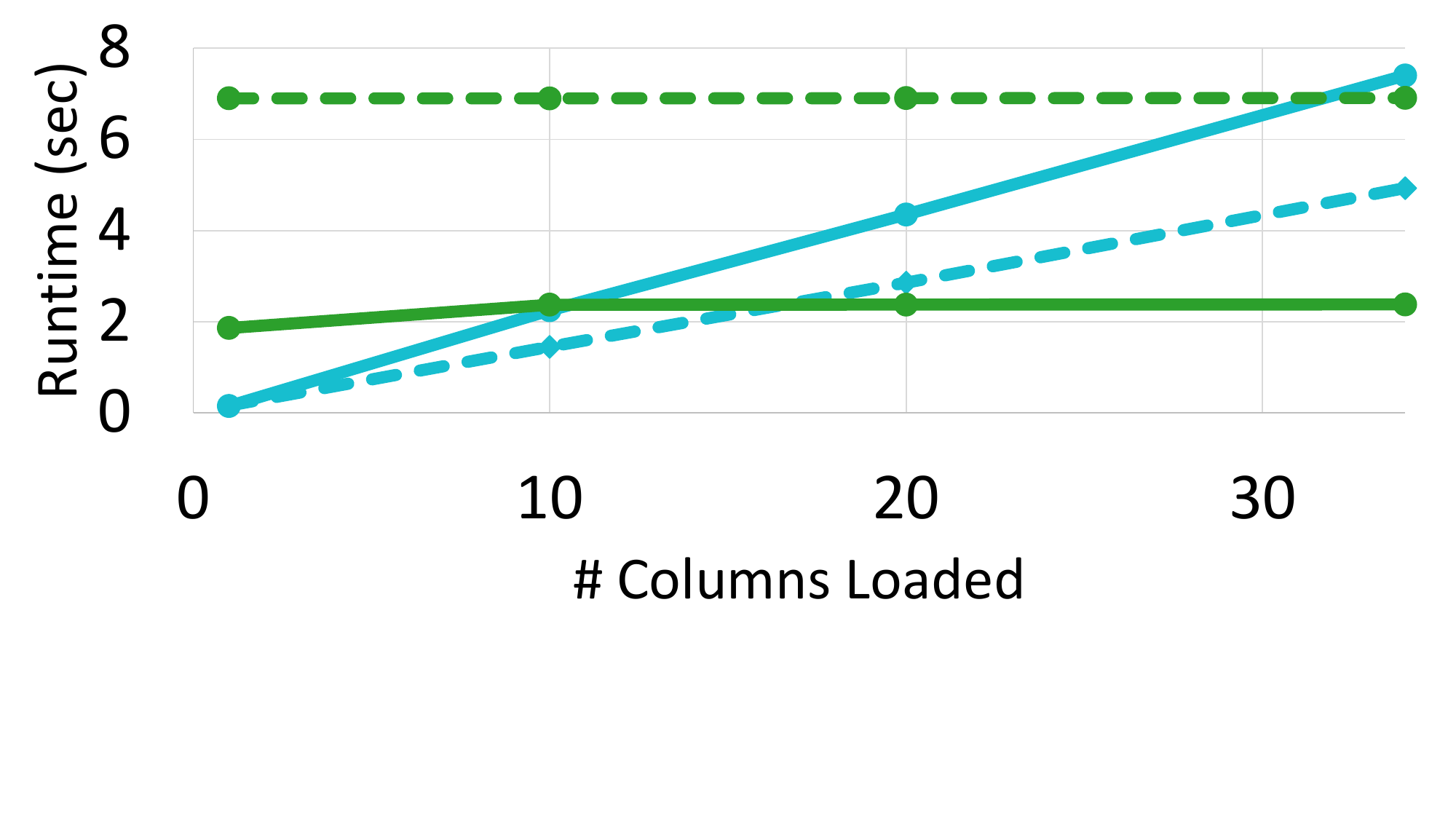}
         \vspace{1em}
         \caption{Cold}
         \vspace{-2.75em}
         \label{fig:allone-cold}
     \end{subfigure}
     \begin{subfigure}[b]{0.33\textwidth}
         \includegraphics[trim={3cm 4.5cm 0 0},clip,height=2.5cm]{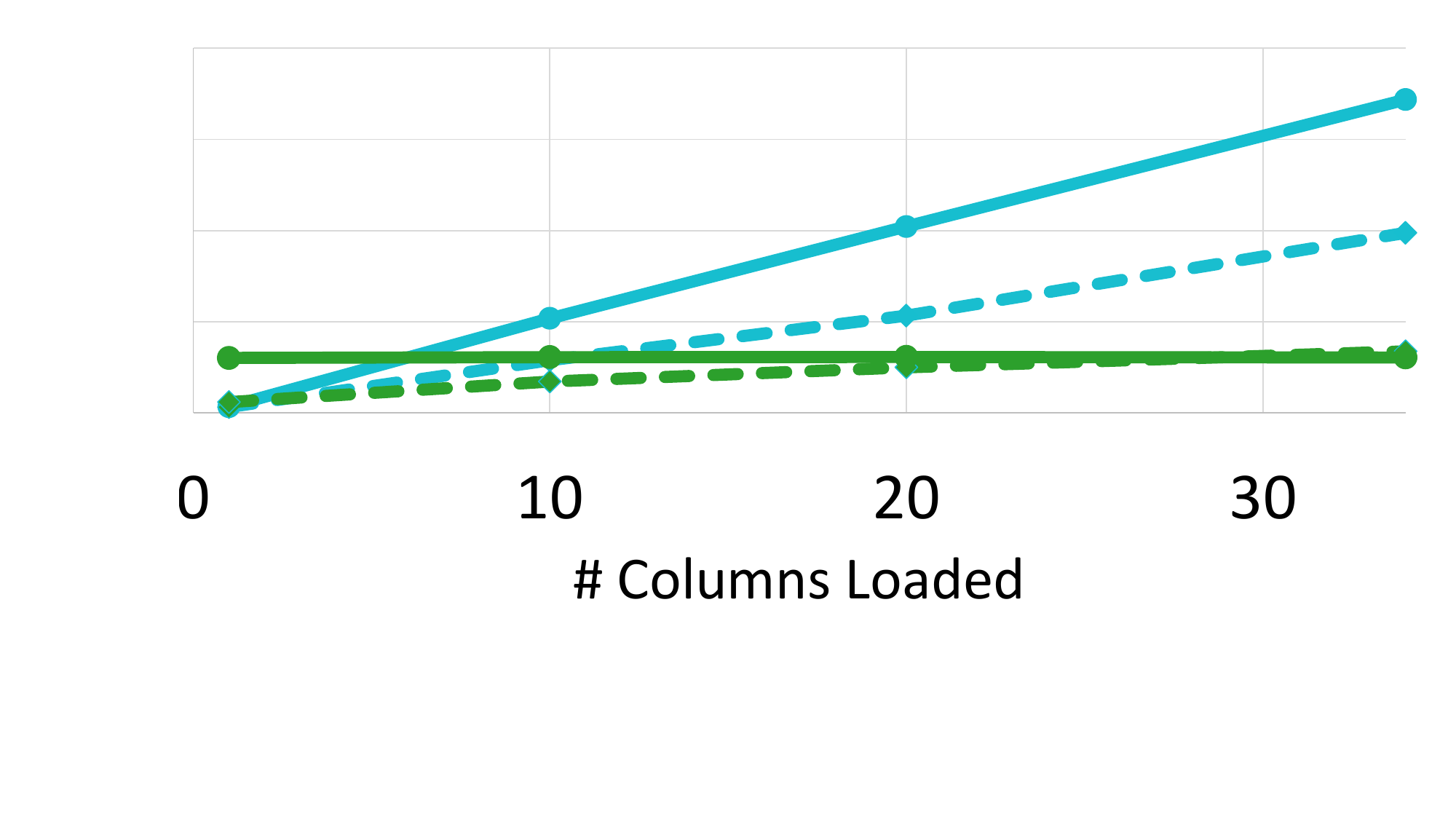}
         \vspace{1em}
         \caption{Warm}
         \vspace{-2.75em}
         \label{fig:allone-warm}
     \end{subfigure}
     \begin{minipage}[t]{0.33\textwidth}
         \includegraphics[trim={0 4.5cm 0 0},clip,width=\textwidth]{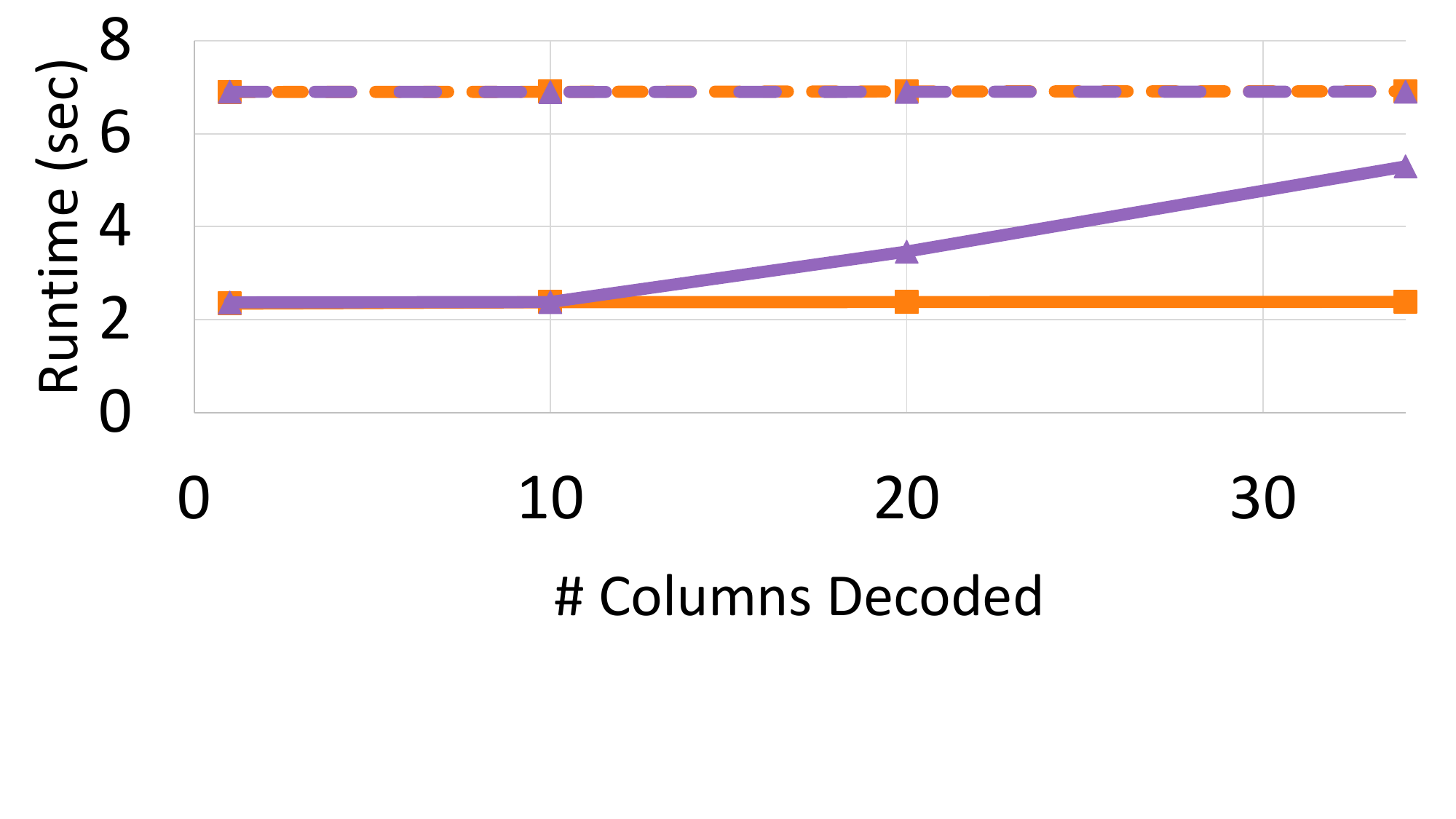}
         \\
         \hspace{2em}
        \includegraphics[trim={0 0 0 17cm},clip,width=\columnwidth]{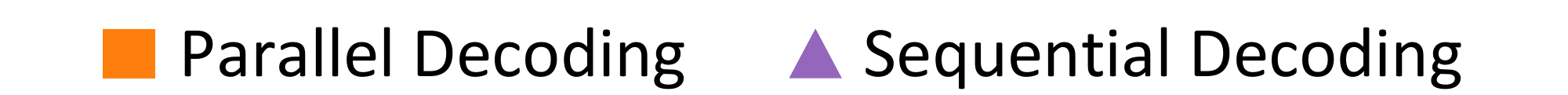}         \vspace{-2.75em}
    \end{minipage}\\
\hspace{-5em}
\includegraphics[trim={0 0 1cm 17cm},clip,width=0.9\columnwidth]{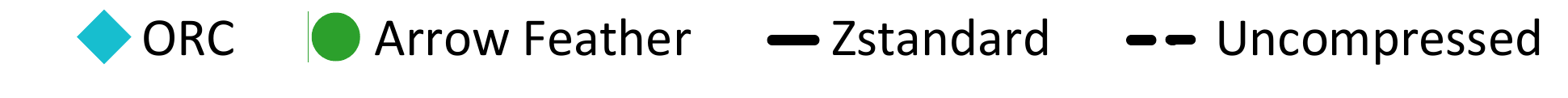}
\\
\hspace{5ex}
\begin{minipage}[t]{0.60\textwidth}
\vspace{-0.75em}
\caption{Profiling single column to full table sequential loading for ORC and Apache Arrow from the {\tt catalog\_sales} table.}
\label{fig:all_one}
\end{minipage}
\hspace{5ex}
\begin{minipage}[t]{0.30\textwidth}
    \vspace{-0.75em}
    \caption{Arrow serial vs parallel (default).}
    \label{fig:sequential_vs_parallel_load}
\end{minipage}
\vspace{-3ex}
\end{figure*}

Having explored the overheads associated with encoding, compression, and scan operations, 
we next evaluate the performance of accessing data in the context of common relational operations found near the leaves of a query plan. 
Specifically, we explore the performance of projecting columns in a dataset (\autoref{sec:eval-project}) and applying filters (\autoref{sec:eval-filter}).
In this and subsequent sections
we only consider Zstd and LZ4 compression since we evaluated the trade-offs of the other compression algorithms in \Cref{sec:eval-compression}.

\vspace{-1ex}
\subsection{Projection}
\label{sec:eval-project}

We first explore projection performance for common data types.
To do so, we: (i) load the data from disk, (ii) decompress and decode the projected columns, and (iii) convert to in-memory Arrow. 


Figures \ref{fig:projection-int} and \ref{fig:projection-double} show the respective runtimes for projecting an integer and double column in the \\{\tt catalog\_sales} table of TPC-DS.
Since {\tt catalog\_sales} does not contain any string columns, 
\Cref{fig:projection-string} instead shows the runtime of projecting a string column drawn from the {\tt customer\_demographic} table. This table is 
narrower than {\tt catalog\_sales} and has
$9$ integer and string columns. The raw data size is 80 MB with ${\sim}2$ million rows.

ORC is the most performant format for projections over integer columns because it applies RLE encoding, which yields a higher compression ratio and lower I/O costs. Parquet is slightly slower as it leverages DICT, which (i) inflates the compressed file size due to dictionary storage overhead; and (ii) slows down the loading process by introducing dictionary lookup overhead. Arrow Feather is by far the worst in this experiment since its API requires parsing the entire byte-array including all columns (not just the projected subset) from disk before column chunks loading can commence (we further discuss this in \Cref{sec:loading}). 

For doubles, Parquet offers the best performance, as we can see from
\Cref{fig:projection-double}.
In this case Parquet's DICT encoding leads to a much smaller compressed size in this dataset (which has low cardinality; see \autoref{sec:background-encoding}).
ORC does not encode doubles and is thus slightly slower than Parquet.
Arrow Feather lags far behind 
for the same reason it did with integer columns.
\eat{Finally, 
String projection runs on $customer\_demographic$ table which
contains fewer rows (1.9 million rows, 77MB for the original file) (\BH{how many rows?}) than the $catalog\_sales$ table used for integer and double experiments (14.4 million rows, 2.9GB for original file) (\BH{how many rows?}).}

\Cref{fig:projection-string} shows the results of projecting a string column.\footnote{The runtimes between the numeric and string experiments are not directly comparable
since {\tt catalog\_sales} has an order of magnitude more records (and ${\sim}5{\times}$ average row size in bytes) than does {\tt customer\_demographic}. Nevertheless, {\tt customer\_demographic} is the largest table in TPC-DS that contains a string column.}
Despite loading all columns into memory before projecting, Arrow Feather outperforms the other formats in this experiment. 
This is because by default Arrow Feather does not dictionary encode its data and is therefore able to entirely avoid the associated lookup overhead.  ORC performs slightly worse than Arrow Feather because ORC separately RLE-encodes each string's lengths, introducing additional decoding overhead. 
Parquet, however, performs the worst because its application of DICT encoding inflates projection time relative to the modest I/O cost of loading the 
{\tt customer\_demographic} table.
Finally, we want to point out that ORC is also faster than Parquet because of its API, which allows for efficiently transforming data into its dedicated in-memory representation, whereas Parquet deserializes data into memory using its relatively slow streaming style API with rudimentary data access control.

\begin{figure*}
\vspace{-0ex}
     \includegraphics[trim={0 0 0 17cm},clip,width=0.65\textwidth]{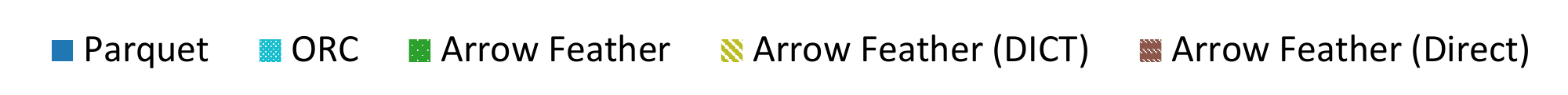}\\
     \centering
     \begin{subfigure}[b]{0.32\textwidth}
         \centering
                  \vspace{-2.8em}
         \includegraphics[trim={0 4cm 0 3.5cm},clip,width=\textwidth]{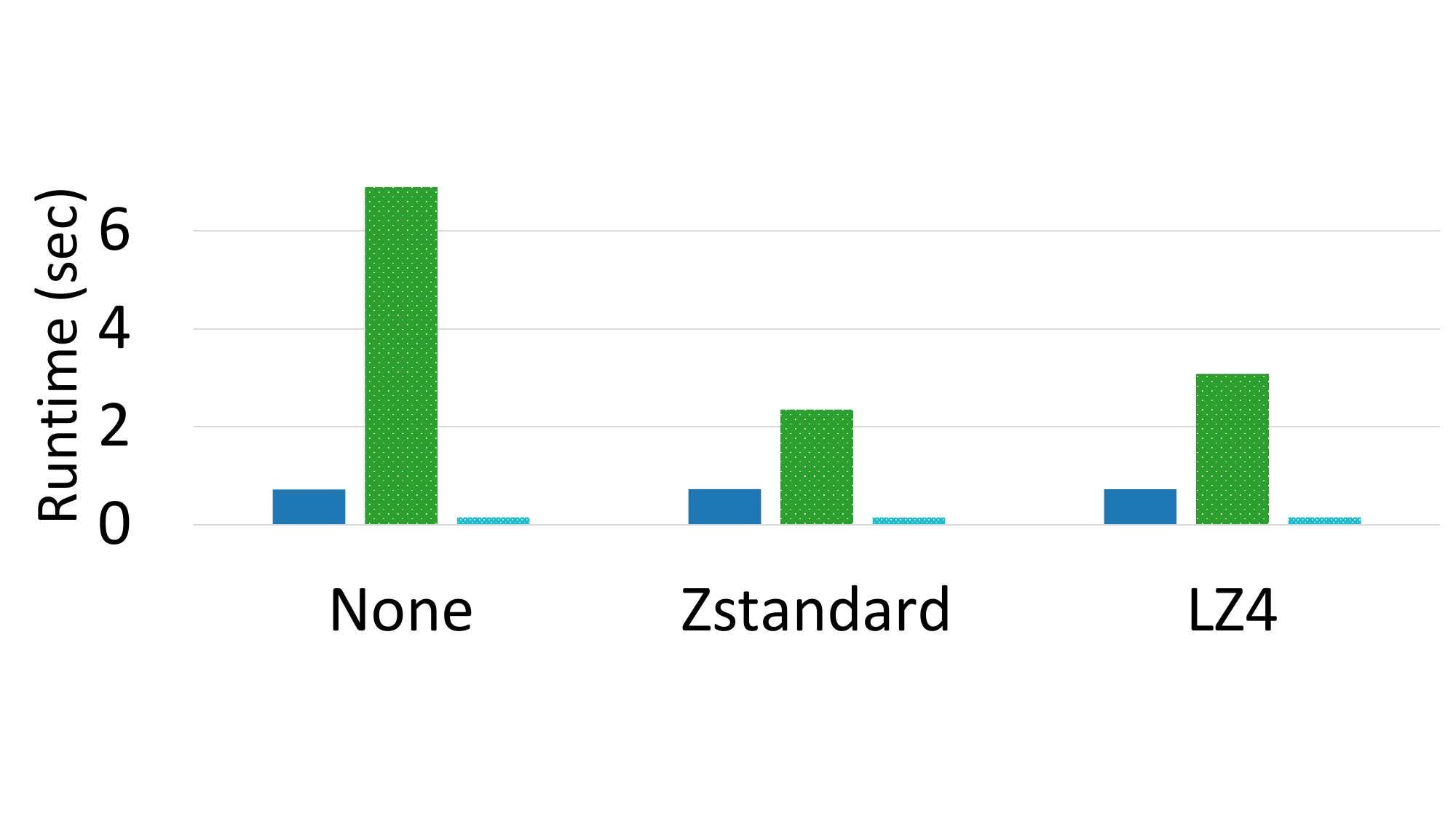}
         \caption{Integers}
         \vspace{-2.25em}
         \label{fig:int-filter}
     \end{subfigure}
     \hfill
     \begin{subfigure}[b]{0.32\textwidth}
         \centering
                  \vspace{-2.8em}
         \includegraphics[trim={0 4cm 0 3.5cm},clip,width=\textwidth]{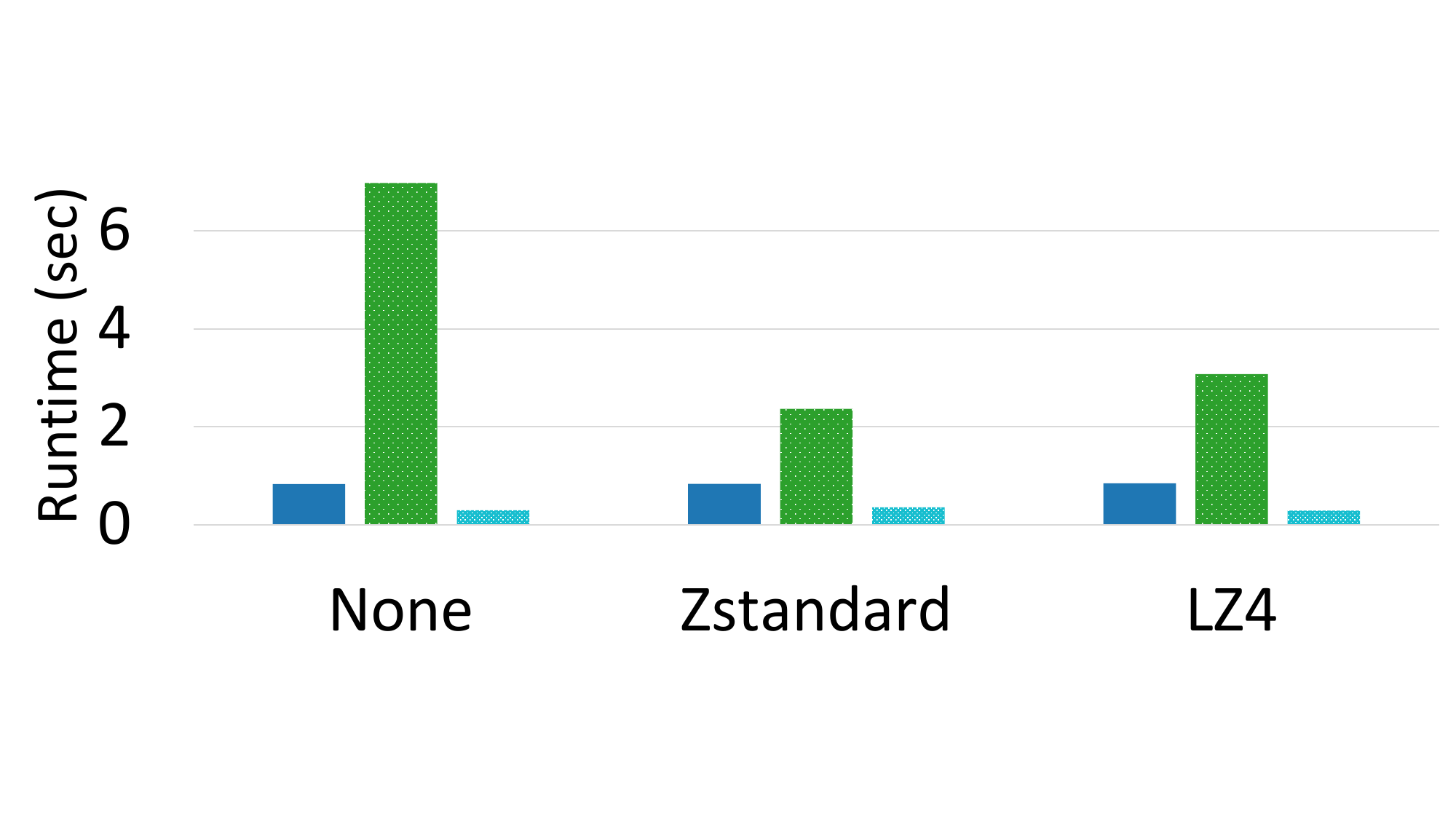}

         \caption{Doubles}
         \vspace{-2.25em}
         \label{fig:double-filter}
     \end{subfigure}
     \hfill
     \begin{minipage}[t]{0.33\textwidth}
         \centering         \includegraphics[trim={0 4cm 0 3.5cm},clip,width=\textwidth]{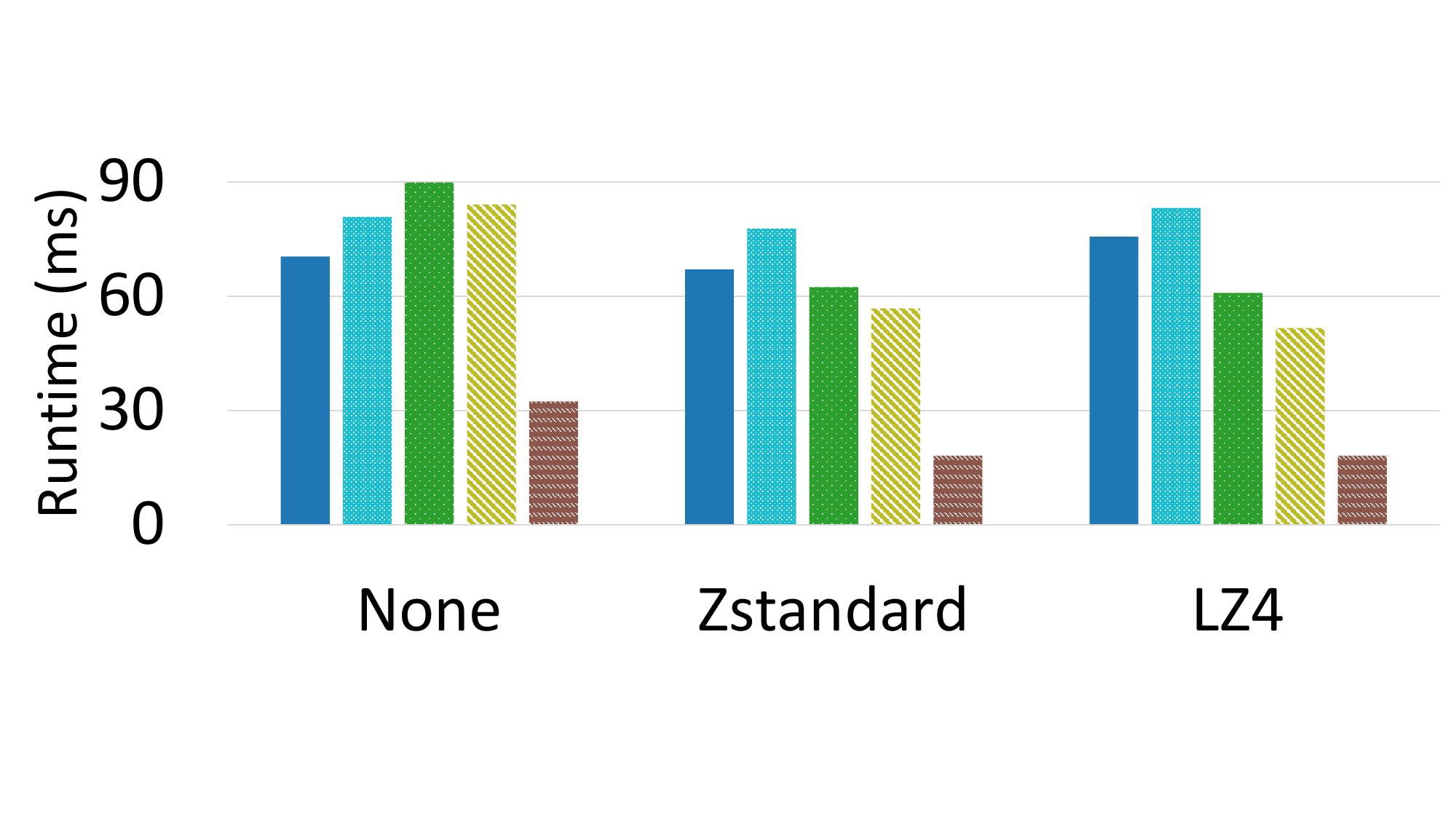}
     \end{minipage}
     \vspace{1em}
     \\
  \begin{minipage}[t]{0.66\textwidth}
    \caption{Filtering numeric types on the {\tt catalog\_sales} table.}
    \label{fig:filtering-numeric}
  \end{minipage}
  \begin{minipage}[t]{0.33\textwidth}
    \caption{Filtering strings on the {\tt customer\_demographic} table.}
    \label{fig:filtering-string}
  \end{minipage}
  \vspace{-3ex}
\end{figure*}

\subsubsection{Profiling Data Loading.} 
\label{sec:loading}
Our previous experiments show that 
Arrow Feather loading is far more expensive than ORC and Parquet. 
We observed that Arrow Feather, even when projecting a single column, requires parsing the entire byte-array.
To better understand these trade-offs, we now explore Arrow's data loading code in deeper depth and contrast it with ORC.

We begin by evaluating
the cost of loading a single column against the cost of loading the whole table.
The results are shown in 
 \Cref{fig:all_one}. 
Overall, ORC performs best when extracting a single column. Relative to Arrow Feather, this occurs for several reasons. 
First, ORC offers the ability to perform fine-grained reads at the column level 
while Arrow Feather requires 
reading, decompressing and decoding the entire row batch before projecting the target column(s).
This means that ORC's runtime is proportional to the number of columns extracted, 
while extracting one single column from an Arrow Feather file is only $2{\times}$ faster than extracting the full table. 
%
Second, upon examining the Arrow Feather deserialization logic, we observed that 
it suffers from substantial synchronization overhead when parsing column chunks within each row batch. In particular, we found that the lock acquisition step consumed ${\sim}80\%$ of the runtime for each row batch.


To better understand synchronization issues,
we finally evaluate Arrow Feather table loading using its native data loading API in both sequential and parallel execution modes. Parallel loading mode leverages a global CPU thread pool to parallelize column decompression.
This potentially results in a performance advantage relative to ORC, which serially decompresses columns.
We see this improvement in \Cref{fig:sequential_vs_parallel_load} where there is a large difference between the two modes
when compression is enabled 
(the Zstd lines in \Cref{fig:sequential_vs_parallel_load}) and no difference when compression is disabled.

\subsection{Filtering}
\label{sec:eval-filter}






 

In this section we evaluate each format's performance when applying filter operations. 
We separately consider 
predicate evaluation (\Cref{sec:eval-condition-evaluation,sec:eval-direct} respectively for numeric and string columns)
and bit-vector evaluation (\Cref{sec:eval-bitvector-evaluation}).

\vspace{-0.1em}
\subsubsection{Numeric predicates.}
\label{sec:eval-condition-evaluation}
We evaluate two predicates over the {\tt customer\_sale} table 
(with respective selectivities of $65\%$ and $30\%$): 

\begin{itemize}[leftmargin=2em]
    \item[] {\sc cs\_ship\_date\_sk >} \texttt{n} (integer column filter)
    \item[] {\sc cs\_wholesale\_cost >} 
\texttt{n} (double column filter)
\end{itemize}

For each format, we 
load the data from disk, decode the target column to its in-memory representation when available (see \autoref{sec:background-formats}), and
evaluate the predicate to generate a bit-vector $x$ (i.e., entry $x_i=\top$ when row $i$ matches the predicate). For Parquet, which does not have a dedicated in-memory representation, we use its native API and interleave decompression 
with predicate evaluation.

The results for each predicate
are shown respectively in \Cref{fig:int-filter,fig:double-filter}. 
%
%
The trends are similar: overall, ORC outperforms both Parquet and Arrow Feather for each data type and compression scheme.
Arrow Feather's performance is 
3--4$\times$ worse than Parquet in the compressed case,
but when uncompressed it further lags (to more than $7\times$)
because the file is $2\times$ larger than Parquet. This is due to the same reasons described in~\Cref{sec:loading}. 
Across all formats and for all expressions, we found that the majority of the time (i.e., ${>}90\%$) is spent on data loading and decoding, whereas the contribution of the execution of the filter condition is minimal.

\subsubsection{String predicates.}
\label{sec:eval-direct}




We next evaluate a predicate 
on a string column in the {\tt customer\_demographic} table with 14\% selectivity: 

\begin{itemize}[leftmargin=2em]
    \item[] {\sc cd\_education\_status =} \texttt{n} (string column filter)
\end{itemize}

The results are shown in \Cref{fig:filtering-string}.
Parquet is faster than ORC while Arrow with plain string encoding 
(``Arrow Feather'') is slower in the uncompressed case, but faster than both  when compression is enabled. This is because the {\tt customer\_demographic} table is small, implying that I/O is not a bottleneck for this experiment. As a result decompression dominates overall cost and Arrow Feather outperforms because it avoids the cost of decoding. 
String filtering on Parquet is faster than on ORC. In fact, ORC's bulk loading data access interface requires more string copying than Parquet because it materializes all strings into memory before filtering. Conversely, Parquet's streaming-style data access API does not require keeping all the strings in memory while filtering.
Enabling Arrow DICT encoding (``Arrow Feather (DICT)'') marginally improves performance because
Arrow decodes everything 
when loading.\footnotemark
\footnotetext{We discuss the ``Arrow Feather (Direct)'' bar in \Cref{sec:arrow-direct}.}

\subsubsection{Bit-vector evaluation.} 
\label{sec:eval-bitvector-evaluation}
The previous sections produced a bit vector mask that indicates which entries match a given predicate.
In this section we look into the performance of applying these bitmaps to produce a result. 
We start with Zstd-compressed data on disk and, for each format, 
load a column $C=\langle c_1, ..., c_n \rangle$ into an 
in-memory representation. 
We then mask $C$ using a randomly-generated bit vector $B=\langle b_1, ..., b_n \rangle$ to produce a result $R=\langle c_i \mid b_i = 1 \rangle$.

\begin{figure}
\centering
\begin{subfigure}[b]{0.49\columnwidth}
  \includegraphics[width=1.03\columnwidth]{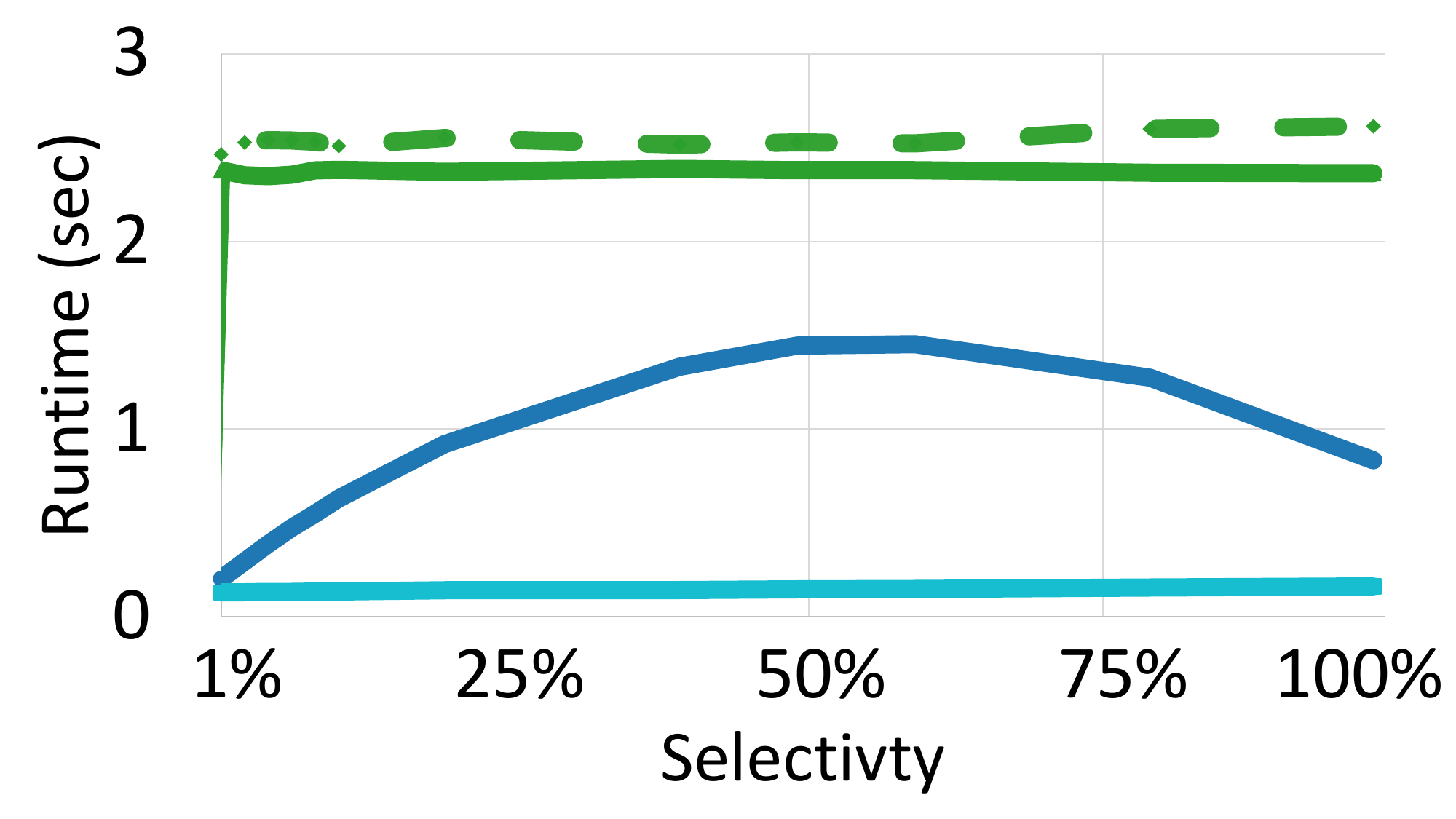}
  \vspace{1em}
  \caption{Performance by selectivity}
  \label{fig:proj-clear}
  \vspace{-3em}
\end{subfigure}
\begin{subfigure}[b]{0.49\columnwidth}
  \includegraphics[width=\columnwidth]{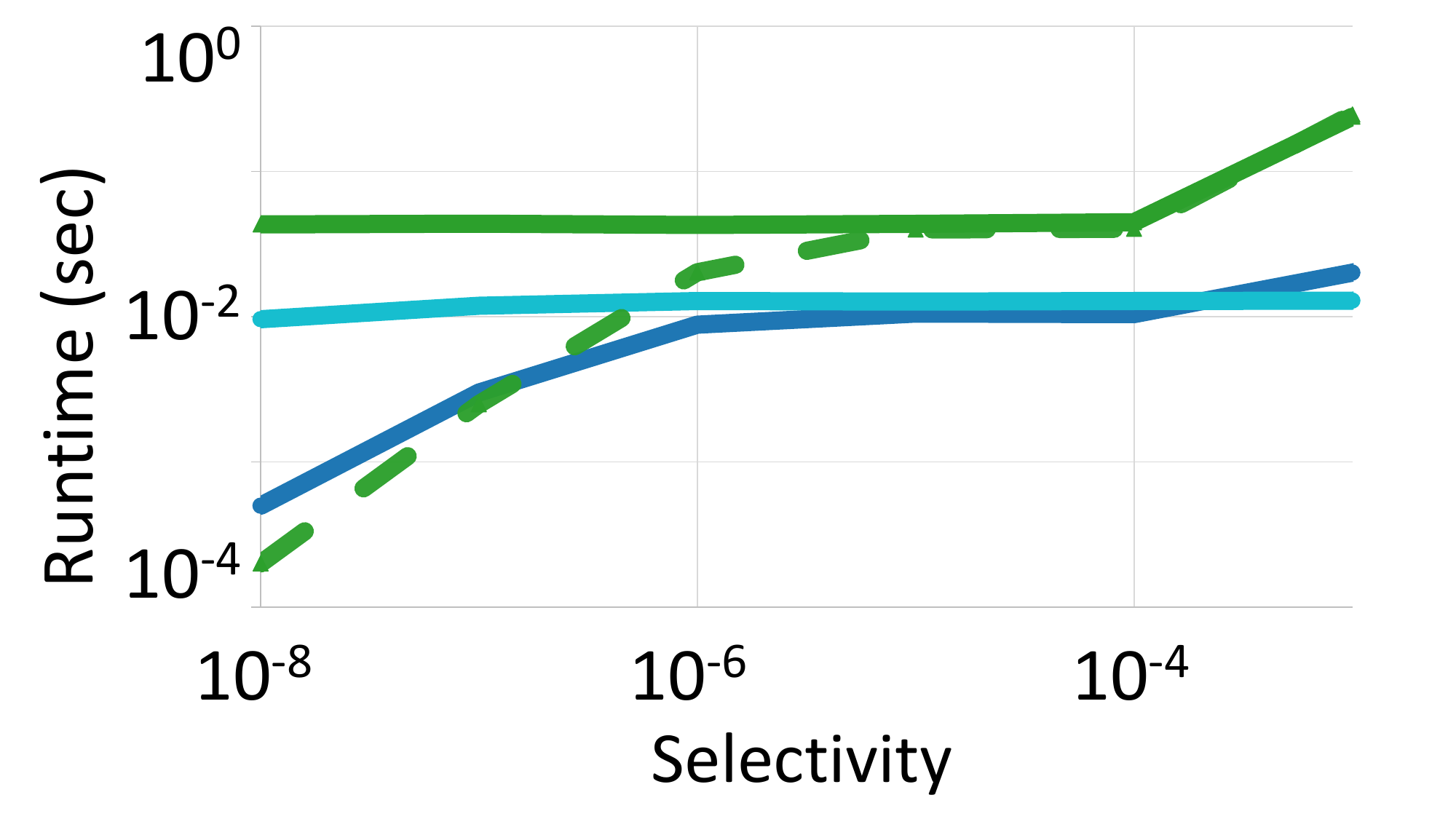}
  \vspace{1em}
  \caption{{Low selectivity values} 
  }
  \label{fig:proj-zoomin}
  \vspace{-3em}
\end{subfigure}
\includegraphics[trim={0 0 0 17cm},clip,width=0.9\columnwidth]{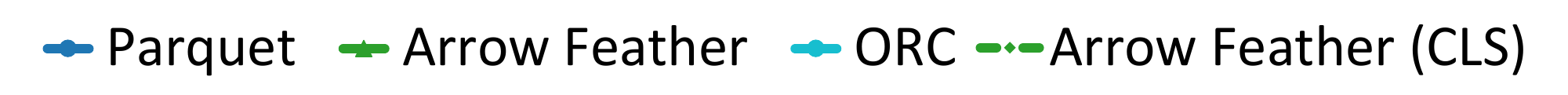}
\vspace{1em}
\caption{Bit-vector application performance by selectivity.} 
\end{figure}

\Cref{fig:proj-clear} shows performance for each format at various selectivity levels (i.e., at selectivity $s$, $\sum b_i = s \cdot n$) on the {\sc cs\_sold\_time\_sk} integer column of the {\tt catalog\_sales} table.
We can see Arrow Feather and ORC runtimes are approximately constant across all selectivity levels (though ORC is far faster). This is because both formats load all data into their in-memory structure before extracting the target records. Conversely, instead of fully loading all data, Parquet ``pushes down'' the operation by decoding only the target records that pass the filter condition.  Because of this, Parquet runtimes vary for different selectivity levels.
However, Parquet runtime is not a simple linear function of selectivity level.
Instead, we observe the highest runtime for this format at $\sim$0.5 selectivity, the point at which the largest number of 
branch mispredictions occur. 

\begin{figure*}
\centering
\begin{minipage}[t]{0.49\textwidth}
\begin{subfigure}[b]{0.49\textwidth}
\includegraphics[width=\textwidth]{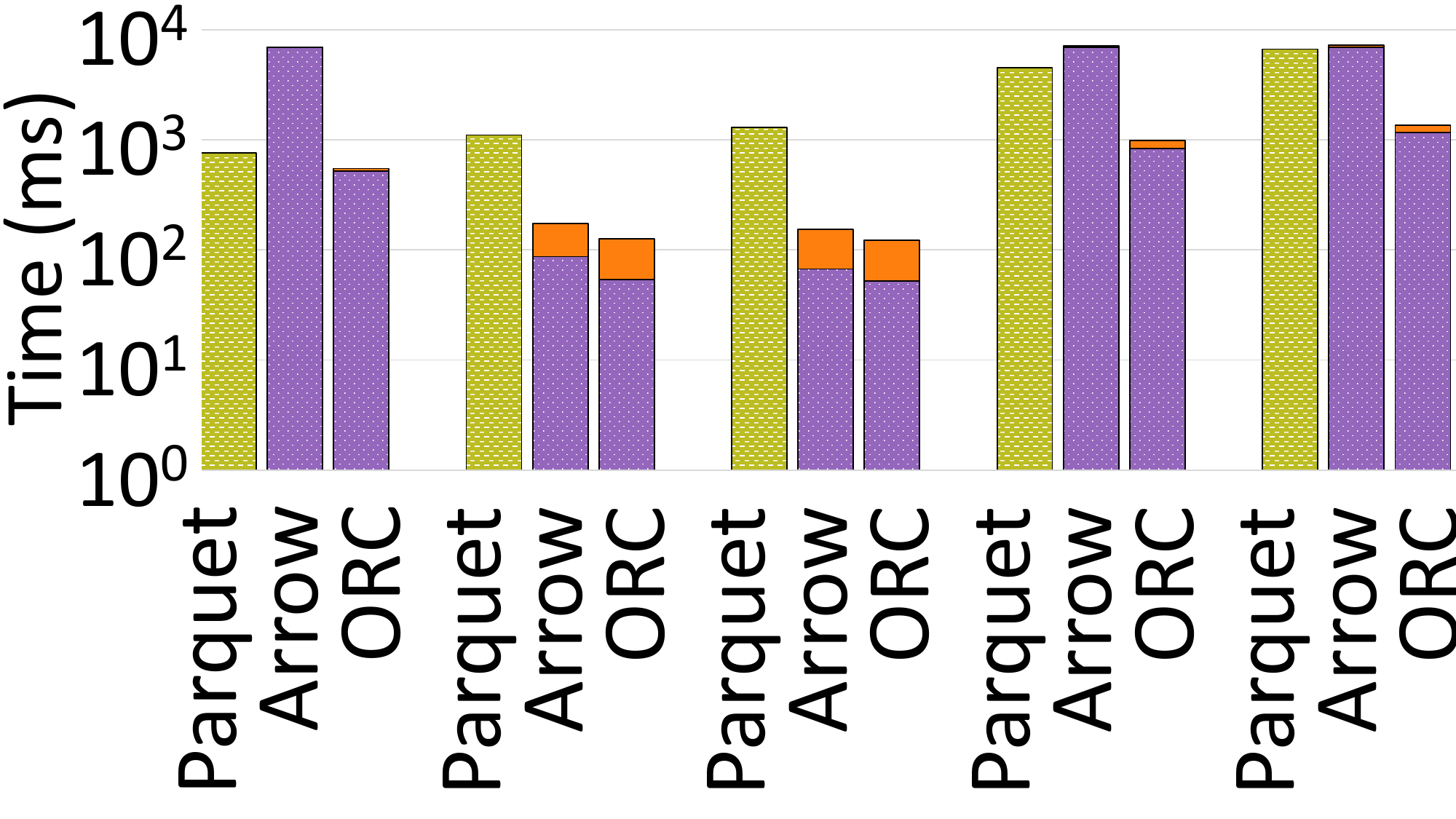}
  \includegraphics[trim={0 0 0 16.4cm},clip,width=\textwidth]{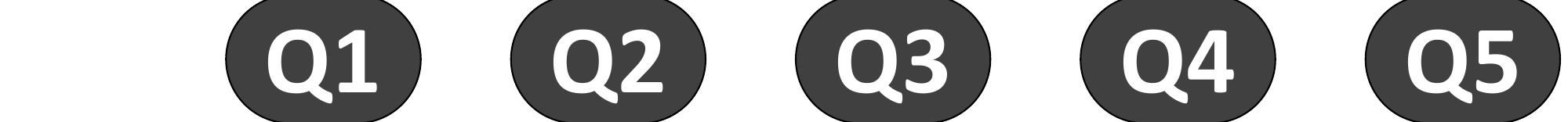}
  \vspace{0.1em}
  \caption{Uncompressed}
\end{subfigure}
\hfill
\begin{subfigure}[b]{0.49\textwidth}
  \includegraphics[width=\textwidth]{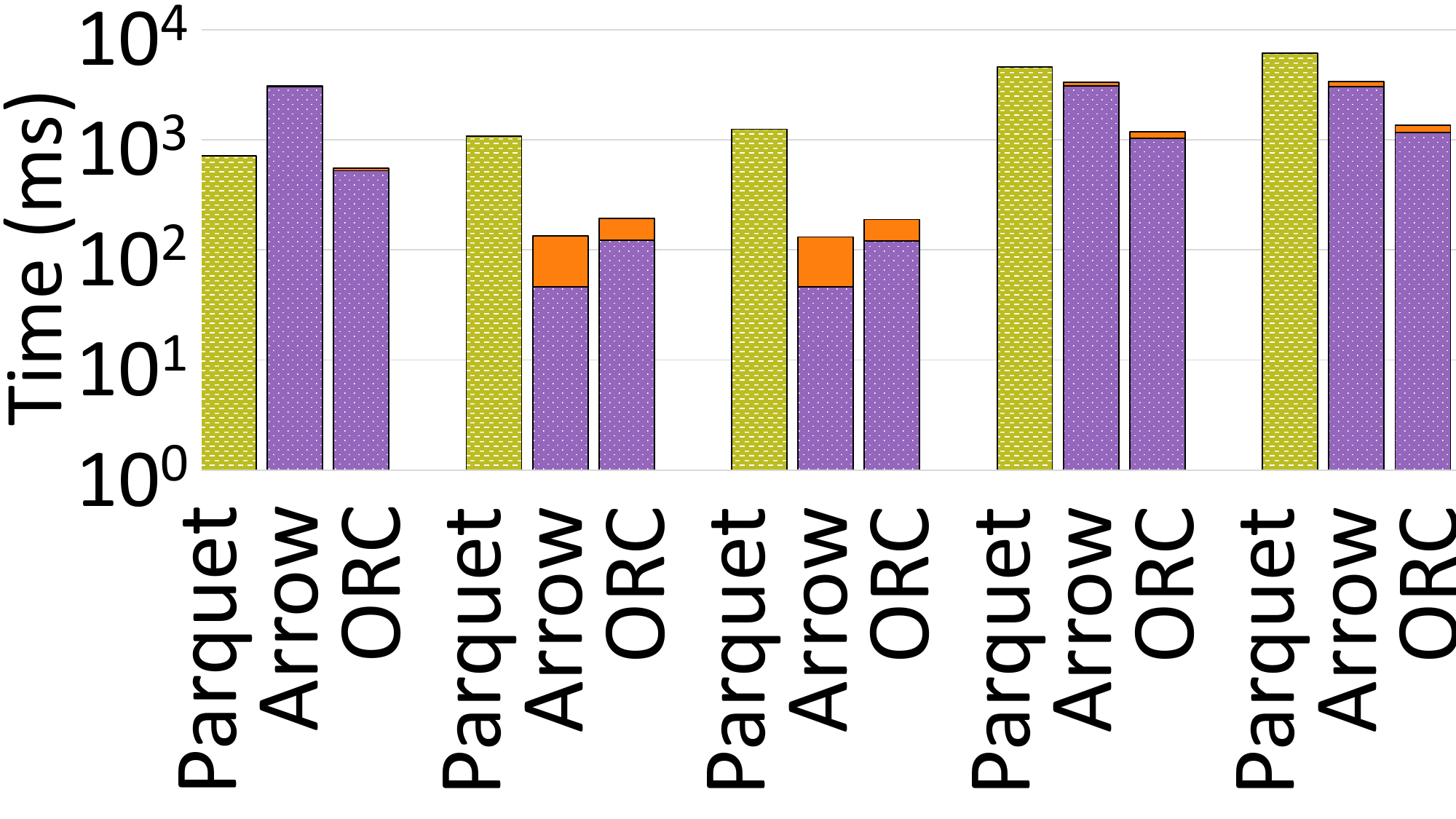}
  \includegraphics[trim={0 0 0 16.4cm},clip,width=\textwidth]{figures-formatted/figure14-labels.pdf}
  \vspace{0.1em}
  \caption{LZ4}
\end{subfigure}
\\
\includegraphics[trim={0 0 0 17cm},clip,width=\textwidth]{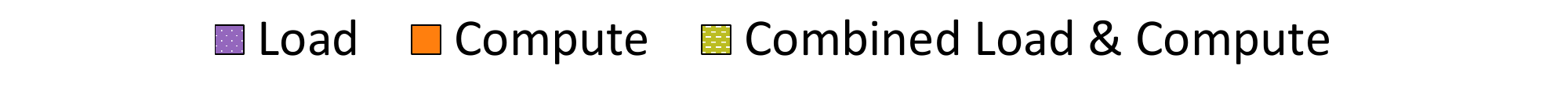}

\caption{Log-scale runtimes by format for \autoref{tab:query-stems} queries with a cold cache. }\label{fig:query_stems_runtime}
\end{minipage}
\hfill
\begin{minipage}[t]{0.49\textwidth}
\begin{subfigure}[b]{0.49\textwidth}
  \includegraphics[width=\textwidth]{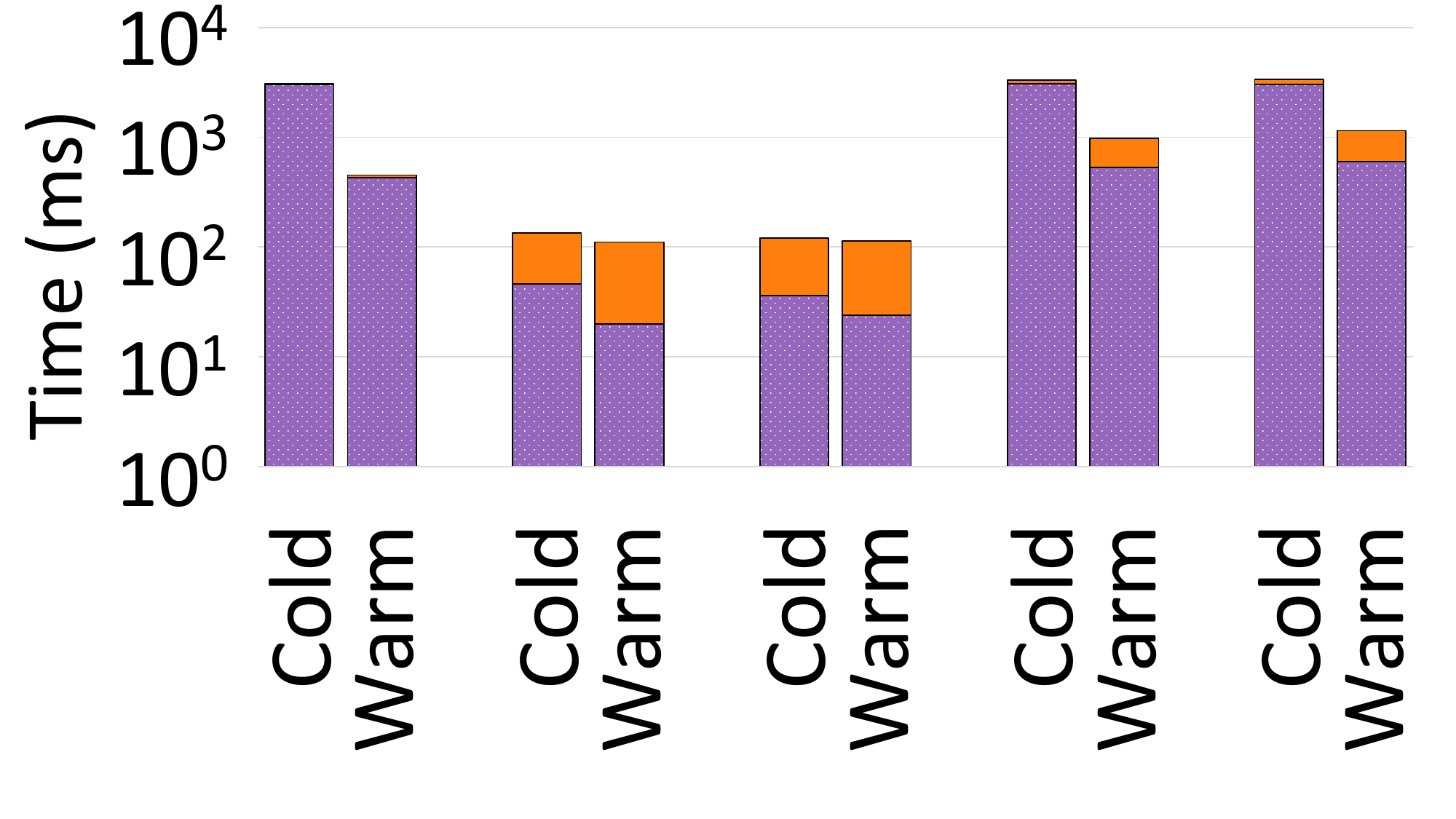}
  \includegraphics[trim={0 0 0 16.4cm},clip,width=\textwidth]{figures-formatted/figure14-labels.pdf}
  \vspace{0.1em}
  \caption{Arrow Feather}
\end{subfigure}
\hfill
\begin{subfigure}[b]{0.49\textwidth}
  \includegraphics[width=\textwidth]{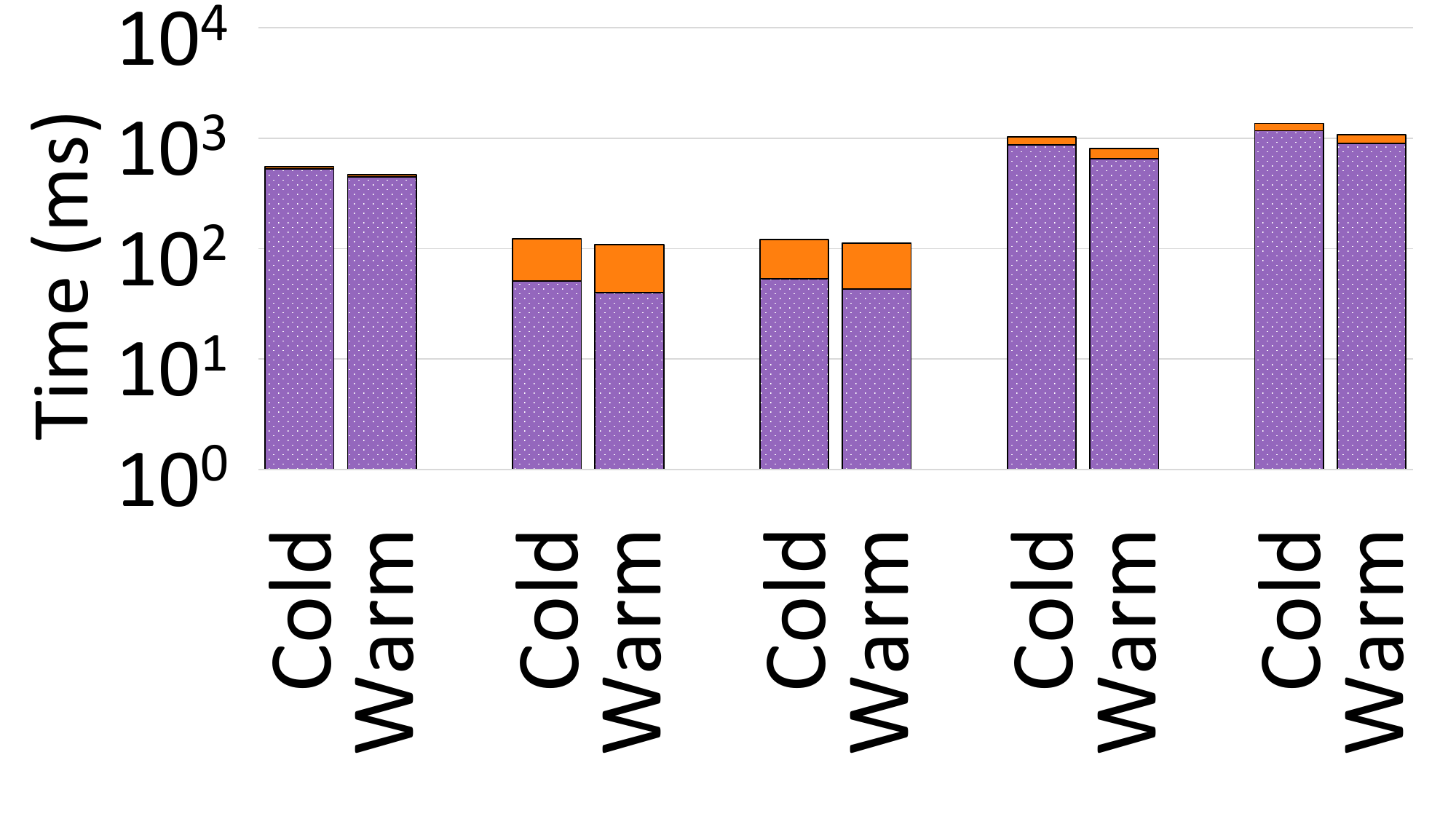}
  \includegraphics[trim={0 0 0 16.4cm},clip,width=\textwidth]{figures-formatted/figure14-labels.pdf}
  \vspace{0.1em}
  \caption{ORC}
\end{subfigure}
\\
\includegraphics[trim={0 0 0 17cm},clip,width=\textwidth]{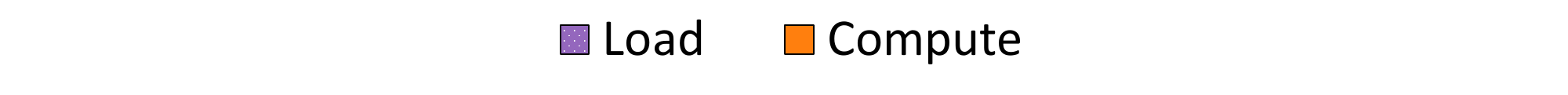}
\caption{\autoref{tab:query-stems} runtimes for cold and warm caches on LZ4 compressed data (Parquet not shown; changes were negligible).}
\label{fig:stems_cold_vs_warm}
\end{minipage}
\end{figure*}

\begin{table}
\scriptsize
\caption{Evaluated TPC-DS SP query subexpressions.}\vspace{-3ex}
\label{tab:query-stems}
\begin{tabular}{ll}
\toprule
\textbf{Q1} & \begin{tabular}[c]{@{}l@{}}SELECT cs\_ship\_date\_sk,   cs\_bill\_customer\_sk  
\\FROM catalog\_sales \\ WHERE cs\_sold\_time\_sk=12032 AND\\
\phantom{WHERE }cs\_sold\_date\_sk=2452653\end{tabular}
\\ 
\midrule
\textbf{Q2} & \begin{tabular}[c]{@{}l@{}}SELECT cd\_demo\_sk, cd\_dep\_college\_count  
\\FROM customer\_demographics \\ WHERE cd\_gender='F' AND\\
\phantom{WHERE }cd\_education\_status = 'Secondary'\end{tabular}   
\\ 
\midrule
\textbf{Q3} & \begin{tabular}[c]{@{}l@{}}SELECT cd\_demo\_sk \\FROM customer\_demographics  
\\WHERE cd\_gender =   'M' AND cd\_marital\_status = 'D' AND \\\phantom{WHERE }cd\_education\_status = 'College'
\end{tabular}                                                              \\ 
\midrule
\textbf{Q4} & \begin{tabular}[c]{@{}l@{}}SELECT cs\_ext\_sales\_price, cs\_sold\_date\_sk, cs\_item\_sk  \\FROM catalog\_sales \\ WHERE cs\_wholesale\_cost\textgreater{}80.0 AND cs\_ext\_tax \textless{} 500.0\end{tabular}                              
            \\ 
\midrule
\textbf{Q5} & \begin{tabular}[c]{@{}l@{}}SELECT cs\_ext\_sales\_price, cs\_sold\_date\_sk, cs\_item\_sk,\\
\phantom{SELECT }cs\_net\_paid\_inc\_tax, cs\_net\_paid\_inc\_ship\_tax, 
\\\phantom{WHERE }cs\_net\_profit \\ FROM   catalog\_sales WHERE cs\_wholesale\_cost \textgreater{} 80\end{tabular}
\\ 
\bottomrule
\end{tabular}
\vspace{-4ex}
\end{table}

Overall, at all but the lowest selectivities ORC performs best.
However, if we ``zoom into'' \autoref{fig:proj-clear} at the very lowest selectivities (i.e., we approach point selection) a different pattern emerges.
As shown in \Cref{fig:proj-zoomin},
at extremely low selectivity levels (i.e., $\le 0.001$), \review{Parquet performs better because it supports fine-grained record level data skipping. Conversely, ORC becomes better than Parquet at slightly higher selectivity (${\sim} 0.01$) since ORC provides a dedicated in-memory representation that efficiently loads batches. Specifically, ORC data loading consumes full data blocks, incurring extra overhead for queries with low selectivity where few entries are evaluated.  On the other hand, this cost is amortized for queries with high selectivity where more data entries pass the predicate.} We also implement an advanced Arrow variant that supports chunk level skipping (CLS), which we discuss in \Cref{sec:a_skip}.

To summarize, trade-offs exist for simple data access operations, with no format being the best in every case. Data skipping is important, but it does not always help. Record-level data access APIs provide flexibility for data skipping but has reduced performance on queries with high selectivity compared to bulk loading APIs. To improve performance, on-disk formats should be more adaptive and co-designed with an in-memory representation.

\section{Leaf Subexpression Evaluation}
\label{sec:eval-api}



In this section we bring together the previous microbenchmarks and explore how select/project (SP) subexpressions found at the leaves of a query plan can be directly evaluated on each storage format. We use the standard API provided by each format. For Parquet, we use its streaming-style API to parse, decompress and decode the data entries while, interleaved, we evaluate the queries. For Arrow Feather and ORC, we load the data into their in-memory representations before applying the query.

We select five representative SP subexpressions from TPC-DS (see \Cref{tab:query-stems}) representing a wide variety of use cases. 
 They project few (Q1, Q3) and many (Q5) columns.  They contain both equality (Q1--Q3) and range predicates (Q4--Q5). They contain predicates on integers (Q1), strings (Q2--Q3), and doubles (Q4--Q5).  Queries have low (Q5), medium (Q2--Q4), and high (Q1) selectivities. 
We execute each subexpression: (i) for each format, (ii) with and without compression (LZ4), and (iii) with and without clearing the system cache (to simulate the cases with repeated queries on the hot data, and infrequent queries on cold files, respectively).

The results are shown in \Cref{fig:query_stems_runtime} (uncompressed vs LZ4), and \Cref{fig:stems_cold_vs_warm} (warm vs cold cache).
Since, both Arrow and ORC provide a custom data loading interface (including parsing, decompressing, and decoding the data into their in-memory representations) before any query evaluation (see~\autoref{sec:background-formats}), we separately report ``Load'' and ``Query'' runtimes for each phase.
Parquet pipelines data loading and computation, so we report only the total runtime for this format.

Overall, ORC performs best in terms of query performance because of its efficient in-memory mapping representation (lower loading time for large files) and more data-skipping opportunities resulting from a smaller row batch size. With the default setting, ORC and Arrow have 14,064 and 228 batches for the {\tt catalog\_sales} table, and 1,876 and 1 batches for the {\tt customer\_}\\{\tt demographic} table, respectively. ORC's finer granularity allows it to skip more entries when no qualified item satisfies the filter condition.
Conversely, Arrow loading time dominates runtime, slowing query evaluation, as we also observed in \Cref{fig:all_one}. Parquet outperforms Arrow for Q1, Q4 and Q5 over large tables since Parquet file has a smaller size (i.e., less I/O) when loading the file. Parquet lags when compression is enabled as Arrow Feather sizes (I/O) decrease.


In \Cref{fig:stems_cold_vs_warm}, we also see that other than Q2 and Q3 (which are both evaluated on the smaller {\tt customer\_}\\{\tt demographic} table) both Arrow Feather and ORC are impacted by the system cache (significantly so for Arrow). This is because loading data from disk and building the required in-memory data structures is expensive (as we also discussed in \Cref{sec:loading}). 

In summary, smaller batches allow for more data skipping, but at the cost of space overhead and increased complexity. Workload-aware data partitioning can be used to balance these factors.
\section{Advanced Optimizations}
\label{sec:eval-opt}

\review{Given that there has been extensive discussion 
about pushing the limits of common open formats 
\cite{influxdatablog,zhang2022compressdb,agarwal2015succinct},
the goal of this section is to 
evaluate the amenability of the Arrow and Parquet formats 
to support more advanced and emerging optimizations of execution engines. }
We will first discuss optimizations related to Arrow (\autoref{sec:optimizing-arrow}) and then turn into Parquet (\autoref{sec:optimizing-parquet}). \review{
While we focus on Arrow and Parquet because of space constraints and their relatively wide adoption, similar
optimizations could also be applied to ORC.} 


\subsection{Optimizing Arrow}
\label{sec:optimizing-arrow}
We first
evaluate the feasibility and effectiveness of integrating new optimizations into Arrow: namely its ability to support direct querying
~\cite{abadi2008column,jiang2021good} (\autoref{sec:arrow-direct}),
and  data skipping (\autoref{sec:a_skip}). 
We then compare 
the performance of this hand-optimized variant 
against Gandiva~\cite{gandiva}, a LLVM-based backend for Arrow (\autoref{sec:arrow-gandiva}). 

\subsubsection{Direct Filtering over String Columns.}
\label{sec:arrow-direct}

We begin by exploring Arrow filtering pushdown (i.e., direct querying) into the encoded space for string columns (we separately evaluate a Parquet variant in \Cref{sec:eval-in-memory-parquet}).
We modify Arrow to implement direct querying as follows.  For each data chunk, we decompress and extract the dictionary from the metadata.
We then map the string constant in the query predicate from the string domain into the encoded integer domain of the extracted dictionary. 
This process allows us to: (i) transform string comparisons into integer comparisons, which can be executed efficiently
; and (ii) decode only the records admitted by the predicate.

We evaluate by repeating the experiment described in \autoref{sec:eval-direct}. 
The result is shown as the ``Arrow Feather (Direct)'' bar in \autoref{fig:filtering-string}.
Our results demonstrate a 2$\times$ to 4$\times$ improvement over other formats.
The approach could be extended to range queries by employing an order-preserving dictionary (e.g., as explored in~\cite{liu2019mostly}).
\begin{figure}
%
\begin{minipage}[t]{1\columnwidth}
  \centering
  \begin{subfigure}[b]{0.49\columnwidth}
  \includegraphics[trim={0 0 0 0},clip,width=\columnwidth]{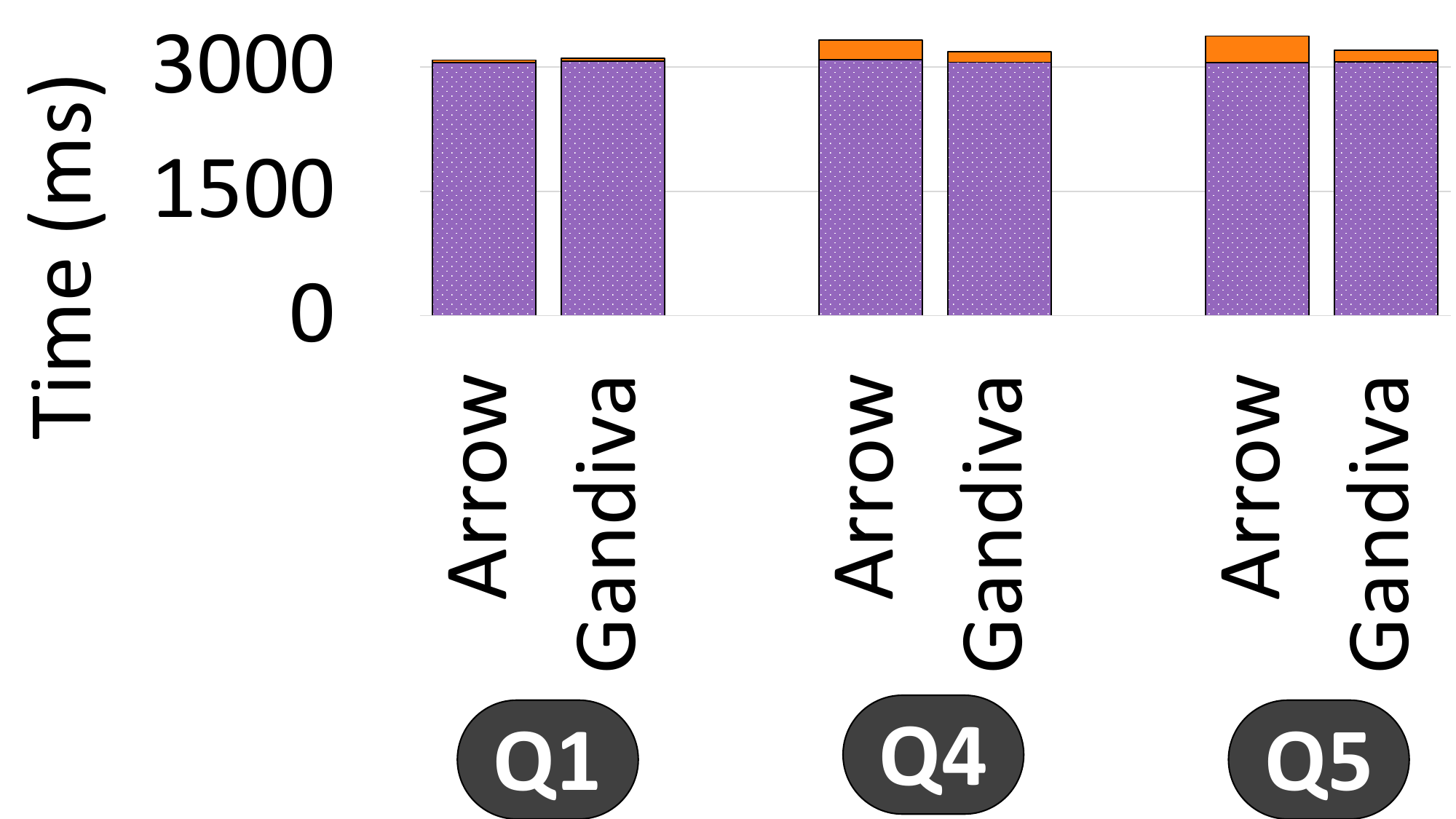}
  \caption{Numeric predicates}
  \label{fig:arrow-numeric}
  \end{subfigure}
  \hfill
  \begin{subfigure}[b]{0.49\columnwidth}
  \includegraphics[trim={0 0 0 0},clip,width=\columnwidth]{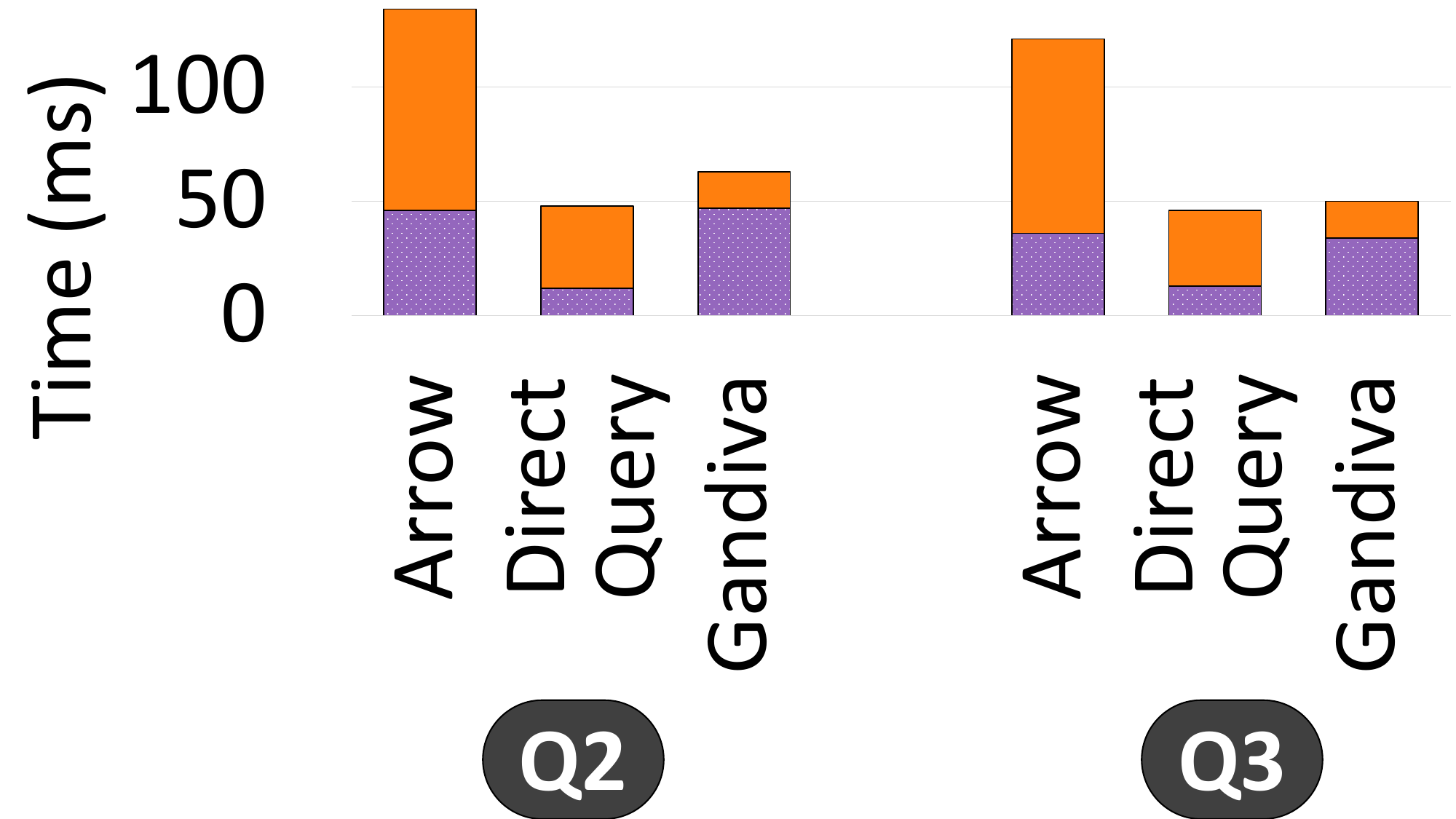}
  \caption{String predicates}
  \label{fig:arrow-strings}
  \end{subfigure}
  \\
  \includegraphics[trim={0 0 0 17cm},clip,width=\textwidth]{figures-formatted/figure15-legend}
  \caption{Arrow Feather runtime with and without direct query (warm cache execution).}
  \label{fig:arrow-direct-query}
\end{minipage}
\end{figure}
\subsubsection{Gandiva.}
\label{sec:arrow-gandiva}
Gandiva~\cite{gandiva} is an LLVM-based execution backend for Apache Arrow.  It is part of the Arrow project, and employs a number of
optimizations (e.g., vectorization) applied via LLVM compiler passes.
Gandiva further improves performance (especially for string and binary data types) by maximizing zero-copy operations.
\footnote{
The Parquet and ORC APIs could benefit from reducing or eliminating redundant string duplication.}

We test Gandiva by executing the queries evaluated in \autoref{sec:eval-api} (listed in \autoref{tab:query-stems}).  To do so, we construct expression trees for each query using \texttt{TreeExprBuilder} instances, which Gandiva transforms into machine code. 
%
As illustrated in \autoref{fig:arrow-direct-query}, 
Gandiva is able to slightly reduce computation time for all queries, while loading time (which is the dominant component of the queries with numeric predicates) remains unchanged. 
For queries with string predicates (i.e., Q2--Q3), we further compare with the optimized Arrow direct query variant described in \autoref{sec:arrow-direct}.
For these queries, the direct query variant achieves 3$\times$ speedup compared to vanilla Arrow and outperforms Gandiva as well. This is because direct query not only reduces compute time, but data loading also improves because the data decoding step is skipped. 

Interestingly, Gandiva fails to generate vectorized versions for any of the evaluated queries.
To test this aspect, we use Gandiva to execute a variant of Q4 that it was able to vectorize: 

\begin{itemize}[leftmargin=1em]
    \item[] {\sc SELECT cs\_ext\_list\_price -}
    \item[] {\sc\phantom{SELECT}
    cs\_ext\_wholesale\_cost - }
    \item[] {\sc \phantom{SELECT }cs\_ext\_discount\_amt +}
    \item[] {\sc \phantom{SELECT}
    cs\_ext\_sales\_price}
    \item[] {\sc FROM catalog\_sales}
\end{itemize}

\noindent Gandiva's use of vectorization (and other LLVM optimizations) resulted in a $1.8\times$ speedup (74ms vs 42ms) relative to normal Arrow.

Finally, we observe that Gandiva's compilation process is expensive relative to execution time.  For the queries on the smaller dataset (i.e., Q2--Q3), the
compilation time exceeded execution time (e.g., for Q2 compilation time was 103ms and runtime was 79ms).

\subsubsection{Data Skipping.}
\label{sec:a_skip}

In \autoref{sec:eval-bitvector-evaluation} we showed how Parquet achieves excellent performance for low-selectivity filters by avoiding decode overhead for unnecessary records. 
We implement a similar technique for Arrow.
Specifically,
we augmented the bulk loading API in Arrow with the data skipping approach leveraged in Parquet. To do so, we modified the Arrow Feather API to support chunk-level skipping (CLS) where we only load the column chunks necessary to answer a given query.  Because of its data layout, CLS is the most granular skipping we can employ for Arrow Feather. 

To evaluate, we repeat the experiment described in \Cref{sec:eval-bitvector-evaluation}, which requests a random set of row IDs.
We show the result 
in \Cref{fig:proj-zoomin} as the ``Arrow Feather (CLS)'' series.
As we see in the figure, this variant initially performs well but quickly degrades to perform similarly to unmodified Arrow.
This is because the input is a bit-vector with random 
row IDs:
even at extremely low selectivities, we quickly select at least one tuple per chunk, obviating any performance advantages.

\subsection{Optimizing Parquet}
\label{sec:optimizing-parquet}

In this section, we build on the lessons learned throughout the paper and augment Parquet with an efficient in-memory representation as well as vectorized instructions
(\autoref{sec:eval-in-memory-parquet}).
Finally, we put everything together and wrap up with an experimental evaluation comparing all optimizations (\autoref{sec:eval-result}).

\subsubsection{In-Memory Parquet and Vectorization.}
\label{sec:eval-in-memory-parquet}

In columnar databases, it is common to encode data for better memory and bandwidth utilization.
Therefore Parquet (and ORC) could potentially be leveraged as a useful
\textit{in-memory} data structure, without transcoding (to Arrow). 
Parquet's existing data access API either (i) fully deserializes data 
(precluding opportunities for fine-grained skipping or direct querying) or (ii) exposes record-level data access, which is 
typically much less efficient than batch loading.
OLAP systems could benefit from an 
access pattern falling in between these two.



One example of this is CodecDB~\cite{jiang2021good}, which introduced a dedicated in-memory representation for Parquet.  In CodecDB, Parquet data is lazily materialized (memory mapped) and fully decoded only when needed.
This enables support for row batch-level, column chunk-level, and record-level skipping.
Lazily-decompressed in-memory Parquet can be further optimized by leveraging vectorized instructions over in-place encoded data, 
as described in SBoost \cite{jiang2018boosting}.
\review{We implemented these optimizations based on the code as-provided
by CodecDB and SBoost. We then compare the scalar performance with direct query, vectorization, and their combinations.}


\subsubsection{Results.}
\label{sec:eval-result}
\begin{figure}
\hspace{-4ex}
  \centering
  \begin{minipage}[t]{0.5\textwidth}
  \includegraphics[trim={0 0 0 17cm},clip,width=\textwidth]{figures-formatted/figure14-legend}
    \includegraphics[width=\columnwidth]{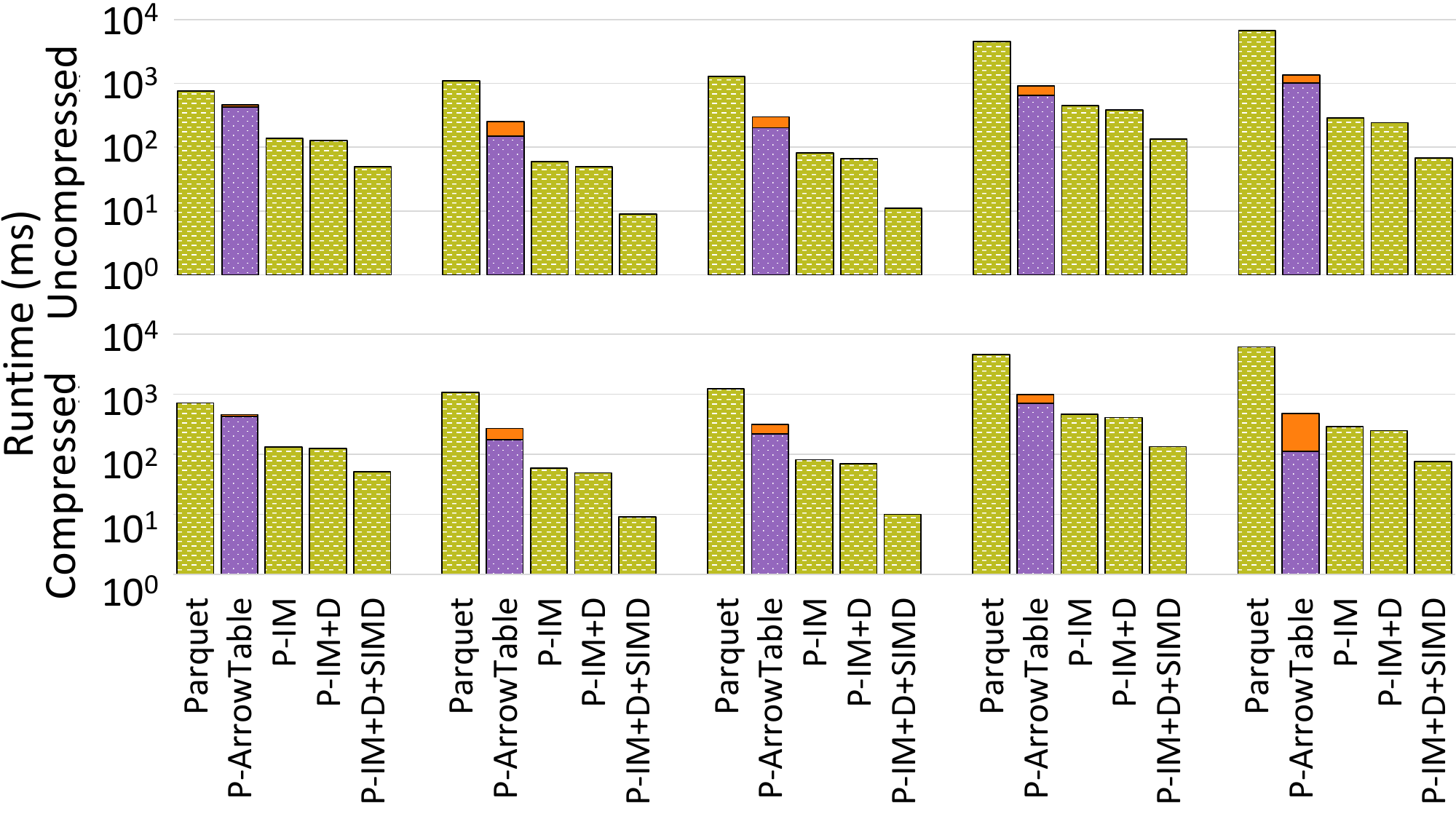}
    \includegraphics[trim={0 0 0 17.25cm},clip,width=\columnwidth]{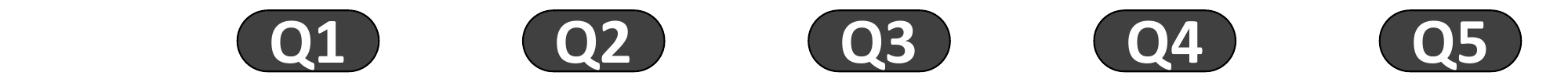}
    \vspace{-5ex}
    \caption{Parquet performance with various optimizations.}
    \label{fig:runtime-parquet}
    \vspace{-3ex}
  \end{minipage}
\end{figure}

We apply the above optimizations to Parquet and evaluate the performance over the queries listed in \Cref{tab:query-stems}. 
The results are in \Cref{fig:runtime-parquet}. As baselines, we show Parquet with its default streaming API (``Parquet'').
We also show a variant in which we load Parquet into Arrow Table before query execution (``P-ArrowTable''); we included this second baseline because in \Cref{sec:eval-api} we observed 
that loading data into an in-memory data structure can be quite efficient.
Next, ``P-IM'' is the Parquet variant augmented with the in-memory format described in \autoref{sec:eval-in-memory-parquet}.  ``P-IM+D'' further adds direct querying in the encoded domain.
Finally, ``P-IM+D+SIMD'' enables  SIMD instructions directly in the encoded domain.

As we can see from the figure, P-ArrowTable outperforms the streaming API. Nevertheless, Arrow still requires fully decoding the data into its table format, which limits the optimization opportunities. In fact, P-IM shows even better performance than P-ArrowTable, and achieves more than one order of magnitude improvement over the Parquet baseline because of its lazy materialization avoiding unnecessary decoding overhead from Parquet into Arrow.
This can be further improved with direct query and avoiding decoding; P-IM+D is up to 60$\times$ faster than the Parquet baseline.
Finally, 
\eat{
We enable the Parquet table as a dedicate in-memory Parquet data structure for Parquet (\textbf{Parquet-T}). We also apply different optimization techniques based on this in-memory Parquet format such as lazy decompression and direct query in the encoded domain for most primitive data types (\textbf{Parquet-TD}) with the support of order preserving dictionary.

As we can see from \Cref{fig:runtime-parquet}, \textbf{Parquet-A} perform better than vanilla Parquet streaming-style access API shown in \Cref{fig:query_stems_runtime}, when we use Arrow Table as a in-memory format for Parquet. Parquet-Arrow Table is overall better than Arrow, since its effective column skipping and only target columns are extracted and parsed. Parquet table is worse than Arrow for Q2 and Q3 where the query stems has filters on string columns in a small table. When we enable dedicated Parquet table (\textbf{Parquet-T}), we can see a huge query speedup because the lazily materialization of Parquet table saves a lot time for unnecessary decoding. And this could be further speeded up when enable the direct query (\textbf{Parquet-TD}) in the encoded space for Parquet table. We can see up to $60\times$ speedup for optimized scalar version of Parquet, when compared with scalar version of Arrow.

We explore the SIMD execution under the encoded domain (\textbf{Parquet-TDV}) with the in-memory Parquet format. 
}
if we enable SIMD execution
(AVX\_512) 
we observe up to $100\times$ speedup compared with the Parquet baseline. 

Overall, there is huge potential for query speedup when we push the query operator further down into the on-disk format and encoded domain, which demonstrates the feasibility of query push-down in the storage format when augmented with a corresponding optimized in-memory representation.




\section{Related Work}

\stitle{Columnar data format trade-offs.}
There has been substantial recent discussion both online (e.g., ~\cite{abadiArrow,wesarrow} and in the research community~\cite{albis,ivanov2020impact} about the benefits and drawbacks of having multiple, often-overlapping open formats for representing columnar data.
Differently than~\cite{albis} our goal is not to propose a new format.
Differently than~\cite{ivanov2020impact} and other SQL-over-Hadoop evaluations~\cite{10.14778/2732977.2733002,8258260} our goal is not to evaluate the end-to-end performance of big data systems, but rather to understand how these format can be leveraged as native formats in analytical DBMSs (and how close they conform to columnar RDBMSs standards)~\cite{abadi2008column,vertipaq}. 
Our takeaways are consistent with the observations of Abadi~\cite{abadiArrow}. 
Nevertheless, we provide an updated view of the trade-offs in these formats. Interestingly, 
we find that several limitations described in \cite{abadiArrow} persist to this day, despite their being highlighted more than six years ago.


\stitle{Other encodings.} In addition to the encoding methods discussed in \autoref{sec:background-encoding}, \emph{delta encoding} is a basic encoding supported by many columnar data formats. Delta encoding works on integer data. Because
differences between adjacent numbers are generally smaller than the numbers themselves, delta encoding greatly reduces data redundancy. Delta encoding works best when the numbers are large, but the value range is small. Direct querying on delta encoded data is challenging because sequential decoding is required to recover a specific record. Delta encoding variations, such as FOR and PFOR~\cite{lemire2015decoding} use a fixed reference
value instead of the previous value, which better serves the direct query on the encoded format. Even though Parquet and ORC support these delta-like encodings, the formats never elected to employ them in our experiments, presumably due to reduced performance or suboptimal encoding selection. 

\stitle{\rev{Text formats.}}
\rev{In addition to the aforementioned binary file formats, text-encoded formats such as CSV and JSON remain highly-utilized.  Such formats offer the advantage of being human-readable,
but are far more limited in terms of compression. 
A comma-separated values (CSV) \cite{csv} file is a delimited text file where each line represents a data record, 
CSV is a common data exchange format supported by many data applications and systems.
Conversely,
JSON \cite{json} 
interleaves schema and data. 
JSON supports nestable and complex data structures.
This interleaving better-supports nesting and complex data layouts, whereas it is less space-efficient relative to CSV because of this schema overhead.
}

\stitle{Binary formats.}
Apache CarbonData \cite{carbondata} is an indexed columnar data format for analytics. Similar to Parquet, it uses compression and encoding to improve efficiency.
Apache Avro \cite{vohra2016apache} is a row-oriented storage format. It stores schema as JSON in the file header, making it an excellent choice for schema evolution tasks and write-heavy data operations such as whole-row consumption and processing. Avro also supports serialization and 
block level compression.
We did not evaluate these formats in the paper since they are either not columnar or rarely employed by OLAP systems.

\stitle{\rev{Domain-specific binary formats.}}
\rev{
Serialization formats such as Pickle, PyTorch's PT, and NumPy's NPY, are also widely used in domain-specific contexts. 
These formats are particularly valuable in Python-based applications and machine learning workflows. 
Pickle is a Python-centric serialization library used to transform Python object instances into a byte stream (and vice versa), which is useful for saving and loading data structures or models directly in Python ~\cite{pickle}. PyTorch's PT format is used to save trained models and their parameters, typically as binary files that contain a model's architecture and weights~\cite{ptformat}. NumPy's NPY format is employed to save arrays of data, storing the array's data, shape, and datatype efficiently~\cite{numpy}. Lance is a recent format introducing specialized optimizations for handling large-blob columns in machine learning workflows, enhancing both random access and scan efficiencies in such specialized data applications \cite{lance}. Lance is built on top of Apache Arrow. Finally, Hierarchical Data Format 5 (HDF5) is a flexible format capable of handling a wide range of storage needs and is extensively used across various scientific disciplines due to its robust handling of large and multidimensional datasets ~\cite{hdf5}.}

\stitle{Other optimizations.}
The latest version of Parquet introduces the concept of an index page that contains statistics for data pages and can be used to skip pages when scanning data in ordered and unordered columns \cite{parquetindex}.
Even with the newly added index page to facilitate the access of data entries, Parquet's random-access operations are still costly in general compared with Arrow.
Velox~\cite{velox} is a recent effort trying to unify the execution layer across analytical engines. As also highlighted in this paper, Velox recognizes that Apache Arrow limited support for encodings is not a good fit for performant analytical engines. Velox therefore proposes improvements on top of Arrow for addressing this gap.
It also highlights the need for a smart partition policy where data layout is organized for a given task (e.g., so that many data blocks can be skipped). 

Other recent workload-driven data partition approaches such as Qd-tree \cite{yang2020qd}, SDCs \cite{madden1self}, Jigsaw \cite{kang2021jigsaw} and Pixels \cite{bian2022pixels} address the partition problem by focusing on data access patterns. 
Differently, in this paper we focus on
workload-agnostic columnar formats without considering any workload-specific techniques. 
Workload-driven data partitioning is an exciting area (further motivated by our experimental results) and could be a promising direction for future work. The co-design of query engine and columnar formats should take workload-driven partitioning into consideration to enhance query performance (e.g., through efficient data skipping and improved compression).
\vspace{-0.8em}




\section{Conclusion}
\vspace{-0.2em}
In this paper, we evaluated three widely-used open columnar formats.  \rev{Our systematic assessment employed micro-benchmarks that included evaluating compression performance across real-world scalar and vector datasets, basic database operations, and end-to-end query subexpression execution.}  We found trade-offs that make each format more or less suitable for use as an internal format in a DBMS. By applying various optimizations, we identified opportunities to more holistically co-design a unified in-memory and on-disk data representation for better query execution in modern OLAP systems.




\balance
\bibliographystyle{abbrv}
\bibliography{ref} 

\end{document}